\documentclass[onecolumn,showpacs,preprintnumbers,elsart]{revtex4}
\usepackage{makeidx}
\usepackage{amssymb}
\usepackage{amsmath}
\usepackage{mathrsfs}
\usepackage{graphicx}
\usepackage{dcolumn}
\usepackage{bm}
\usepackage[center]{subfigure}
\usepackage{color}

\begin{document}

\title{Matter-wave solitons supported by field-induced dipole-dipole
repulsion with a spatially modulated strength}
\author{Yongyao Li$^{1,2,3}$, Jingfeng Liu$^{1}$, Wei Pang$^{4}$, and Boris
A. Malomed$^{2}$}
\email{malomed@post.tau.ac.il}
\affiliation{$^{1}$Department of Applied Physics, South China Agricultural University,
Guangzhou 510642, China \\
$^{2}$Department of Physical Electronics, School of Electrical Engineering,
Faculty of Engineering, Tel Aviv University, Tel Aviv 69978, Israel\\
$^{3}$Modern Educational Technology Center, South China Agricultural
University, Guangzhou 510642, China\\
$^{4}$ Department of Experiment Teaching, Guangdong University of
Technology, Guangzhou 510006, China.}

\begin{abstract}
We demonstrate the existence of one and two-dimensional \emph{%
bright} solitons in the Bose-Einstein condensate with
\emph{repulsive} dipole-dipole interactions induced by a combination
of dc and ac polarizing fields, oriented perpendicular to the plane
in which the BEC is trapped, assuming that the strength of the
fields grows in the radial ($r$) direction faster than $r^{3}$.
Stable tightly confined 1D and 2D fundamental solitons, twisted
solitons in 1D, and solitary vortices in 2D are found in a numerical
form. The fundamental solitons remain robust under the action of an
expulsive potential, which is induced by the interaction of the
dipoles with the polarizing field. The confinement and scaling
properties of the soliton families are explained analytically. The
Thomas-Fermi approximation is elaborated for fundamental solitons.
The mobility of the fundamental solitons is limited to the central
area. Stable 1D even and odd solitons are also found in the setting
with a double-well modulation function, along with a regime of
Josephson oscillations.
\end{abstract}

\pacs{03.75.Lm; 42.65.Tg; 47.20.Ky; 05.45.Yv}
\maketitle

\section{Introduction and the setting}

The transition of ultracold dipolar atomic gases into the Bose-Einstein
condensate (BEC) has been demonstrated in chromium \cite%
{Griesmaier1,Beaufils2}, dysprosium \cite{LuM3}, and erbium \cite{Aikawa4}.
Also promising for experiments in this direction is the use of CO \cite%
{Bethlem5}, ND$_{3}$ \cite{Hendrick6}, and OH \cite{Bochinski7,VDMeerakker8}
molecular gases. Unlike the usual contact nonlinearity, which represents
effects of collisions between atoms, dipole-dipole interactions (DDIs) give
rise to long-range anisotropic forces. The DDIs account for a number of
remarkable phenomena in ultracold Bose gases \cite{review0}-\cite{review10},
such as various pattern-formation scenarios \cite%
{Saito9,Nath10,Maluckov11,Klawunn12,Lakomy13}, fractional domain walls \cite%
{Larson}, \textit{d}-wave collapse \cite{Lahaye14,Metz15}, specific
possibilities for precision measurements \cite%
{Vengalattore16,Fattori17,Pollack18}, stabilization of the dipolar BEC by
optical lattices \cite{Muller19,Cuevas22}, the Einstein - de Haas effect
\cite{Gawryluk20}, \textit{etc}. Dipolar BECs can be also used as \textit{%
matter-wave simulators }\cite{simulator}, to emulate, in particular, the
creation of multi-dimensional solitons via the nonlocal nonlinearity---a
subject which has also drawn much attention in optics, where nonlocal
interactions of other types (with different interaction kernels) occur too
\cite{Conti,optics,Desyatnikov}. In fact, the dipolar condensates not only
emulate the situation known in optics, but also make it possible to predict
the existence of solitons with novel properties. Recently, one- and
two-dimensional (1D and 2D) fundamental and vortical solitons in dipolar BEC
have been predicted in various continuous and discrete settings \cite%
{Cuevas22,Sinha21,Pedri23,Tikhonenkov24,Lashkin25,Tikhonenkov26,Koberle27,Eichler28,Nath29,Nath30,LYY31,Zhihuan}%
. A similar mechanism can create 1D solitons in the Tonks-Girardeau gas with
attractive DDIs between particles \cite{TG}.

The formation of bright solitons, which was previously demonstrated in BEC\
experimentally \cite{Randy}, and studied in detail theoretically \cite%
{soliton-theory1,soliton-theory2}, requires the presence of self-attraction.
However, in models with local interactions it has been recently demonstrated
that \emph{bright} solitons may be supported by the \emph{repulsive} cubic
nonlinearity in the $D$-dimensional geometry, provided that the nonlinearity
strength is modulated in space, growing from the center to periphery at any
rate faster than $R^{D}$, where $R$ is the radial coordinate \cite{Olga}-%
\cite{B4}. A similar result was obtained for the local self-repulsive
quintic nonlinearity, in which case the nonlinearity strength must grow
faster than $r^{2D}$ \cite{B5}. A generalization for bright solitons in the
1D optical model with a self-defocusing nonlocal thermal nonlinearity, whose
strength grows at $|x|\rightarrow $ $\infty $ through the corresponding
modulation of the density of absorbing dopants, was very recently elaborated
in Ref. \cite{Yingji}.

The use of the spatially profiled repulsive nonlinearities for the creation
of multidimensional solitons is more than an exploration of an exotic
possibility. Indeed, 2D and 3D solitons supported by usual self-attractive
cubic terms are subject to the instability against the critical or
supercritical collapse, which makes their stabilization a great challenge
\cite{Wise}. In the case if the self-repulsion, the collapse is ruled
out---in fact, the fundamental 2D solitons and simplest vortices are
automatically stable in that case, if they exist \cite{Olga}-\cite{B4}.

The subject of the present work is to predict the creation of stable bright
solitons in nearly-2D or 1D dipolar condensates, which are trapped,
respectively (by means of an appropriate optical potential), in a thin layer
close to $z=0$ (or in a ``cigar" around axis $x$), with the local strength
of the \emph{repulsive} DDI growing fast enough at $r\equiv \sqrt{x^{2}+y^{2}%
}\rightarrow \infty $ (or at $|x|\rightarrow \infty $). This situation can
be implemented in the case when the atoms or molecules do not carry
permanent electric or magnetic dipole moments, but rather ones induced by
external electric or magnetic field \cite{induced1}-\cite{induced5}. To the
best of our knowledge, the formation of solitons or other nonlinear modes in
the gas of dipoles induced by inhomogeneous external fields was not
investigated previously in any setting.

We consider a combination of dc and ac external fields directed along the $z$
direction:
\begin{equation}
\mathbf{G}(r)=F(r)\left[ f_{\mathrm{dc}}+f_{\mathrm{ac}}\cos \left( \omega
t\right) \right] \mathbf{e}_{z}.  \label{G}
\end{equation}%
Then, the the local dipolar moment $\mathbf{g}=g(t)\mathbf{e}_{z}$ of the
atom or molecule is determined by the intrinsic equation of motion,
considered here in the classical approximation \cite{dynamical}:\textbf{\ }%
\begin{equation}
\ddot{g}\mathbf{+}\omega _{0}^{2}g+\gamma \dot{g}=F(r)\left[ \lambda (0)f_{%
\mathrm{dc}}+\lambda (\omega )f_{\mathrm{ac}}\cos \left( \omega t\right) %
\right] ,  \label{dyn}
\end{equation}%
where $\omega _{0}$\textbf{\ }is the intrinsic eigenfrequency and $\gamma $\
is the damping coefficient, $\lambda (0)$\ and $\lambda (\omega )$ being
effective static and dynamical susceptibilities. We will also consider a
model combining permanent and induced moments, see Eq. (\ref{d}) below.

In the off-resonance situation, when ac frequency, $\omega $, is not too
close to $\omega _{0}$, the small dissipative term in Eq. (\ref{dyn}) may be
neglected, which gives rise to an obvious solution,%
\begin{equation}
g_{\mathrm{off}}(r,t)=F(r)\left[ \frac{\lambda (0)}{\omega _{0}^{2}}f_{%
\mathrm{dc}}+\frac{\lambda (\omega )}{\omega _{0}^{2}-\omega ^{2}}f_{\mathrm{%
ac}}\cos \left( \omega t\right) \right] .  \label{g}
\end{equation}%
On the other hand, the ac drive close to the resonance yields%
\begin{equation}
g_{\mathrm{res}}(r,t)=\frac{\lambda (\omega _{0})}{\gamma \omega _{0}}%
F(r)\sin \left( \omega _{0}t\right) .  \label{sin}
\end{equation}%
These results lead to the following time-averaged DDI strengths,%
\begin{eqnarray}
\left\langle g_{\mathrm{off}}(r_{1},t)g_{\mathrm{off}}(r_{2},t)\right\rangle
&=&F(r_{1})F(r_{2})\left[ \frac{\lambda ^{2}(0)}{\omega _{0}^{4}}f_{\mathrm{%
dc}}^{~2}+\frac{\lambda ^{2}(\omega )}{2\left( \omega _{0}^{2}-\omega
^{2}\right) ^{2}}f_{\mathrm{ac}}^{~2}\right] ,  \label{offoff} \\
\left\langle g_{\mathrm{res}}(r_{1},t)g_{\mathrm{res}}(r_{2},t)\right\rangle
&=&F(r_{1})F(r_{2})\frac{\lambda ^{2}(\omega _{0})}{2\gamma ^{2}\omega
_{0}^{2}}.  \label{resres}
\end{eqnarray}

In addition to the DDIs, in the off-resonance situation the field-induced
dipole moments give rise to the effective averaged potential of the
dipole-field interaction:%
\begin{equation}
V(r)=-\left\langle \mathbf{g}_{\mathrm{off}}\mathbf{\cdot G}\right\rangle
=-F^{2}(r)\left[ \frac{\lambda (0)}{\omega _{0}^{2}}f_{\mathrm{dc}}^{~2}+%
\frac{\lambda (\omega )}{2\left( \omega _{0}^{2}-\omega ^{2}\right) }f_{%
\mathrm{ac}}^{~2}\right] \equiv -\chi F^{2}(r),  \label{exp}
\end{equation}%
where $\chi $ is the effective average polarizability. On the contrary, in
the resonant situation the substitution of expression (\ref{sin}) yields $%
V(r)=0$. With the spatially growing modulation function $F(r)$, potential (%
\ref{exp}) is \emph{expulsive} ($\chi >0$), at $\omega ^{2}<\Omega ^{2}$,
with $\Omega ^{2}$ defined by equation%
\begin{equation}
\frac{\Omega ^{2}}{\omega _{0}^{2}}=1+\frac{\lambda (\Omega )}{2\lambda (0)}%
\frac{f_{\mathrm{ac}}^{~2}}{f_{\mathrm{dc}}^{~2}},  \label{Omega}
\end{equation}%
and trapping ($\chi <0$) at $\omega ^{2}>\Omega ^{2}$. Obviously, the
expulsive potential (EP) hampers the possibility of inducing self-trapping
of localized modes, while the trapping one makes it rather trivial. Below,
we chiefly focus on the setting with the self-trapping determined by the
DDIs in the ``pure" form, when EP (\ref{exp}) vanishes. This may correspond
to $\omega =\Omega $, or to the resonance, $\omega =\omega _{0}$, see above.
Nevertheless, it will also be demonstrated that the spatially modulated DDI
may support the self-trapping even in the presence of EP (\ref{exp}),
provided that its strength is weak enough.

The electric field subject to the appropriate spatial modulation may be
created by charged grids forming a lens-like capacitor, as shown in Fig. \ref%
{Fig1}. Such capacitors can be built using techniques developed for
ion-holding microtraps \cite{ion1,ion2}. In particular, suitable separable
solutions of the wave equation for the ac electric field ($\omega \neq 0$),
or Laplace equation for the dc field ($\omega =0$) are%
\begin{equation}
\left\{
\begin{array}{c}
E_{z}^{(\mathrm{2D})}\left( x,z\right)  \\
E_{z}^{(\mathrm{1D})}\left( r,z\right)
\end{array}%
\right\} =E_{0}\left\{
\begin{array}{c}
\cos \left( \sqrt{1/x_{0}^{2}+\omega ^{2}/c_{0}^{2}}z\right) I_{0}\left(
r/x_{0}\right)  \\
\cos \left( \sqrt{1/x_{0}^{2}+\omega ^{2}/c_{0}^{2}}z\right) \cosh \left(
x/x_{0}\right) ,%
\end{array}%
\right\} \cos \left( \omega t\right)   \label{phi}
\end{equation}%
for the nearly-2D and 1D condensates. Here, $E_{0}$ is the field amplitude, $%
x_{0}$ is an arbitrary length scale, $c_{0}$\ is the light velocity in
vacuum, and $I_{0}$ is the modified Bessel function. The shape of the
electrodes creating such dc fields is determined by respective equipotential
surfaces:%
\begin{equation}
\sin \left( z/x_{0}\right) =\pm \frac{1}{2}U\left\{
\begin{array}{c}
\left[ I_{0}\left( r/x_{0}\right) \right] ^{-1},~\mathrm{in~2D~}, \\
\mathrm{sech}\left( x/x_{0}\right) ,~\mathrm{in~1D~},%
\end{array}%
\right.   \label{equipot}
\end{equation}%
where $U$ is the voltage applied to the capacitor. Equation
(\ref{equipot}) demonstrates that, for a given modulation scale
$x_{0}$ (a natural range of values is $x_{0}\sim 10$ $\mathrm{\mu
}$m), the distance between the electrodes may be made large enough,
if this is required by the design of the experimental setup. On the
other hand, it is relevant to mention that available technologies
make it possible to build capacitors with the separation between the
electrodes $\sim $ a few $\mathrm{\mu }$m,
while the lateral size of the capacitor may be measures in hundreds of $%
\mathrm{\mu }$m \cite{capacitor}.
\begin{figure}[tbp]
\centering\subfigure[]{
\includegraphics[scale=0.45]{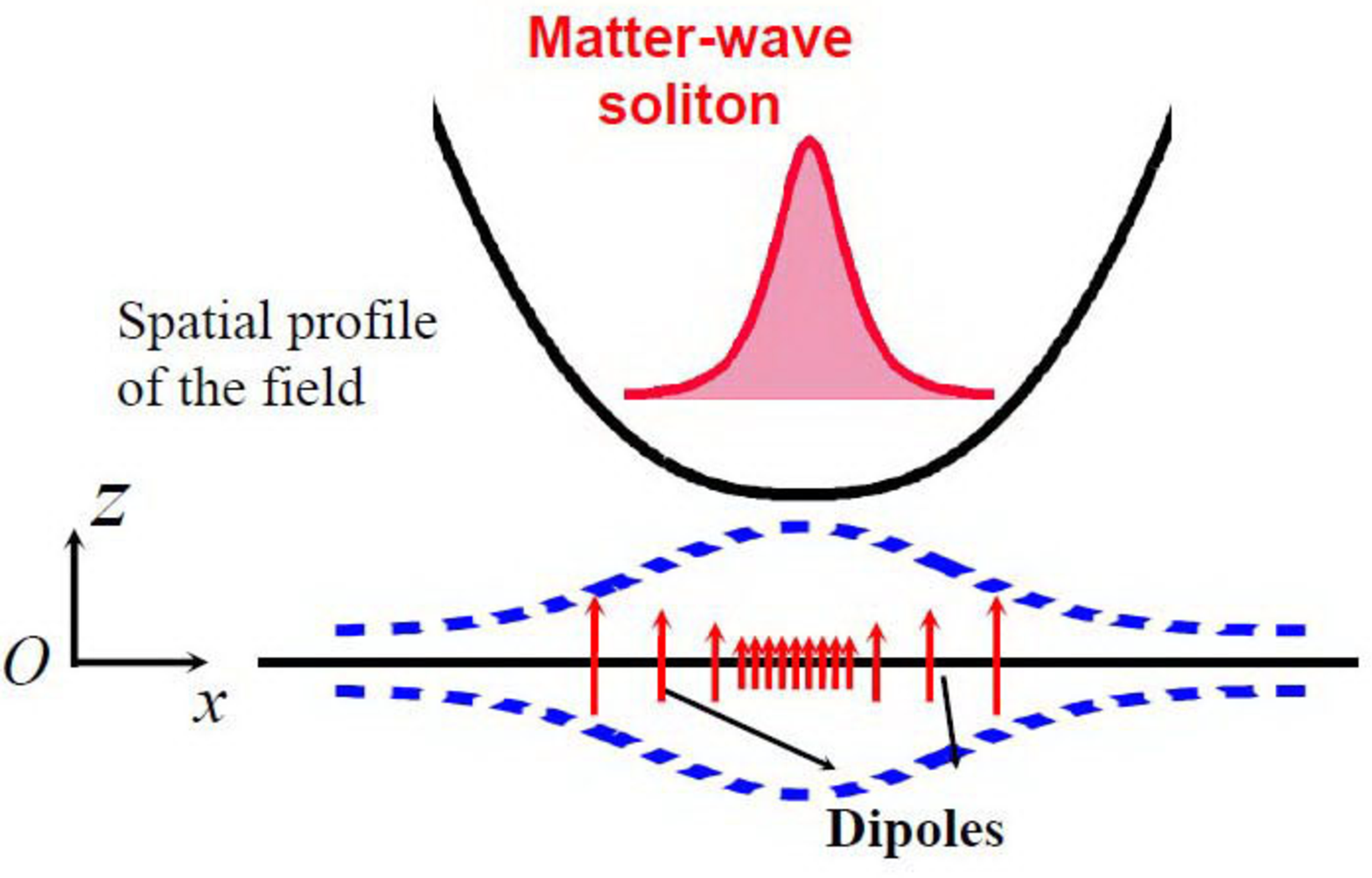}}
\subfigure[]{
\includegraphics[scale=0.38]{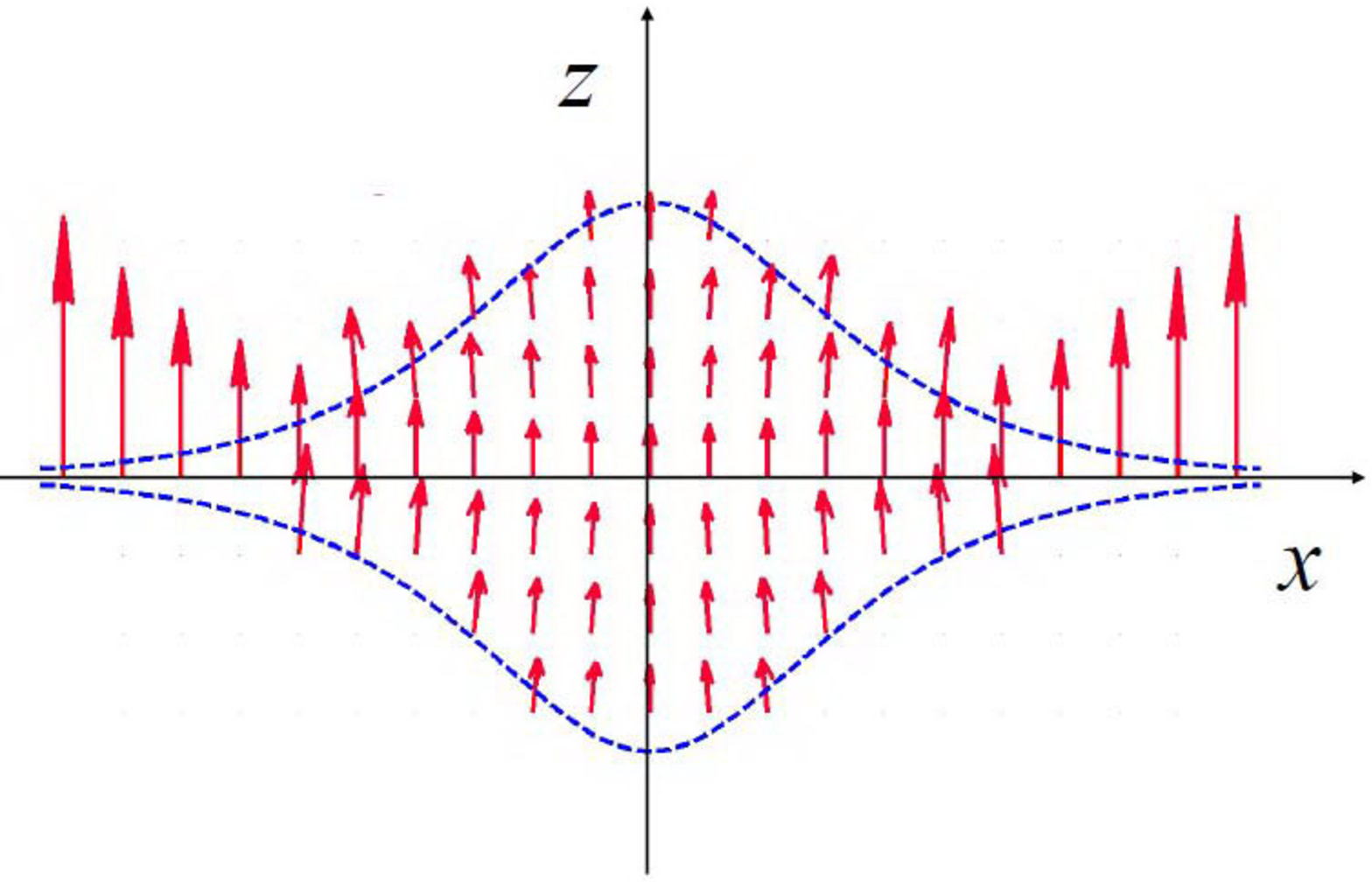}}
\caption{(Color online) (a) The setting for the condensate trapped in the $%
(x,y)$ plane. The polarizing field, with the strength increasing
along the radial coordinate, $r=\protect\sqrt{x^{2}+y^{2}}$, is
directed parallel to the $z$-axis, inducing the local dipole moments
oriented in the same direction. The dashed curves designate
lens-shaped charged grids (electrodes), which may create the
necessary electrostatic field, such as the one given by Eq.
(\protect\ref{phi}). In the latter case, the shape of the electrodes
is determined by Eq. (\protect\ref{equipot}), and the full
distribution of the electrostatic field
$\mathbf{E}^{\mathrm{(2D)}}$, as given by Eq. (\protect\ref{phi}),
is displayed in panel (b), in the plane of $y=0$ (the scale is
arbitrary).} \label{Fig1}
\end{figure}

The time-averaged potential of the DDI between two dipole moments, \textbf{$%
g $}$_{1}$ and \textbf{$g$}$_{2}$, placed at positions $\mathbf{r}_{1}$ and $%
\mathbf{r}_{2}$, is
\begin{equation}
V_{\mathrm{dd}}(\mathbf{R}_{12})=C_{\mathrm{dd}}{\frac{\left\langle \mathbf{g%
}_{1}(r_{1})\cdot \mathbf{g}_{2}(r_{2})\right\rangle
r_{12}^{2}-3\left\langle (\mathbf{g}_{1}(r_{1})\cdot \mathbf{r}_{12})(%
\mathbf{g}_{2}(r_{2})\cdot \mathbf{r}_{12})\right\rangle }{r_{12}^{5}}},
\label{Vdd}
\end{equation}%
where $r_{12}=|\mathbf{r}_{1}-\mathbf{r}_{2}|\mathbf{,}$ the time averaging
, $\left\langle ...\right\rangle $, is realized as in Eqs. (\ref{offoff})
and (\ref{resres}), and $C_{\mathrm{dd}}=1/\left( 4\pi \epsilon _{0}\right) $
or $C_{\mathrm{dd}}=\mu _{0}/4\pi $ for the electric and magnetic dipole
moments, respectively. Because the dipole moments in the setting displayed
in Fig. \ref{Fig1} are parallel to the $z$ axis, and the condensate is
confined to the $(x,y)$ plane, expression (\ref{Vdd}) simplifies to $V_{%
\mathrm{dd}}=C_{\mathrm{dd}}\left\langle
g_{1}(r_{1})g_{2}(r_{2})\right\rangle /r_{12}^{3}$. In the mean-field
approximation, the Hamiltonian of this 2D setting is \cite{review}
\begin{equation}
\mathrm{H}=\frac{\hbar ^{2}}{2m}\int \mathrm{d}\mathbf{r}|\nabla \psi (%
\mathbf{r})|^{2}+{\frac{1}{2}}C_{\mathrm{dd}}\int \int \mathrm{d}\mathbf{r}%
\mathrm{d}\mathbf{r^{\prime }}f(\mathbf{r}-\mathbf{r^{\prime }})g(r)|\psi (%
\mathbf{r})|^{2}g(r\mathbf{^{\prime }})|\psi (\mathbf{r^{\prime }}%
)|^{2}-\chi \int \mathrm{d}\mathbf{r}F^{2}(r)|\psi (\mathbf{r})|^{2},
\label{Ham}
\end{equation}%
where $\psi (\mathbf{r})$ is the single-particle wave function, $m$ is the
mass of the particle, $\chi $ is the strength of EP (\ref{exp}) (if it is
present), and the kernel in the integral term is taken as
\begin{equation}
f(\mathbf{r}-\mathbf{r^{\prime }})={\left( \varepsilon ^{2}+|\mathbf{r}-%
\mathbf{r^{\prime }}|^{2}\right) ^{-3/2},}  \label{f}
\end{equation}%
where the regularization scale $\varepsilon $ is provided by the thickness
of the confined layer in the $z$ direction. The scaled Gross-Pitaevskii
equation (GPE), Eq. (\ref{scaledGPE}), is written below so that $x=1$
corresponds to physical distance $\sim 10$ $\mathrm{\mu }$m. Accordingly, we
set $\varepsilon =0.1$ in Eq. (\ref{f}). We do not include local (contact)
interactions here, to focus on the possibility of inducing the self-trapping
solely by means of the spatially modulated \emph{repulsive} DDI.

Hamiltonian (\ref{Ham}) gives rise to the 2D nonlocal GPE,
\begin{equation}
i{\frac{\partial \psi (\mathbf{r})}{\partial t}}=-{\frac{\hbar ^{2}\nabla
^{2}}{2m}}\psi (\mathbf{r})+C_{\mathrm{dd}}g(\mathbf{r})\psi (\mathbf{r}%
)\int \mathrm{d}\mathbf{r^{\prime }}f(\mathbf{r}-\mathbf{r^{\prime }})g(%
\mathbf{r^{\prime }})|\psi (\mathbf{r^{\prime }})|^{2}-\chi F^{2}(r)\psi (%
\mathbf{r}).  \label{GPE}
\end{equation}%
Obviously, Eq. (\ref{GPE}) has three dynamical invariants, namely the
Hamiltonian, the total number of particles, which is proportional to the
norm of the wave function,%
\begin{equation}
P=\int |\psi (\mathbf{r})|^{2}\mathrm{d}\mathbf{r},  \label{norm}
\end{equation}%
and the $z$-component of the angular momentum,%
\begin{equation}
M=i\int \psi ^{\ast }\left( y\psi _{x}-x\psi _{y}\right) \mathrm{d}\mathbf{r,%
}
\end{equation}%
where $\psi ^{\ast }$ stands for the complex conjugate.

As shown in Fig. \ref{Fig1}, the magnitude of the polarizing field grows at $%
r\rightarrow \infty $, which results in a growing local value of the dipole
moment. With the repulsive DDI, fundamental and vortical solitons may
self-trap in a finite isotropic area around the center, due to the greater
strength of the DDI-mediated repulsion in the outer area. The 1D version of
the system is described by the obvious 1D reduction of Eqs. (\ref{GPE}) and (%
\ref{f}).

In reality, the indefinite growth of the polarizing field at $r\rightarrow
\infty $ is not necessary. As shown below, solitons emerging in such
settings are well localized modes, hence the supporting profile of the
external field should be actually created in a finite area, as the presence
of the field at large distances from the center, to which the soliton
extends no tangible tail, is not needed. The same argument is relevant as
concerns the possible presence of EP (\ref{exp}). Formally speaking, trapped
modes cannot exist in the presence of the expulsive potential which
indefinitely grows at $r\rightarrow \infty $. However, as is shown below,
the system considered in a finite area of a reasonably large size may
readily overcome the destructive effect of the EP.

Thus, our objective is to demonstrate the self-trapping of localized modes
(of the fundamental and vortical types alike) in the present model via the
action of the repulsive DDI, assuming that the local dipole moments are
induced by the external field according to Eq. (\ref{G}), with the field
growing from the center to periphery as $r^{\alpha }$. As we demonstrate
below analytically, an important difference of the present nonlocal model
from its local counterparts \cite{Olga}-\cite{B5} is tight \emph{%
super-exponential} localization of the solitons, see Eq. (\ref{far}) below,
in contrast with the loose (power-law) localization in the local models,
which is determined by the Thomas-Fermi approximation (TFA) \cite{Olga},
\begin{equation}
\left\vert \psi (r)\right\vert ^{2}\sim r^{-\alpha }.  \label{TFA}
\end{equation}%
Obviously, the sharp localization should help to observe solitons in the
experiment.

Another significant difference is that the minimum value of the growth rate $%
\alpha $, above which normalizable self-trapped modes exist in the present
model, is $\alpha _{\mathrm{\min }}^{\left( \mathrm{dd}\right) }=3$, and it
does not depend on spatial dimension $D$, see Eq. (\ref{cr}) below (we
actually use $\alpha =4$), unlike the above-mentioned minimum value in the
local model with the cubic nonlinearity:
\begin{equation}
\alpha >\alpha _{\mathrm{\min }}^{\left( \mathrm{local}\right) }=D,
\label{cr=D}
\end{equation}%
which actually follows from Eq. (\ref{TFA}) \cite{Olga}. Also drastically
different from the local model are scalings which characterize dependences
between the solitons' norm and chemical potential, as Eqs. (\ref{2/7}) and (%
\ref{2/3}) demonstrate in the following sections.

To estimate a range of physical parameters relevant to the setting
considered here, we note that the intrinsic nonlinearity, induced by the
magnetic \cite{review0} or electric \cite{Saka} DDIs, may be roughly
estimated as the contact interaction with an effective scattering length, $%
a_{s}\sim mg^{2}/\hbar ^{2}$. With characteristic values of the molecular
electric polarizability relevant to experiments with ultracold gases, $\chi
\sim 100$ $\overset{\circ }{\mathrm{A}}^{3}$\cite{molecular}, and the
corresponding molecular weight, $\sim 100$, the magnitude of the effective
scattering length sufficient\ for the formation of localized modes, $%
a_{s}\sim 0.1$ nm \cite{Randy}, may be emulated by the polarizing dc
electric field in a range of $E\sim 10$ kV/cm, which is definitely
accessible to the experiment. Further, results for the 1D and 2D settings,
presented in Figs. \ref{1Dfundamental} and \ref{2DFundamental},
respectively, along with the modulation profile (\ref{d}) adopted below,
demonstrate that, within the area of the actual localization of the
solitons, the field increases from the center to periphery by a factor $%
\lesssim $ $20$, which is compatible with the above-mentioned range of the
values of $E$. Because the density of the condensate is very low ($\sim
10^{15}$ cm$^{-3}$, in the most typical case), and the contact of molecules
with the field-inducing grids is prevented by the optical trap, see Fig. \ref%
{Fig1}, the electric breakdown of the low-density gas is not a severe danger
either in this setting.

As concerns the role of EP (\ref{exp}), an estimate suggests that it can be
made negligible in comparison with the DDI, in the region where the
self-trapped mode is localized, if the condensate density is raised to
values $\sim 10^{18}~$cm$^{-3}$ (then, the number of molecules expected in
the nearly-1D soliton will be $\sim 10^{6}$, instead of the most typical
value $\sim 10^{3}$ \cite{Randy}). Alternatively, the same result may be
achieved by bringing the ac drive to a proximity of the resonance with the
relative detuning $\left\vert \omega _{0}-\omega \right\vert /\omega
_{0}\sim 10^{-3}$, see Eq. (\ref{dyn}).

Lastly, it is relevant to mention that a similar situation may be expected
in BEC with long-range interactions induced by the resonant laser
illumination \cite{light-induced-DD}. However, the consideration of that
setting is beyond the scope of the present work.

The rest of the paper is structured as follows. In Sec. II, analytical and
numerical results are reported for basic types of stable self-trapped modes
which can be supported by the spatially growing nonlocal repulsion, namely,
1D and 2D fundamental solitons, twisted (spatially odd) modes in 1D, and
solitary vortices in 2D. The phenomenology of the soliton modes is
summarized by means of dependences of their chemical potentials and spatial
size on the norm. In most cases, these dependences can be explained by means
of a simple analysis of scaling in Eq. (\ref{scaledGPE}) (with $\chi =0$).
Stability of the modes in the presence of EP (\ref{exp}) is considered too,
as well as the TFA for the 1D and 2D fundamental solitons. In Sec. III,
motion of shifted and/or kicked 1D and 2D fundamental solitons around the
center is considered. In Sec. IV, we change the 1D setting from the
single-well modulation of the polarizing field to a double-well
configuration, and study properties of solitons in that case (configurations
of this type were not studied previously even in models which maintain
bright solitons by means of the spatially growing local repulsive
nonlinearity). The paper is concluded by Sec. V.

\section{Solitons supported by the field-induced repulsive dipole-dipole
interaction}

\subsection{Analytical considerations}

Stationary solutions to Eq. (\ref{GPE}) with chemical potential $\mu $ are
looked for as $\psi (t,\mathbf{r})=e^{-i\mu t}\phi (\mathbf{r})$. Setting,
by means of an obvious rescaling, $\hbar =m=C_{\mathrm{dd}}=1$, and, as said
above, scaling the distances so that $x=1$ corresponds to physical length $%
\sim 10$ $\mathrm{\mu }$m, the corresponding equation for the (generally,
complex) stationary wave function $\phi $ is derived in the following form:
\begin{equation}
\mu \phi (\mathbf{r})+{\frac{1}{2m}}\nabla ^{2}\phi (\mathbf{r})-g(\mathbf{r}%
)\phi (\mathbf{r})\int \mathrm{d}\mathbf{r^{\prime }}f(\mathbf{r}-\mathbf{%
r^{\prime }})g(\mathbf{r^{\prime }})|\phi (\mathbf{r^{\prime }})|^{2}+\chi
F^{2}(r)\phi (\mathbf{r})=0.  \label{scaledGPE}
\end{equation}%
As said here, we chiefly solved Eq. (\ref{scaledGPE}) with $m=1$, but
coefficient $m$ is kept as a free one for the consideration of the TFA (see
below), which corresponds to dropping the kinetic-energy term in the
equation, i.e., setting $m\rightarrow \infty $. The above-mentioned physical
estimates imply that values $P\sim 1$ of the scaled norm (\ref{norm})
correspond to the numbers of particles $N\sim 10^{3}$ and $10^{4}$ in the 1D
and 2D solitons displayed below, see Figs. \ref{1Dfundamental} and \ref%
{2DFundamental}.

Multiplying Eq. (\ref{scaledGPE}) by $\phi ^{\ast }(\mathbf{r})$ and
integrating the result over the space, it is easy to prove that the equation
may give rise to localized solutions only with $\mu >0$ (this proof is
similar to that in the model with the spatially modulated strength of the
local self-repulsive nonlinearity \cite{B2}), while the usual bright
solitons, in the uniform space with self-attractive nonlinearities, always
have $\mu <0$.

The tightness of self-trapping of the 2D modes is characterized by their
effective area,%
\begin{equation}
\mathrm{A_{eff}}=P^{2}\left( \int \left\vert \phi (\mathbf{r})\right\vert
^{4}\mathrm{d}\mathbf{r}\right) ^{-1},  \label{Aeff}
\end{equation}%
where $P$ is the norm introduced in Eq. (\ref{norm}).\ The 1D counterpart of
$\mathrm{A_{eff}}$ measures the effective width of the 1D mode.

To introduce the spatial modulation of the local dipole moment, we assume
that the strength of the polarizing field and, accordingly, the local moment
[see Eqs. (\ref{g}) and (\ref{sin})] grow with $r$ as
\begin{equation}
g(r)=r^{\alpha }+g_{0},  \label{d}
\end{equation}%
with $g_{0}\geq 0$. Two interpretations of this modulation profile are
possible: (i) the constant term, $g_{0}$, may be a permanent part of the
particle's dipole \ moment, while $r^{\alpha }$ is, in the appropriately
scaled notation, the addition induced by the external field whose strength
grows as $r^{\alpha }$, or (ii) the field profile is patterned as in Eq. (%
\ref{d}), the entire dipole moment being induced by the field.

Solitons with a convergent norm exist if the growth rate $\alpha $ in Eq. (%
\ref{d}) exceeds a certain critical value, $\alpha _{\mathrm{\min }}$. In
the local model with the strength of the cubic self-repulsive term growing
as $r^{\alpha }$, the TFA [see Eq. (\ref{TFA})] readily demonstrates that
the self-trapped modes are normalizable for $\alpha >D$, as stated in Eq. (%
\ref{cr=D}). In fact, this result is an exact one, which is not predicated
on the validity of the TFA \cite{Olga}.

In the present nonlocal model, another approximation makes it possible to
identify $\alpha _{\mathrm{\min }}$. Indeed, assuming that the soliton is
represented by an axisymmetric localized solution of Eq. (\ref{scaledGPE}), $%
\phi (r)$, or by its 1D counterpart $\phi \left( x\right) $, in the limit of
$r\rightarrow \infty $ (or $|x|\rightarrow \infty $, at $D=1$), the
asymptotic form of Eqs. (\ref{scaledGPE}), (\ref{d}) and (\ref{f}) with $%
\chi =0$ yields%
\begin{equation}
\frac{d^{2}\phi }{dr^{2}}+\frac{D-1}{r}\frac{d\phi }{dr}+2\mu \phi
-4r^{\alpha -3}\phi (r)\int_{0}^{\infty }\left\vert \phi (r^{\prime
})\right\vert ^{2}g(r^{\prime })\left( \pi r^{\prime }\right)
^{D-1}dr^{\prime }=0.  \label{asympt}
\end{equation}%
For $\alpha >3$, Eq. (\ref{asympt}) takes the form of the 1D linear Schr\"{o}%
dinger equation, with coordinate $r$, and an effective potential growing as $%
r^{\alpha -3}$. Accordingly, the asymptotic form of the relevant solution to
this equation is
\begin{equation}
\phi (r)=\phi _{0}\exp \left( -\frac{4\sqrt{Q_{D}}}{\alpha -1}r^{\frac{%
\alpha -1}{2}}\right) ,  \label{far}
\end{equation}%
where constants $\phi _{0}$ and $Q_{D}\equiv \int_{0}^{\infty }\phi
^{2}(r)g(r)\left( \pi r^{\prime }\right) ^{D-1}dr$ are characteristics of
the corresponding global solution. Thus, at
\begin{equation}
\alpha >\alpha _{\mathrm{\min }}^{\left( \mathrm{dd}\right) }=3,  \label{cr}
\end{equation}%
the super-exponentially localized self-trapped states exist for either
dimension, $D=1$ or $2$. Furthermore, an analysis of Eq. (\ref{scaledGPE}),
with regard to Eqs. (\ref{d}) and (\ref{f}), suggests that, at $\alpha >3$,
the solitons exists for all values of norm (\ref{norm}). In particular, at $%
P\rightarrow 0$ the soliton becomes broad, and Eq. (\ref{scaledGPE}) gives
rise to the following scaling relations between $P$, peak density $\phi
_{0}^{2}$, and a characteristic radial size of the soliton, $r_{0}$:%
\begin{equation}
r_{0}\sim P^{-\frac{1}{2\alpha -1}},~\phi _{0}^{2}\sim P^{\frac{2\alpha +D-1%
}{2\alpha -1}}.  \label{scaling}
\end{equation}%
On the other hand, at $\alpha <3$ Eq. (\ref{asympt}) simplifies, in the
lowest approximation, to $\phi ^{\prime \prime }+2\mu \phi =0$, which,
obviously, cannot have localized solutions with $\mu >0$. Detailed analysis
of the critical case, $\alpha =3$, is beyond the scope of the present work.
Below, we report numerical results with $\alpha =4,$ for $D=1$ and $2$ alike.

\subsection{One-dimensional solitons}

\begin{figure}[tbp]
\centering\subfigure[] {\label{fig2a-a}
\includegraphics[scale=0.3]{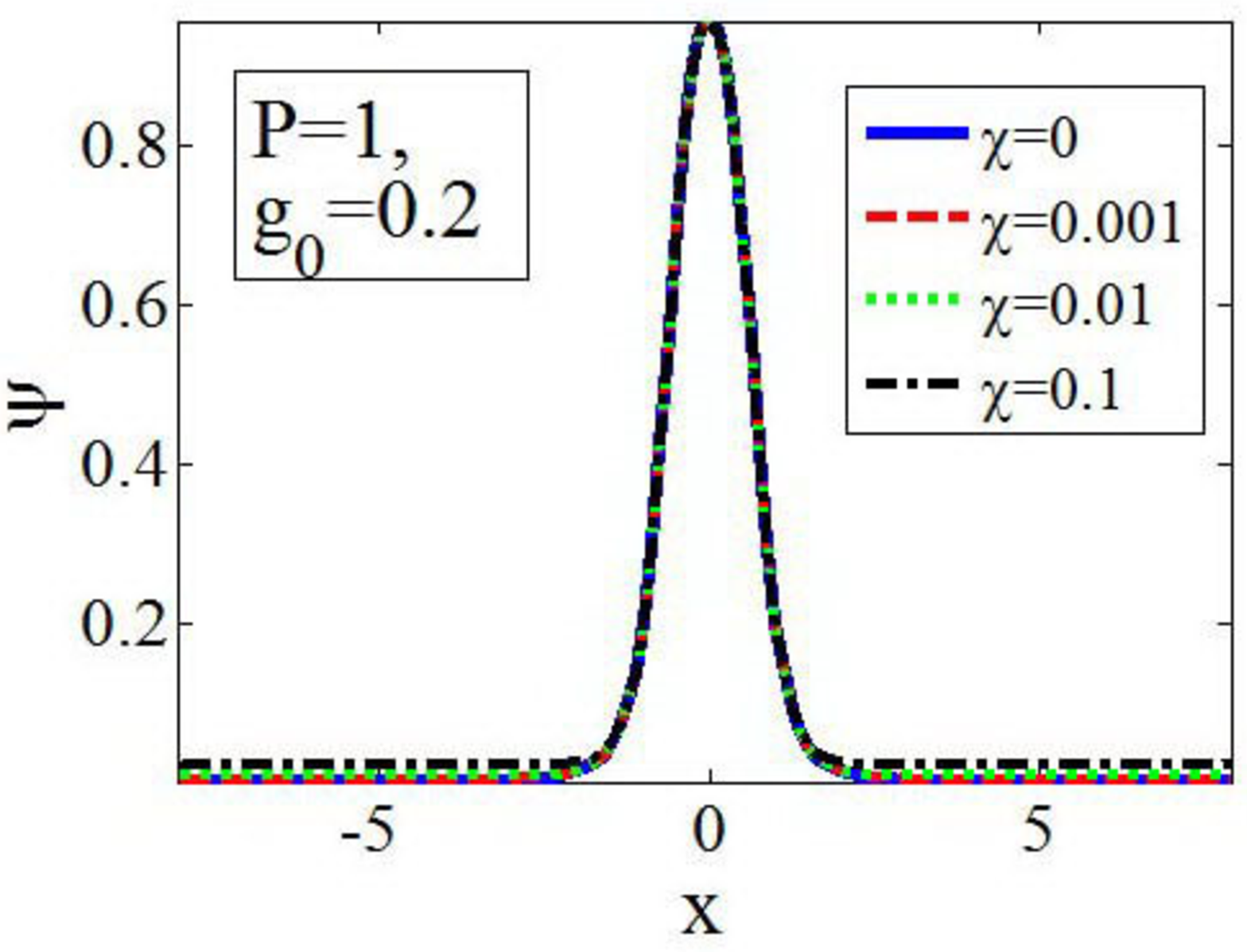}}%
\subfigure[] {\label{fig2b-a}
\includegraphics[scale=0.3]{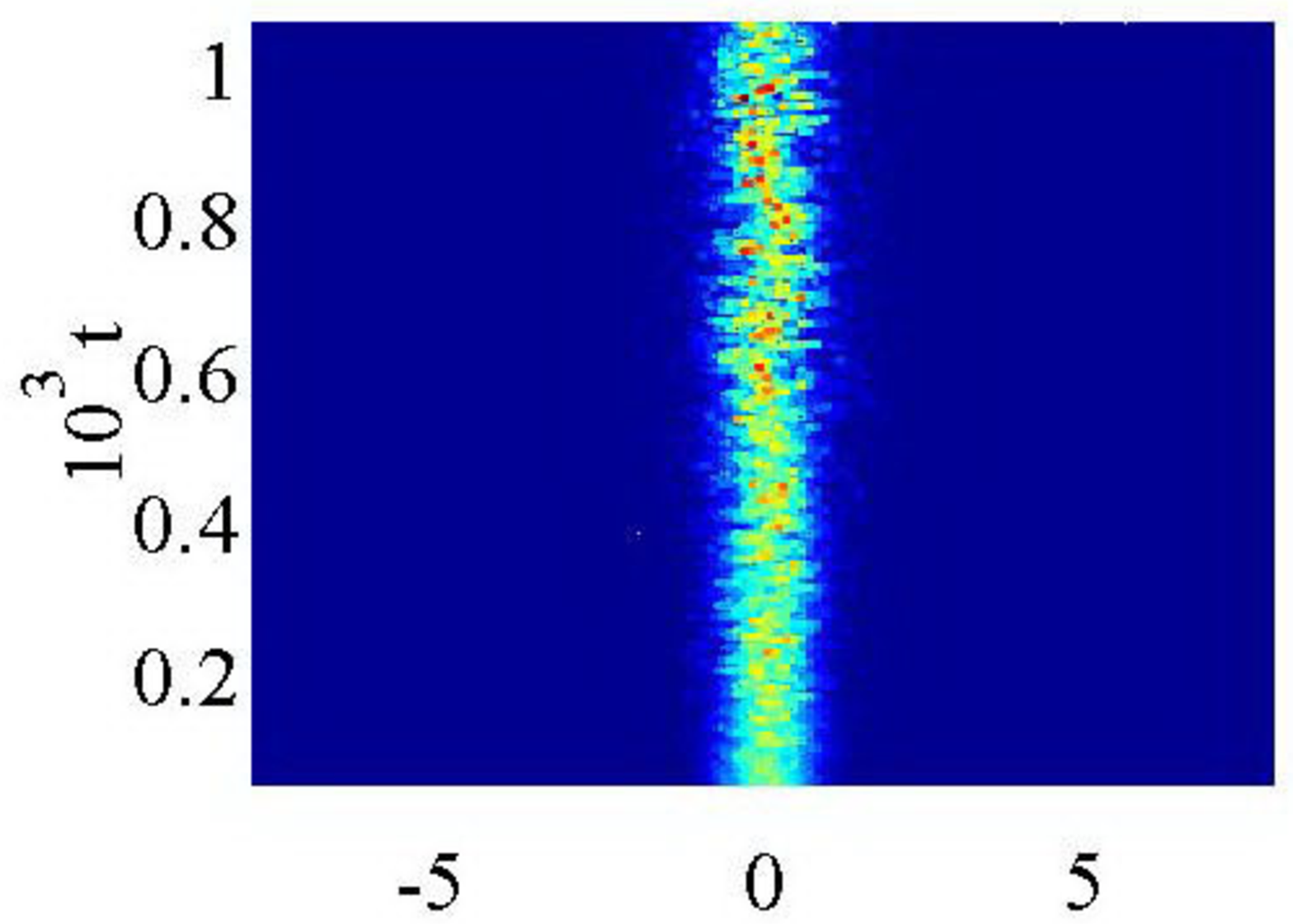}}
\subfigure[] {\label{fig2c-a}
\includegraphics[scale=0.3]{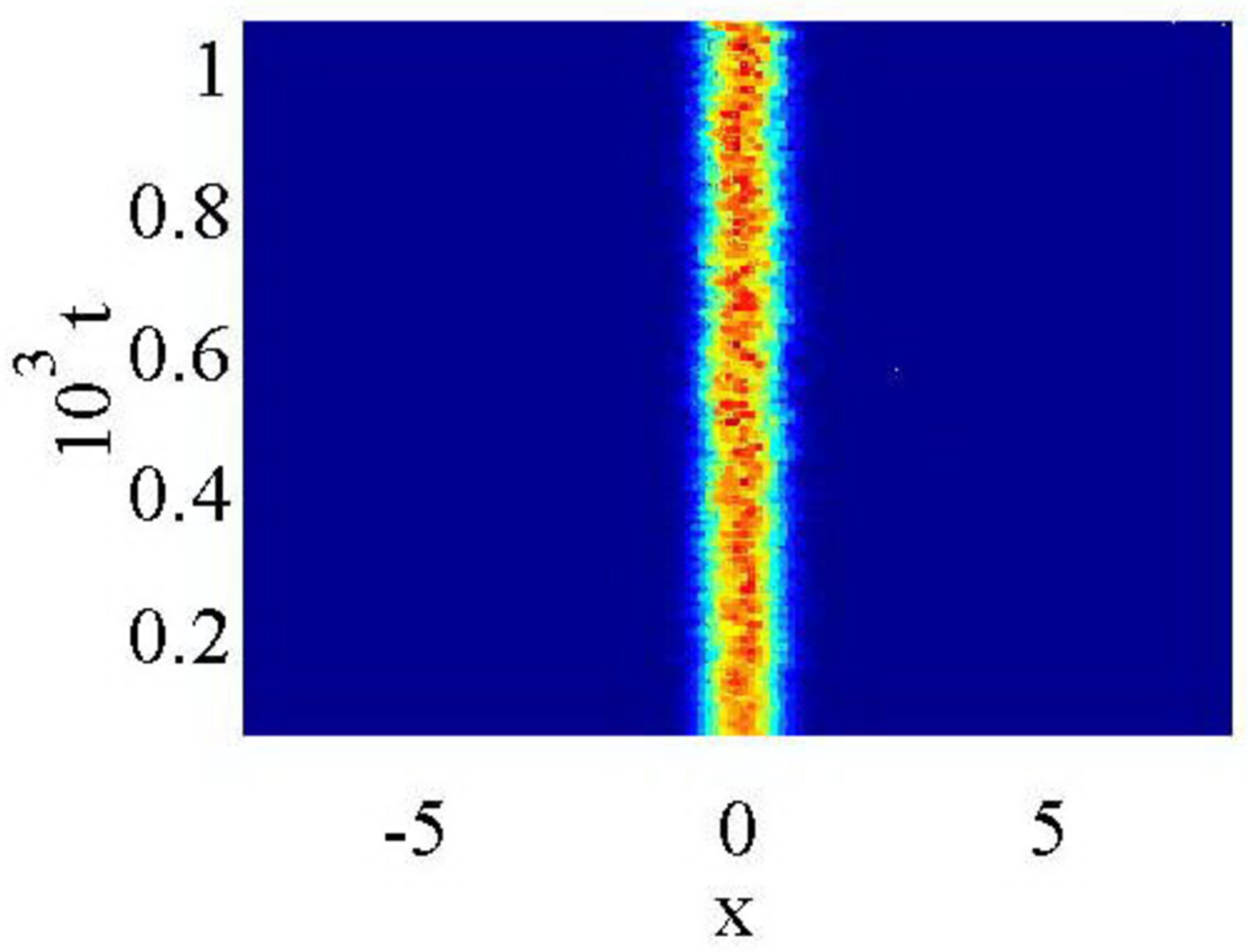}}
\subfigure[] {\label{fig2d-a}
\includegraphics[scale=0.3]{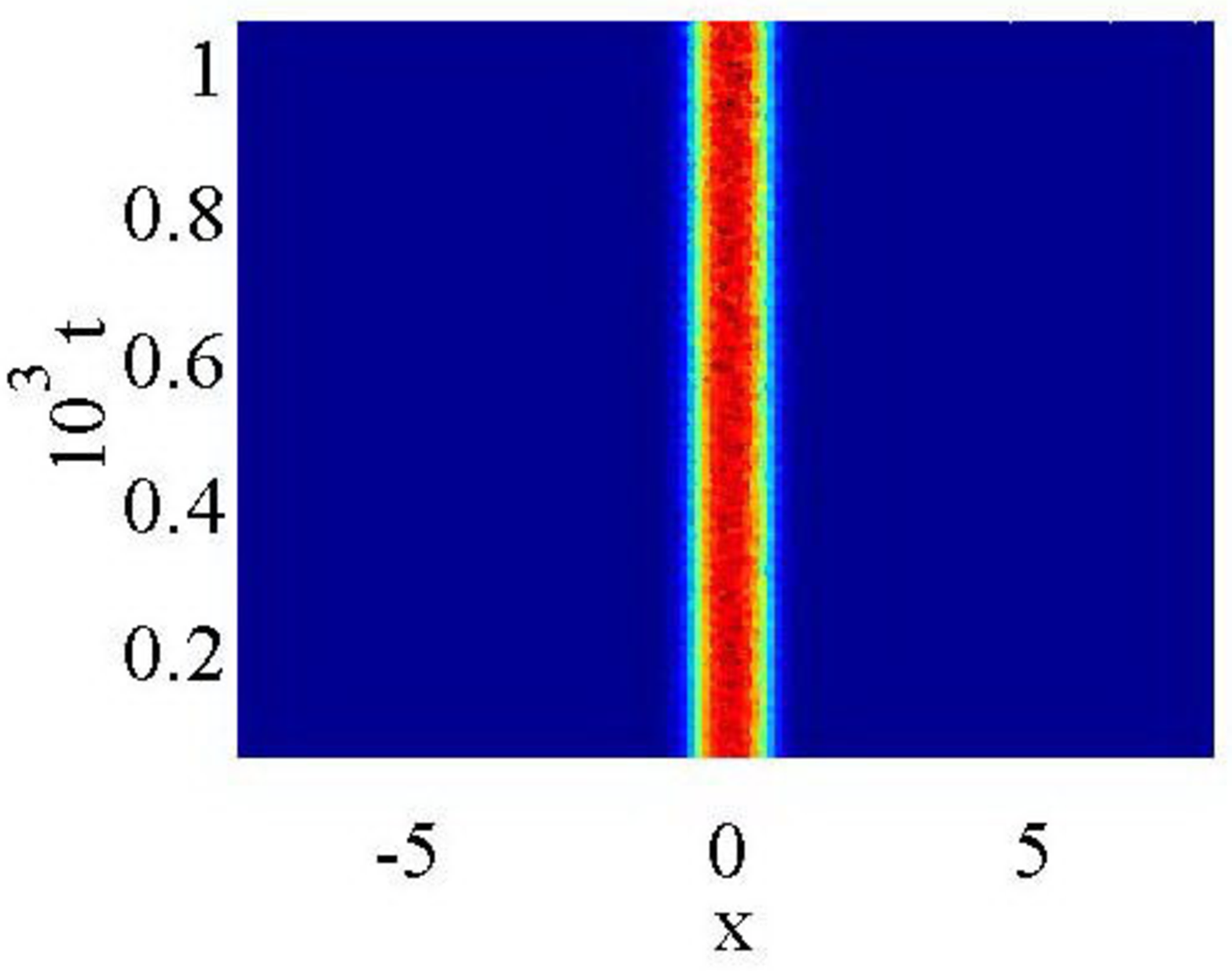}}
\caption{(Color online) (a) Self-trapped modes found for different values of
strength $\protect\chi $ of the expulsive potential (\protect\ref{exp}), as
indicated in the panel, for scaled norm $P=1$ and $g_{0}=0.2$ in Eq. (%
\protect\ref{d1D}). Perturbed evolution of the self-trapped modes is shown
in panel (b) for $P=0.1$, in (c) for $P=1$, and in (d) for $P=5$. In the
latter three panels, $\protect\chi =0.1$ and $g_{0}=0.2$. The evolution is
unstable for the weakly and moderately nonlinear modes in (b,c), and stable
for the strongly nonlinear one in (d).}
\label{1DfundamentalVexp}
\end{figure}

The numerical solution of the 1D version of Eq. (\ref{scaledGPE}) with
\begin{equation}
g(x)=x^{4}+g_{0}  \label{d1D}
\end{equation}%
was carried out by means of numerical code PCSOM elaborated in Refs. \cite%
{YJK1,YJK2}. First, in Fig. \ref{1DfundamentalVexp} we present basic results
obtained in the model including EP (\ref{exp}) with strength $\chi $. Panel
(a) demonstrates that, if $\chi $ is not small enough, the EP generates a
nonvanishing tail, which breaks the self-trapped character of the mode. For
the same case, panel \ref{1DfundamentalVexp}(c) shows that the presence of
the EP makes the soliton unstable in direct simulations, which were
performed by adding small random perturbations to the initial configuration.
The instability is, naturally, still stronger for a smaller value of the
norm, as shown in panel \ref{1DfundamentalVexp}(b). On the other hand, the
increase of the norm makes the soliton robust due to the stronger
nonlinearity, in accordance with the estimate given at the end of Section I.

\begin{figure}[tbp]
\centering\subfigure[] {\label{fig2a}
\includegraphics[scale=0.3]{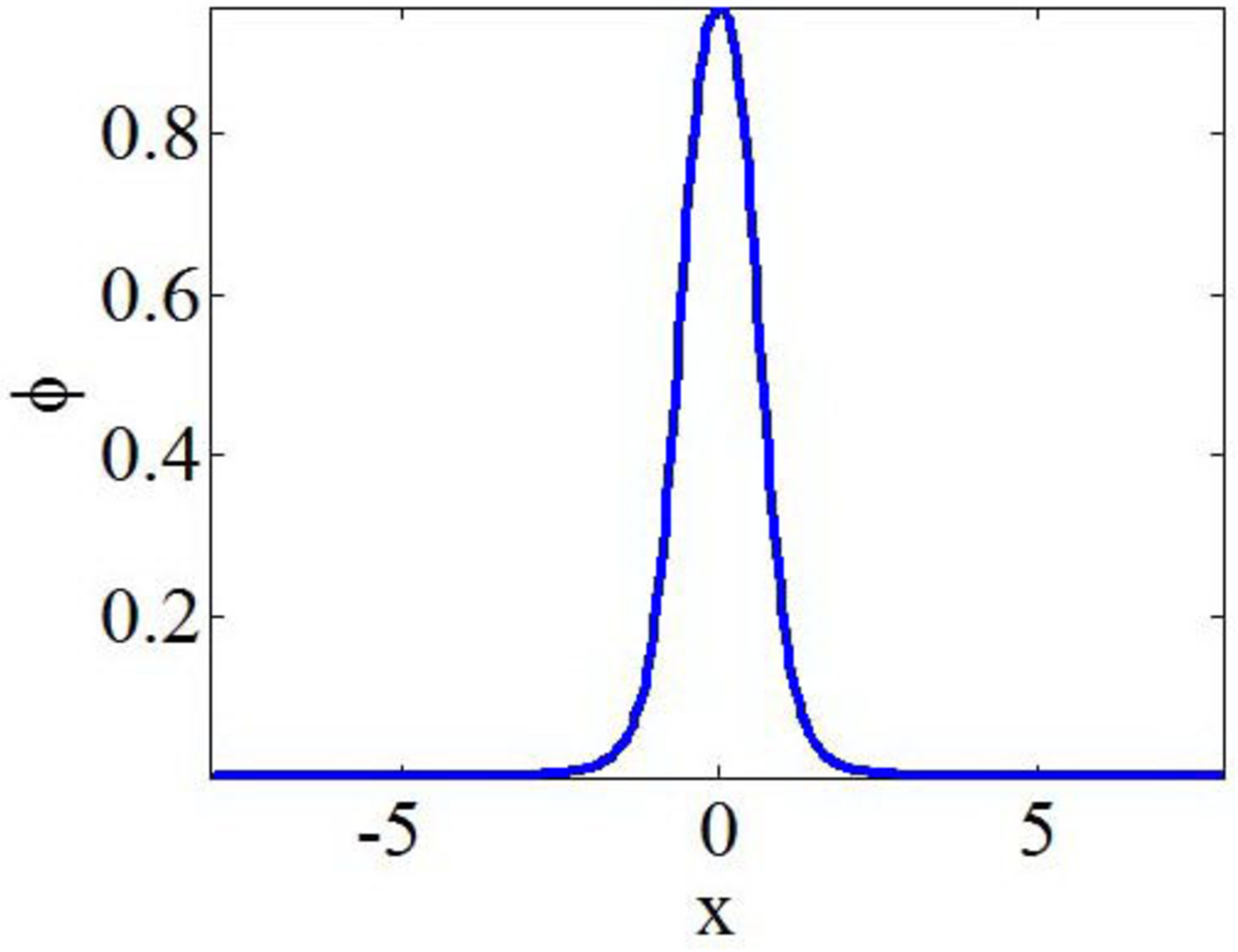}}%
\subfigure[] {\label{fig2b}
\includegraphics[scale=0.4]{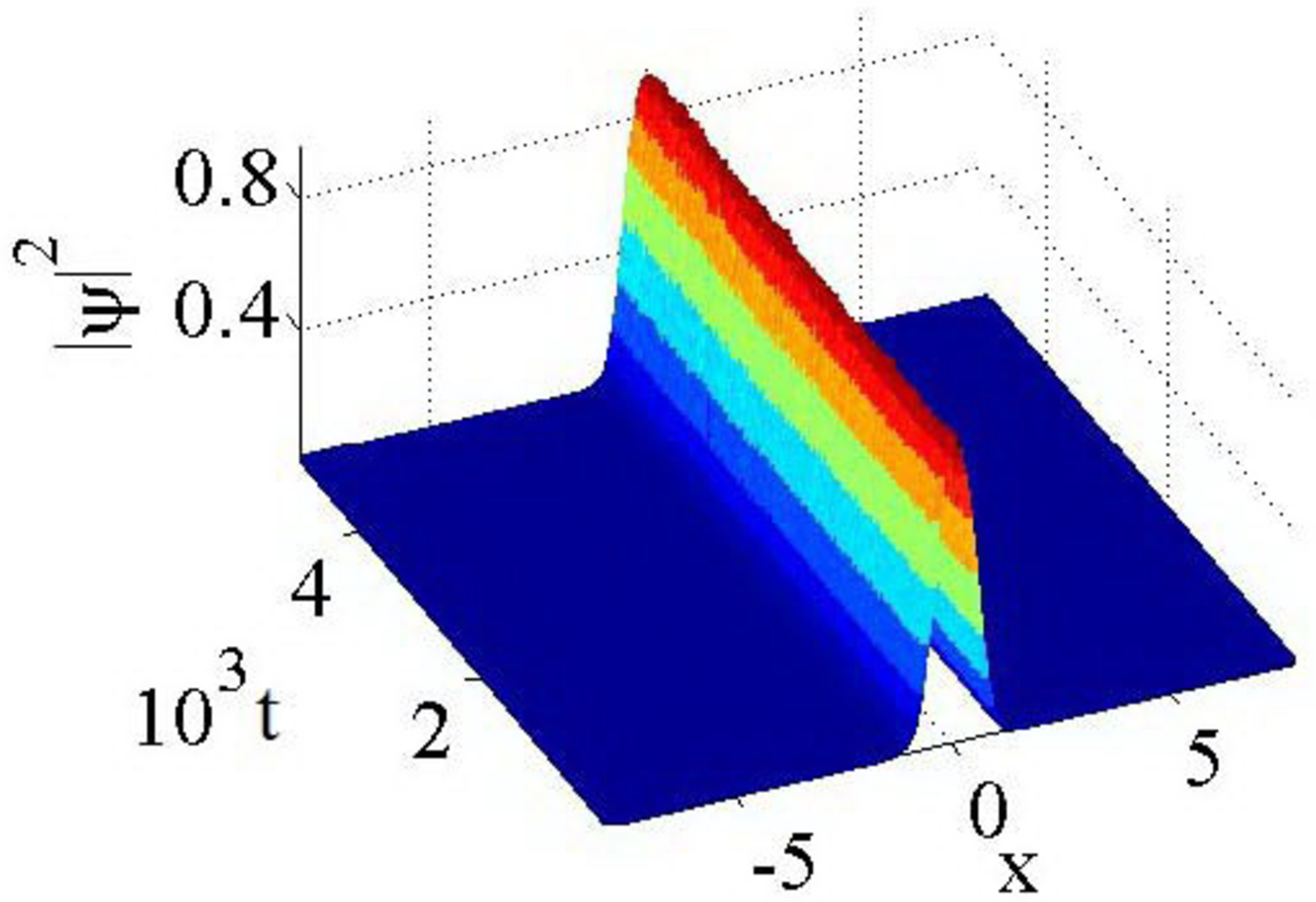}}
\subfigure[] {\label{fig2c}
\includegraphics[scale=0.22]{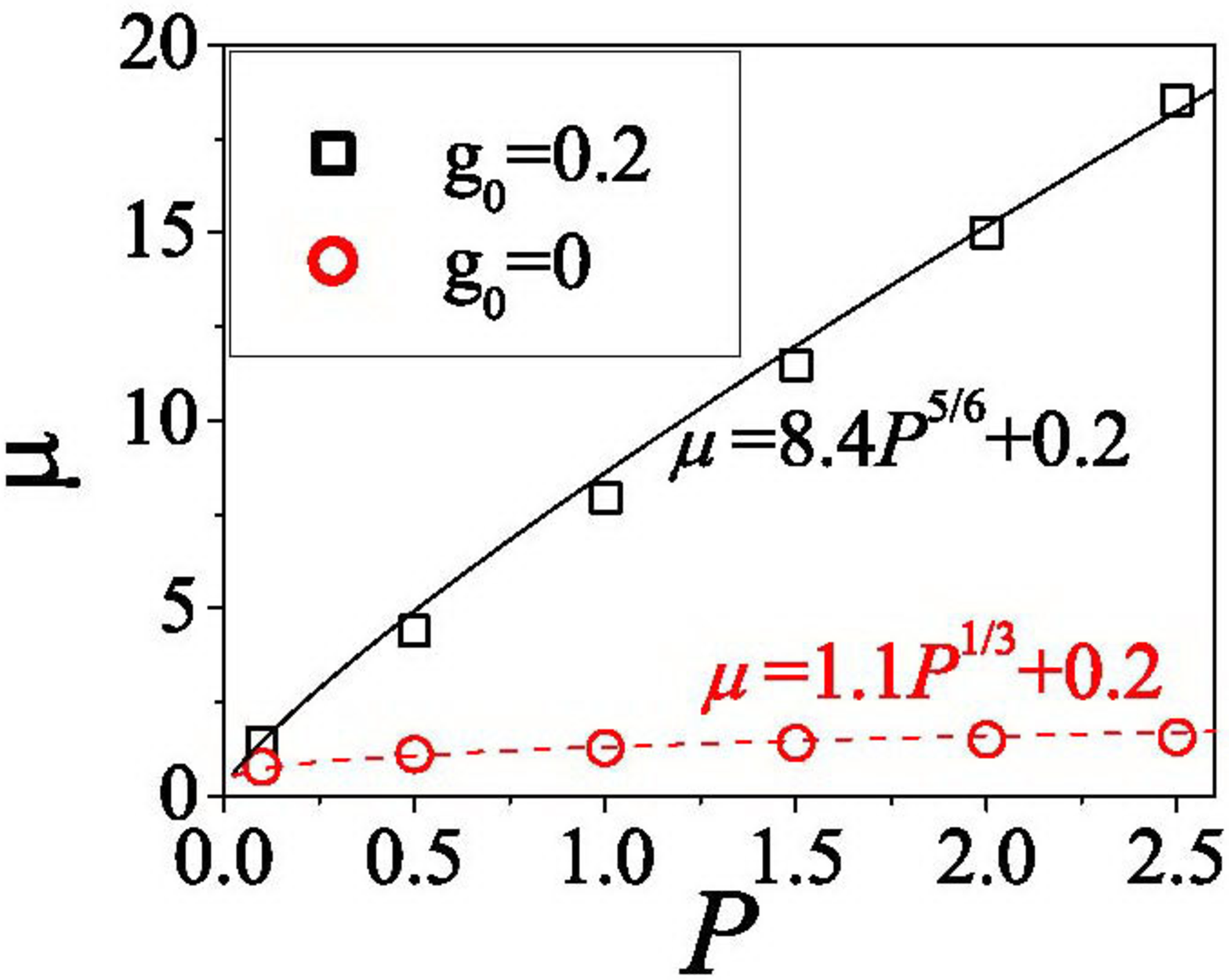}}
\subfigure[] {\label{fig2d}
\includegraphics[scale=0.22]{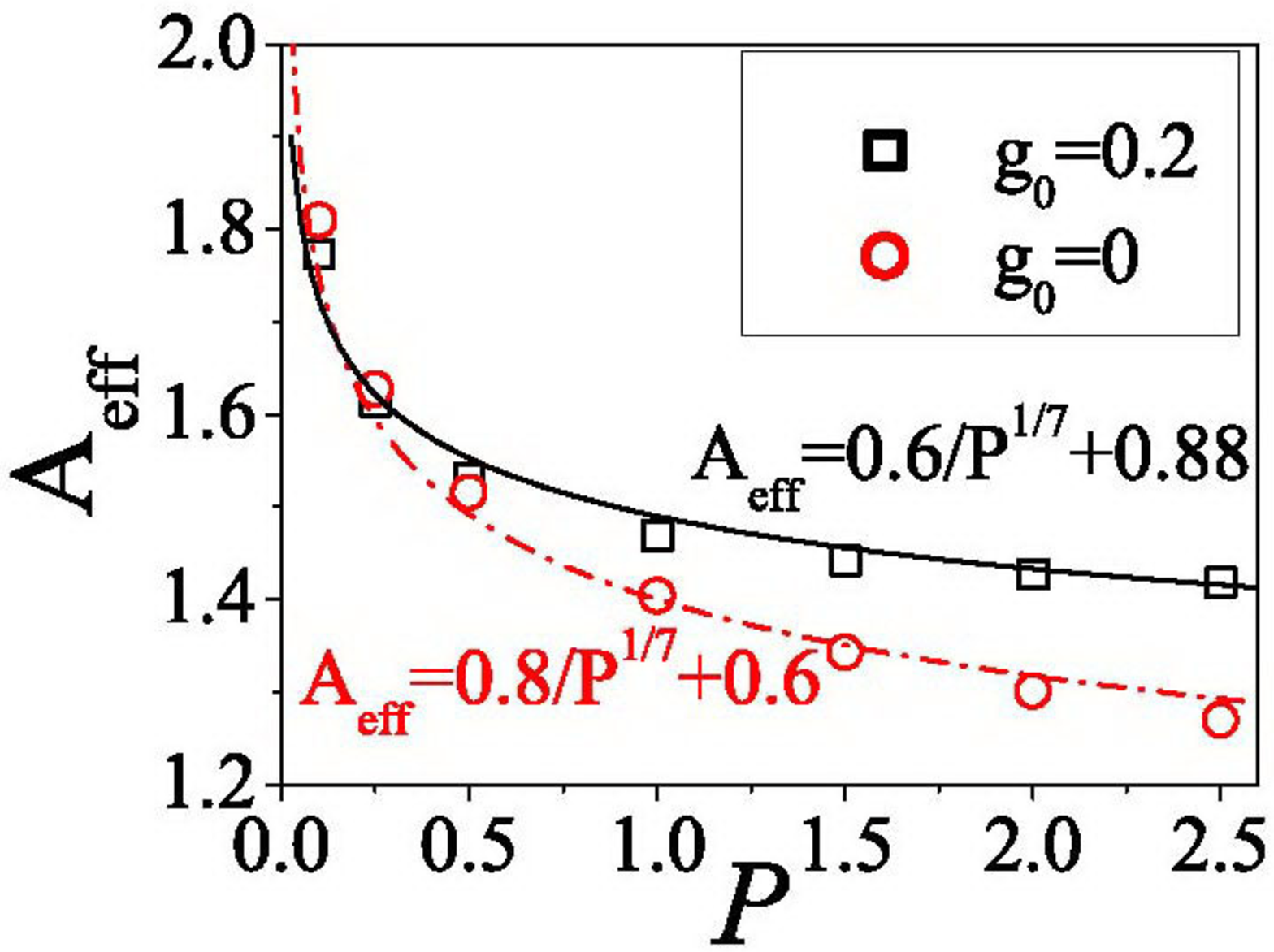}}
\subfigure[] {\label{fig3a}
\includegraphics[scale=0.3]{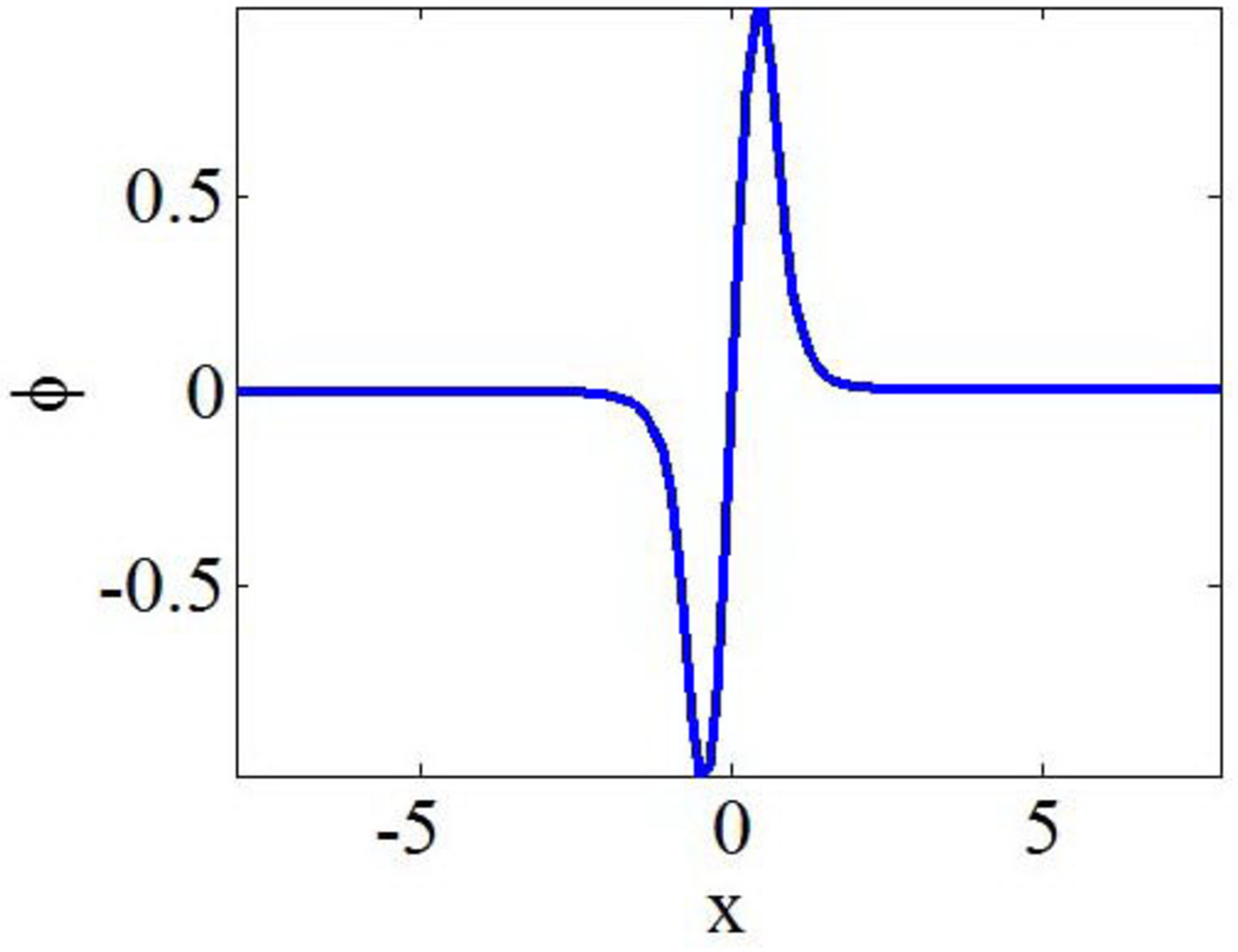}}%
\subfigure[] {\label{fig3b}
\includegraphics[scale=0.4]{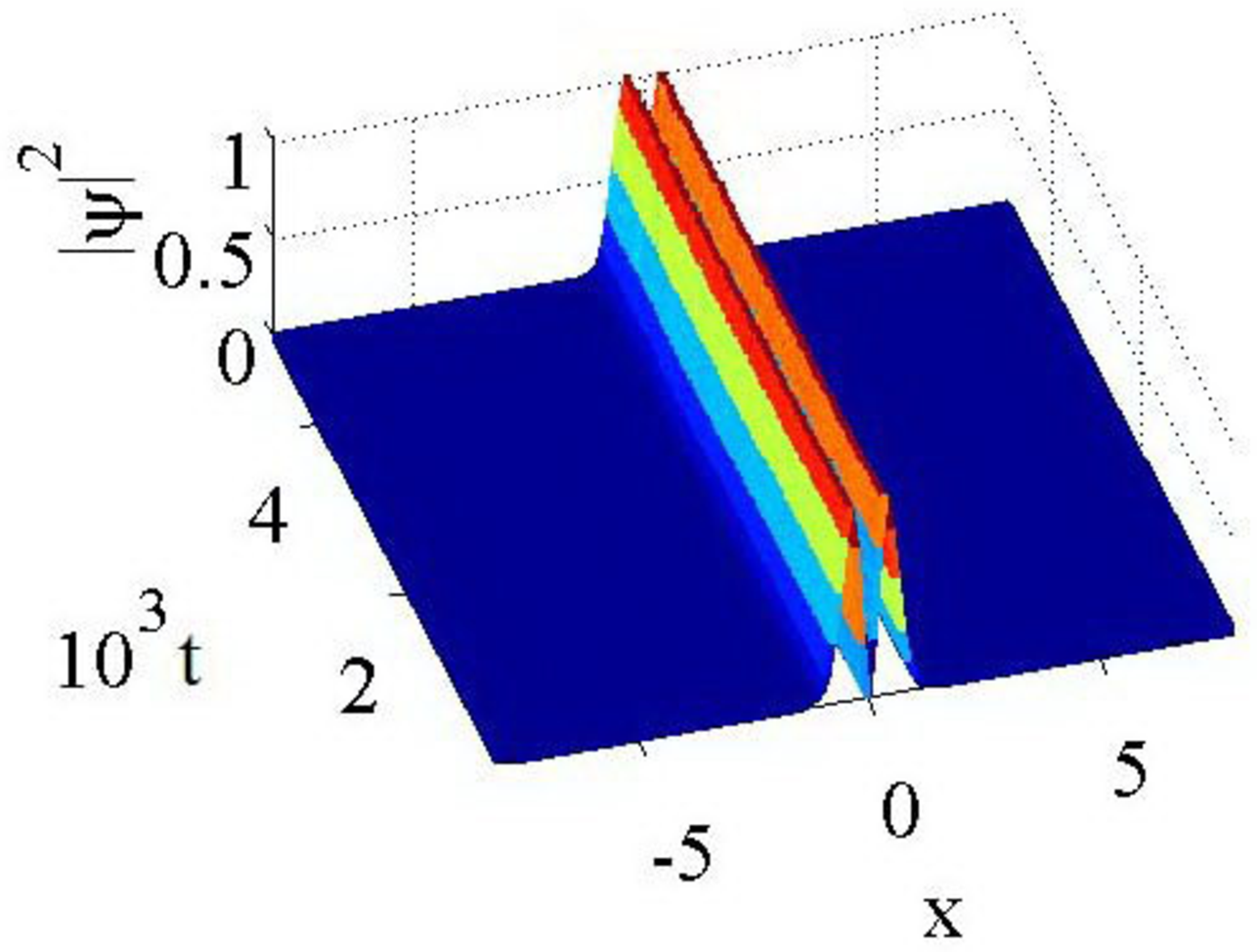}}
\subfigure[] {\label{fig3c}
\includegraphics[scale=0.22]{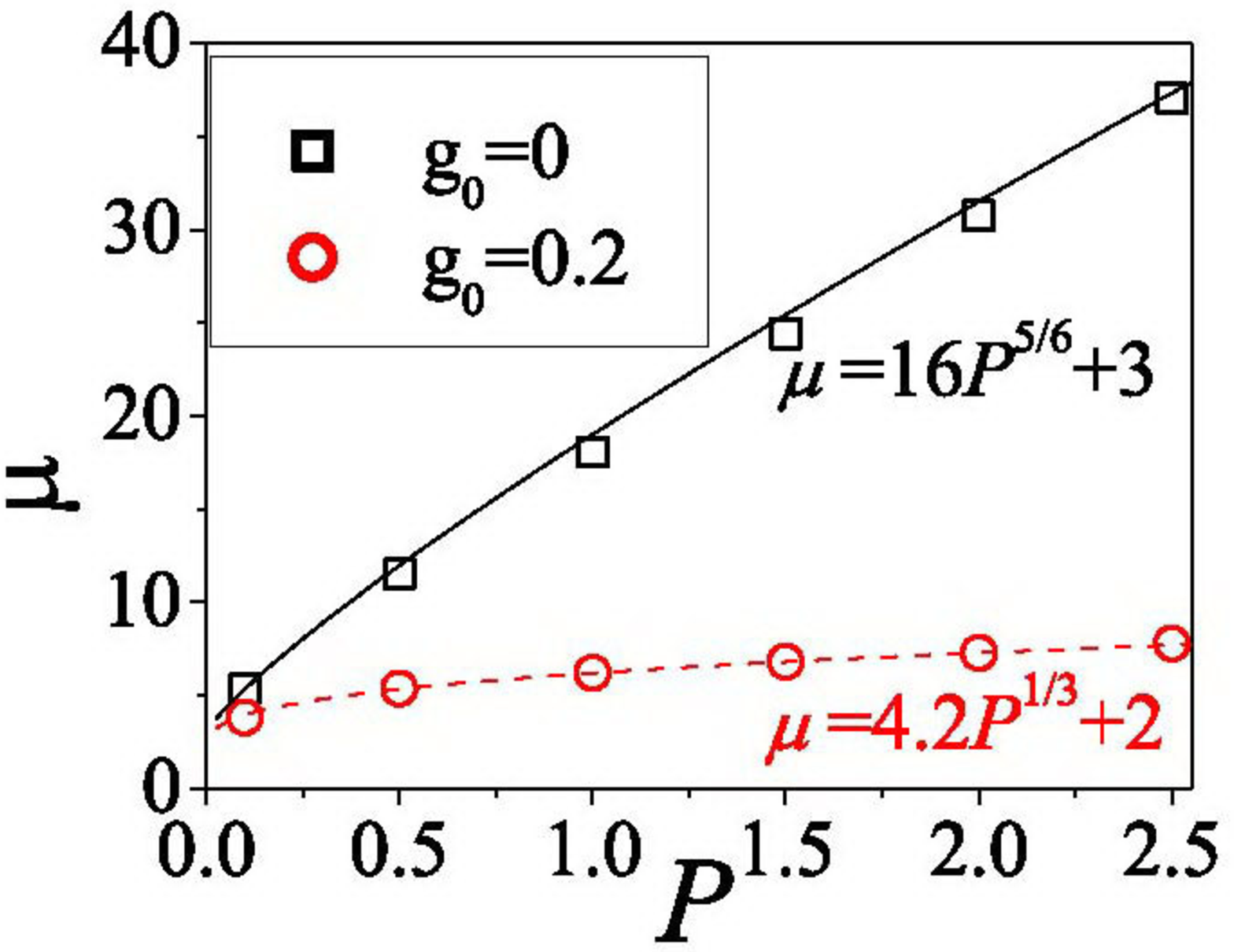}}
\subfigure[] {\label{fig3d}
\includegraphics[scale=0.22]{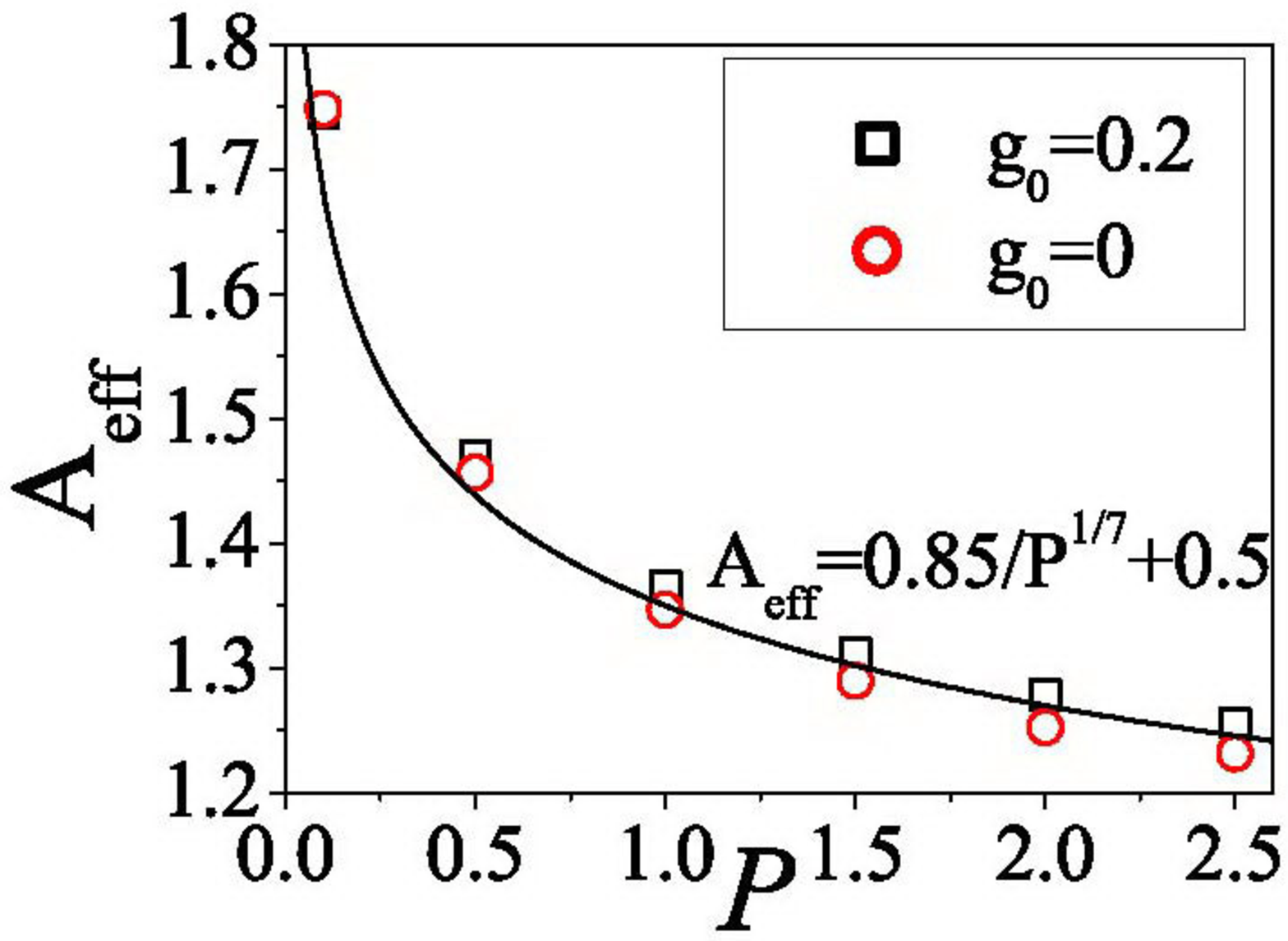}}
\caption{(Color online) (a) An example of a 1D fundamental soliton with $P=1$
and $g_{0}=0.2$. (b) Stable perturbed evolution of this soliton. (c) The
chemical potential versus the total norm for 1D fundamental-soliton families
with different values of $g_{0}$ [see Eq. (\protect\ref{d})]. (d) The
spatial width, $\mathrm{A_{eff}}$, defined by the 1D counterpart of
expression (\protect\ref{Aeff}), versus $P$ for the same families of the
fundamental solitons. In panel (c), continuous curves display a fit of the
numerical results to power-law approximations. (e)-(h) The same as in panel
(a)-(d), but for twisted 1D solitons. In particular, panel (e) pertains to $%
P=1$ and $g_{0}=0.2$.}
\label{1Dfundamental}
\end{figure}

Next, we report the results obtained in the basic model with $\chi =0$, to
which the more general one may be reduced as discussed in Section I. In this
case, the numerical solution of the 1D version of Eq. (\ref{scaledGPE})%
\textbf{\ }produces families of fundamental (spatially even) and twisted
(odd) solitons. Typical examples are displayed in Fig. \ref{1Dfundamental}
[higher-order (\textit{multipole}) localized 1D modes can be easily found
too]. In particular, Figs. \ref{fig2c} and \ref{fig3c} show that the soliton
families satisfy the so-called anti-Vakhitov-Kolokolov criterion, $d\mu
/dP>0 $, which plays the role of a necessary stability condition for bright
solitons in self-repulsive media \cite{anti}. Indeed, direct simulations of
the perturbed evolution of the solitons, performed in the framework of Eq. (%
\ref{GPE}), confirm that the entire families of the fundamental and twisted
solitons are stable, see examples of the stability test in Figs. \ref%
{1Dfundamental}(b) and (f).

For $g_{0}=0$, a simple analysis of Eq. (\ref{scaledGPE}) demonstrates that $%
\mu $ scales as $r_{0}^{-2}$, whereas the effective self-trapping size, $%
r_{0}$ (provided that it is essentially larger than the transverse
thickness, $\varepsilon $), scales with the total norm, $P$, exactly as in
Eq. (\ref{scaling}). From here, the following scaling can be predicted for $%
\alpha =4$ and $g_{0}=0$:%
\begin{equation}
\mu \left( g_{0}=0\right) \sim P^{\frac{2}{2\alpha -1}}\equiv P^{2/7},
\label{2/7}
\end{equation}%
which pertains to the fundamental and twisted modes alike, and is quite
close to the scaling exponent, $\approx 1/3$, found as the best fit of the
numerically found dependences, $\mu (P)$ (for both the fundamental and
twisted solitons), to power-law functions in Figs. \ref{fig2c} and \ref%
{fig3c}. This scaling is specific to the solitons in the nonlocal model,
while in the local one it is completely different, $\mu \sim P$, for $D=1$
and $2$ alike \cite{Olga}.

The presence of $g_{0}>0$ in Eq. (\ref{d1D}) does not affect very broad
solitons corresponding to small $P$, hence the curves corresponding to $%
g_{0}=0$ and $g_{0}=0.2$ in Figs. \ref{fig2c} and \ref{fig3c} start from the
same point at the smallest value of $P$. On the other hand, for narrow
solitons with large $P$, one can still use the asymptotic equation (\ref%
{asympt}) for the soliton's tail (at $|x|\rightarrow \infty $), while inside
the integral one may substitute $g(x)\approx g_{0}$, as suggested by Eq. (%
\ref{d1D}). This means that, for the narrow solitons, scaling relations are
obtained in the form of Eqs. (\ref{scaling}) and (\ref{2/7}), but with $%
2\alpha $ replaced by $\alpha $. In particular, for $\alpha =4$ Eq. (\ref%
{2/7}) is replaced by%
\begin{equation}
\mu \left( g_{0}>0\right) \sim P^{\frac{2}{\alpha -1}}\equiv P^{2/3},
\label{2/3}
\end{equation}%
which is reasonably close to the empirically found fitting exponent $5/6$
quoted in Figs. \ref{fig2c} and \ref{fig3c}.

As mentioned above, the derivation of the scaling relations (\ref{2/7}) and (%
\ref{2/3}) does not depend on the type of the self-trapped mode
(fundamental/twisted), in agreement with the numerical results presented in
Figs. \ref{fig2c} and \ref{fig3c}. Higher-order multipole modes, which are
not considered here, are expected to feature the same scaling too. On the
other hand, the local model with the spatially growing strength of the
self-repulsion \cite{Olga,B2} suggests that instability may appear in
families of higher-order modes.

Further, Figs. \ref{fig2d} and \ref{fig3d} show that, quite naturally, the
spatial size of the fundamental and twisted modes decreases with the
increase of the total norm, cf. Eq. (\ref{scaling}). In this connection, Eq.
(\ref{scaling}) predicts, for $\alpha =4$, $A_{\mathrm{eff}}\sim P^{-1/7}$
in 1D, which is in accordance with the empirically found scaling exponents
in Figs. \ref{fig2d} and \ref{fig3d}.

\subsection{The Thomas-Fermi approximation for 1D fundamental solitons}

As said above, the TFA very accurately predicts properties of fundamental
solitons self-trapped in the model with the local strength of the
self-repulsive contact nonlinearity growing as $r^{\alpha }$ \cite{Olga}.
This fact suggests to try the same approximation in the present model, which
implies the consideration of the limit of $m\rightarrow \infty $ in Eq. (\ref%
{scaledGPE}). Unlike the local case, the TFA for the nonlocal equation
cannot be solved analytically.

\begin{figure}[tbp]
\centering\subfigure[] {\label{fig4a}
\includegraphics[scale=0.3]{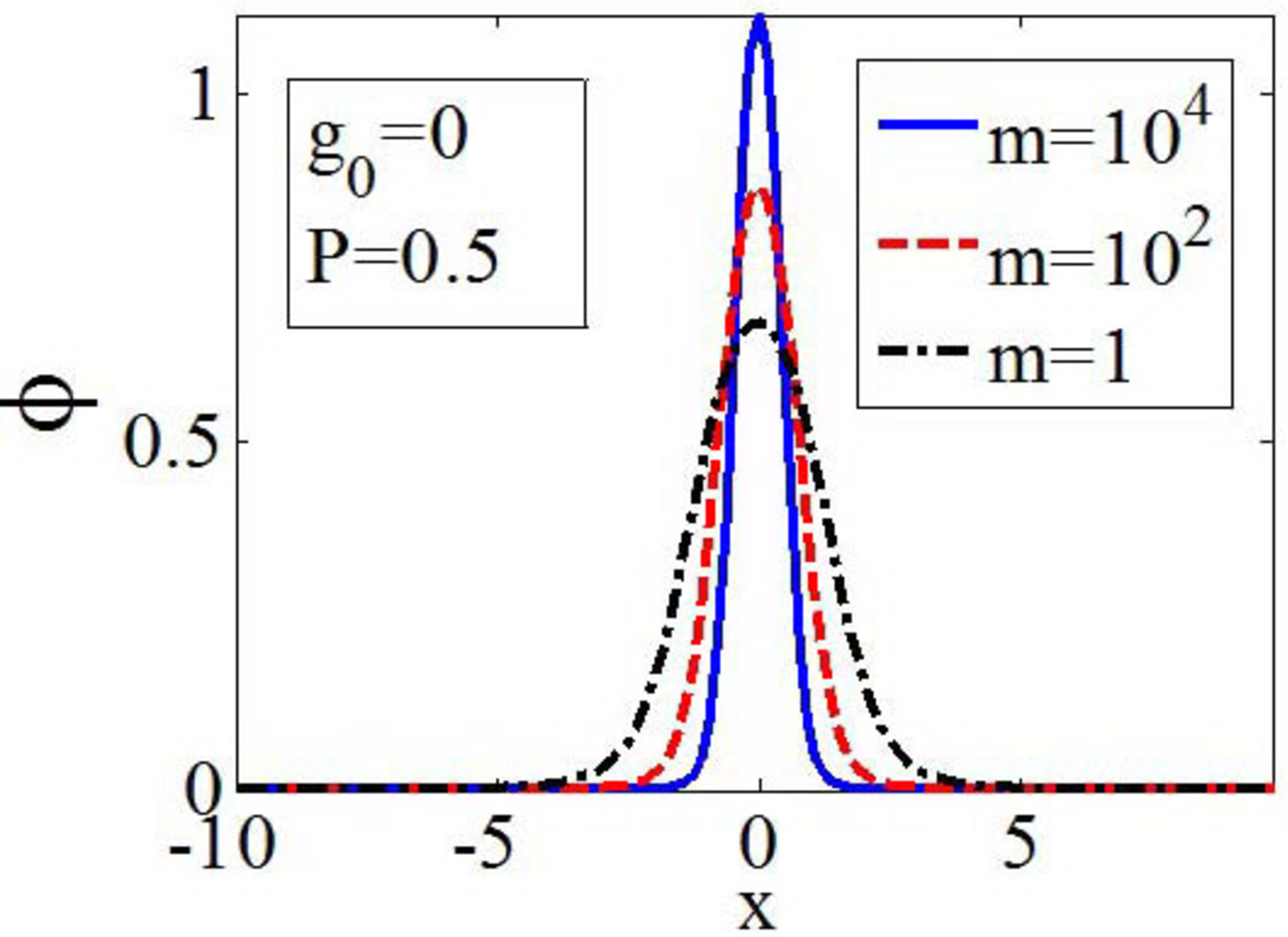}}%
\subfigure[] {\label{fig4b}
\includegraphics[scale=0.3]{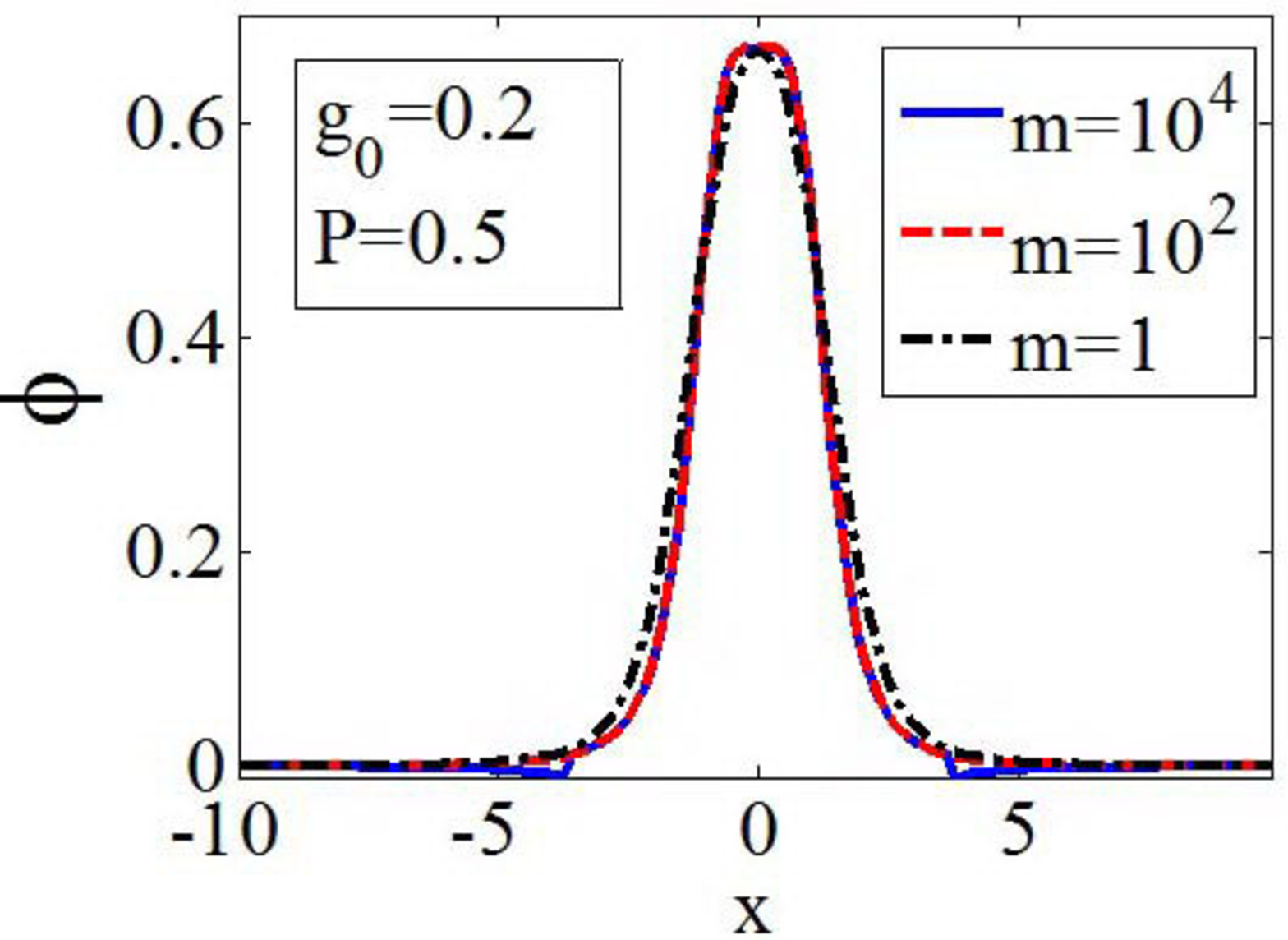}}
\subfigure[] {\label{fig4c}
\includegraphics[scale=0.22]{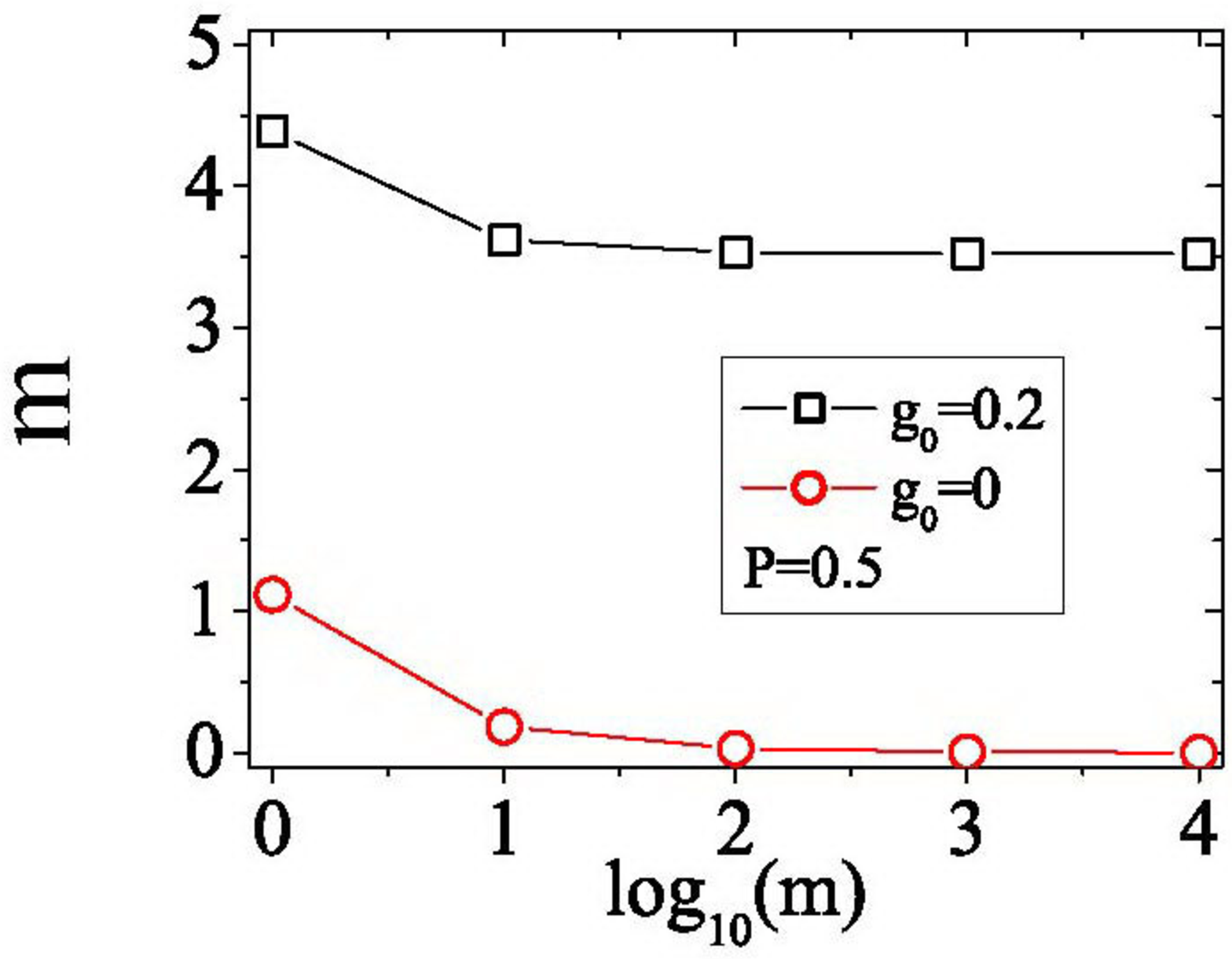}}
\subfigure[] {\label{fig4d}
\includegraphics[scale=0.22]{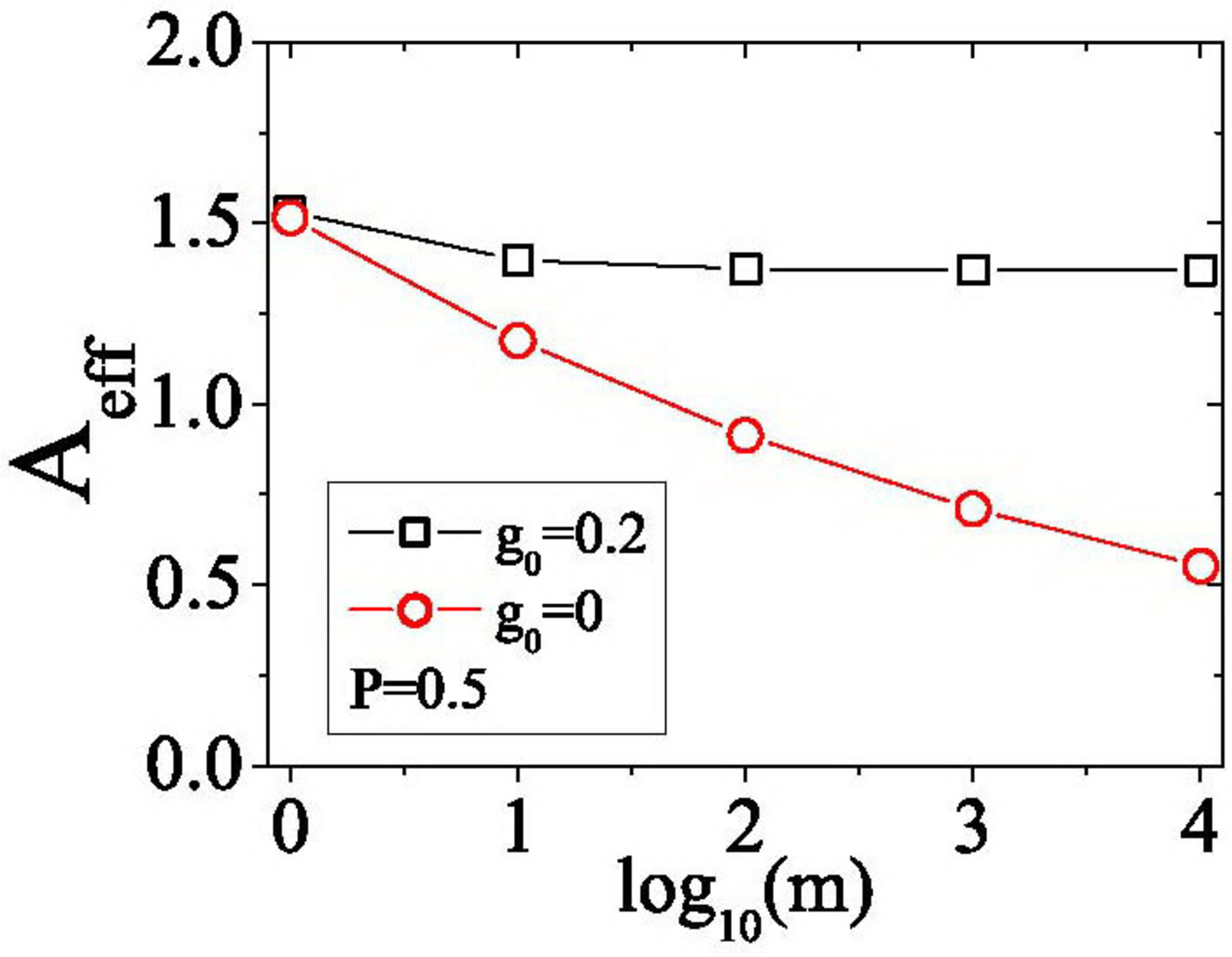}}
\caption{(Color online) (a) Profiles of 1D fundamental solitons for a
gradually increasing mass parameter, $m$ [see Eq. (\protect\ref{scaledGPE}%
)], and $g_{0}=0$. (b) The same for $g_{0}=0.2$. (c,d) The chemical
potential and spatial size of the soliton versus $m$ for different values of
$g_{0}$. The case of large $m$ represents the Thomas-Fermi approximation in
the model with the long-range interactions.}
\label{1DTFA}
\end{figure}

Figure \ref{1DTFA} shows a set of profiles of 1D fundamental solitons,
produced by the numerical solution of Eq. (\ref{scaledGPE}) for increasing $%
m $, at different values of $g_{0}$, along with the chemical potential and
effective size of the solitons as functions of $m$. The results demonstrate
that the TFA is very accurate at $g_{0}>0$, but it fails for $g_{0}=0$. This
conclusion is not surprising, as the validity of the TFA is predicated on
the presence of a nonvanishing self-repulsive nonlinearity.

\subsection{Numerical results for two-dimensional solitons}

\begin{figure}[tbp]
\centering\subfigure[] {\label{fig5a-a}
\includegraphics[scale=0.3]{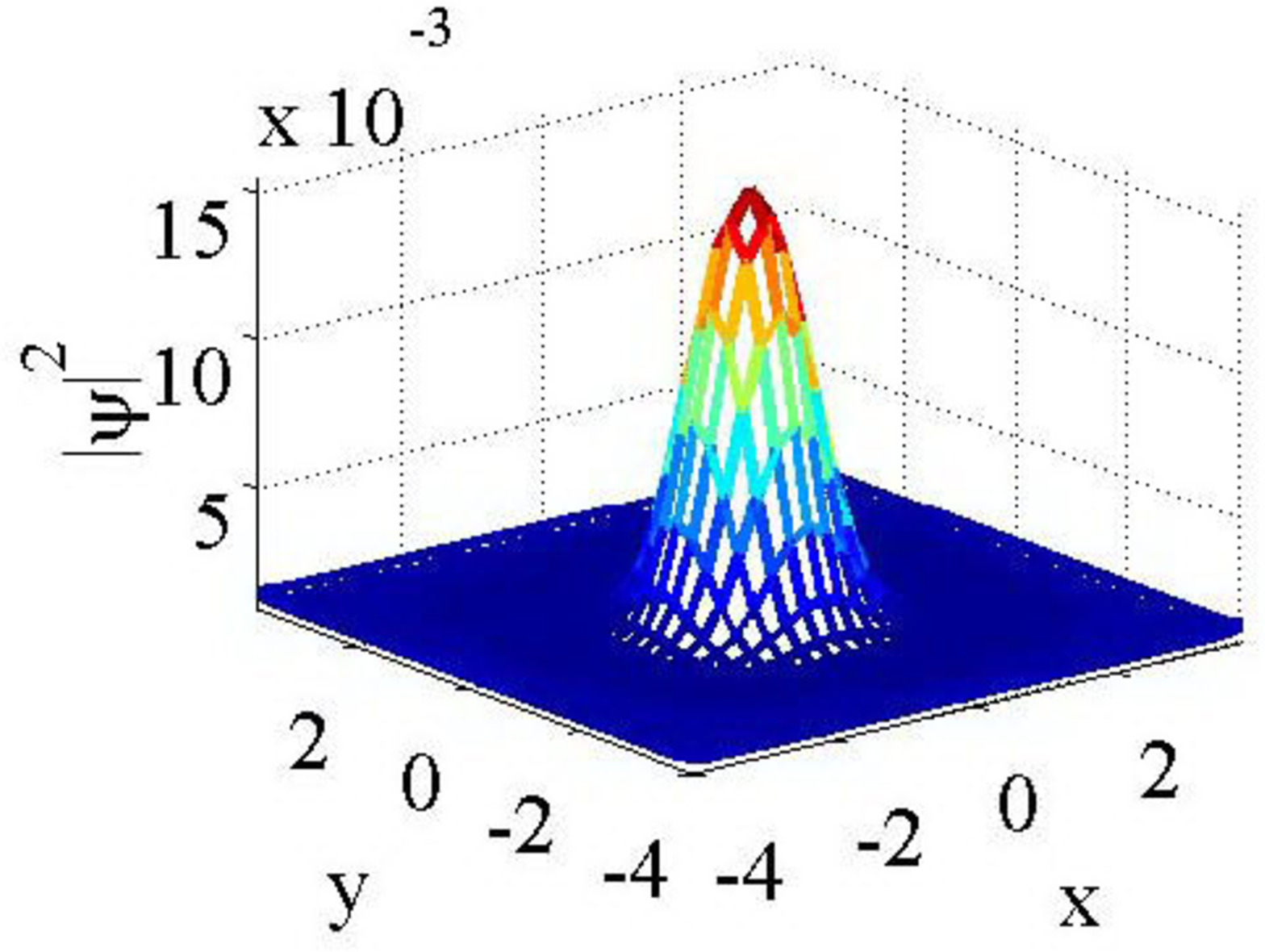}}%
\subfigure[] {\label{fig5b-a}
\includegraphics[scale=0.3]{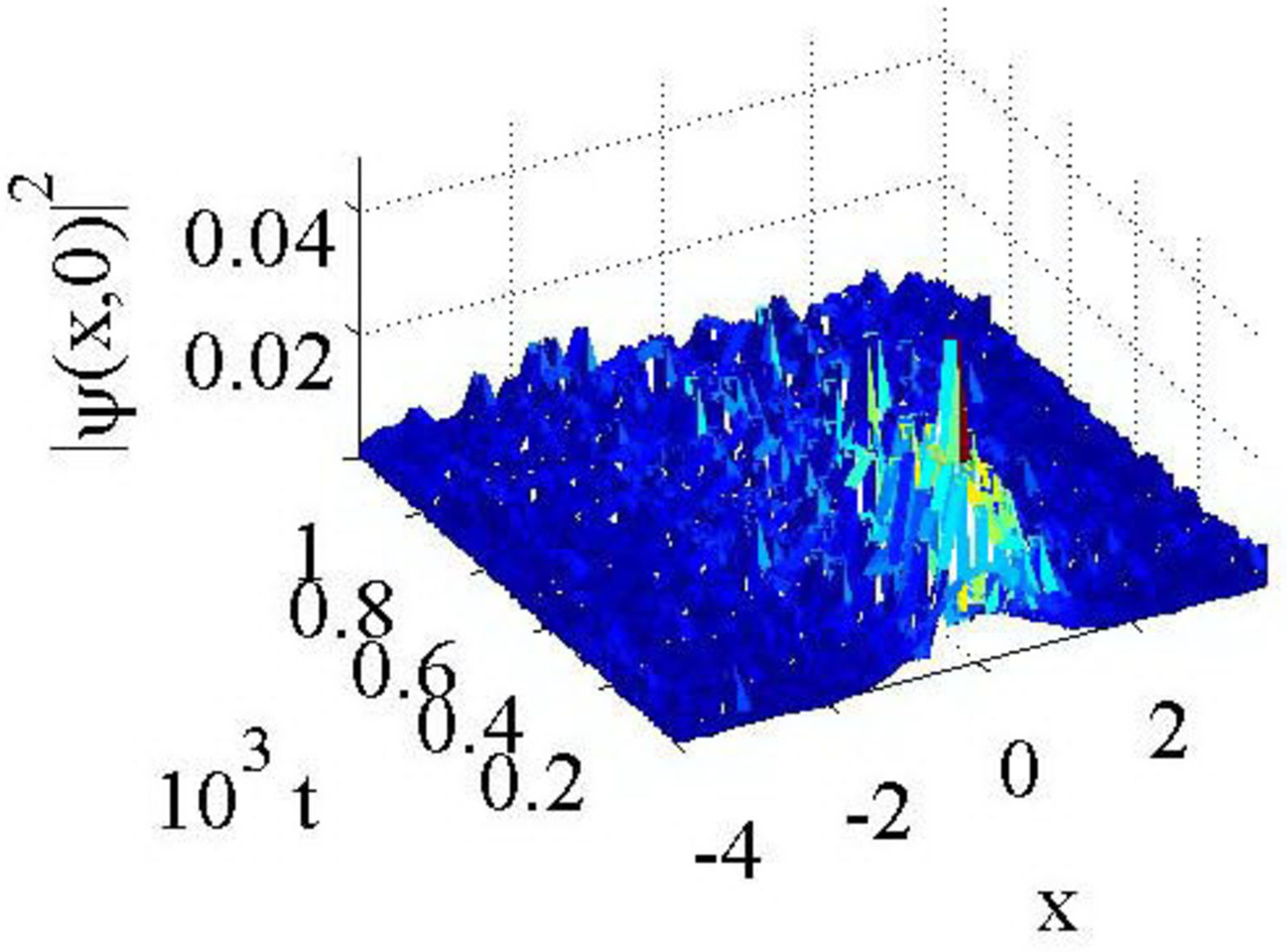}}
\subfigure[] {\label{fig5c-a}
\includegraphics[scale=0.3]{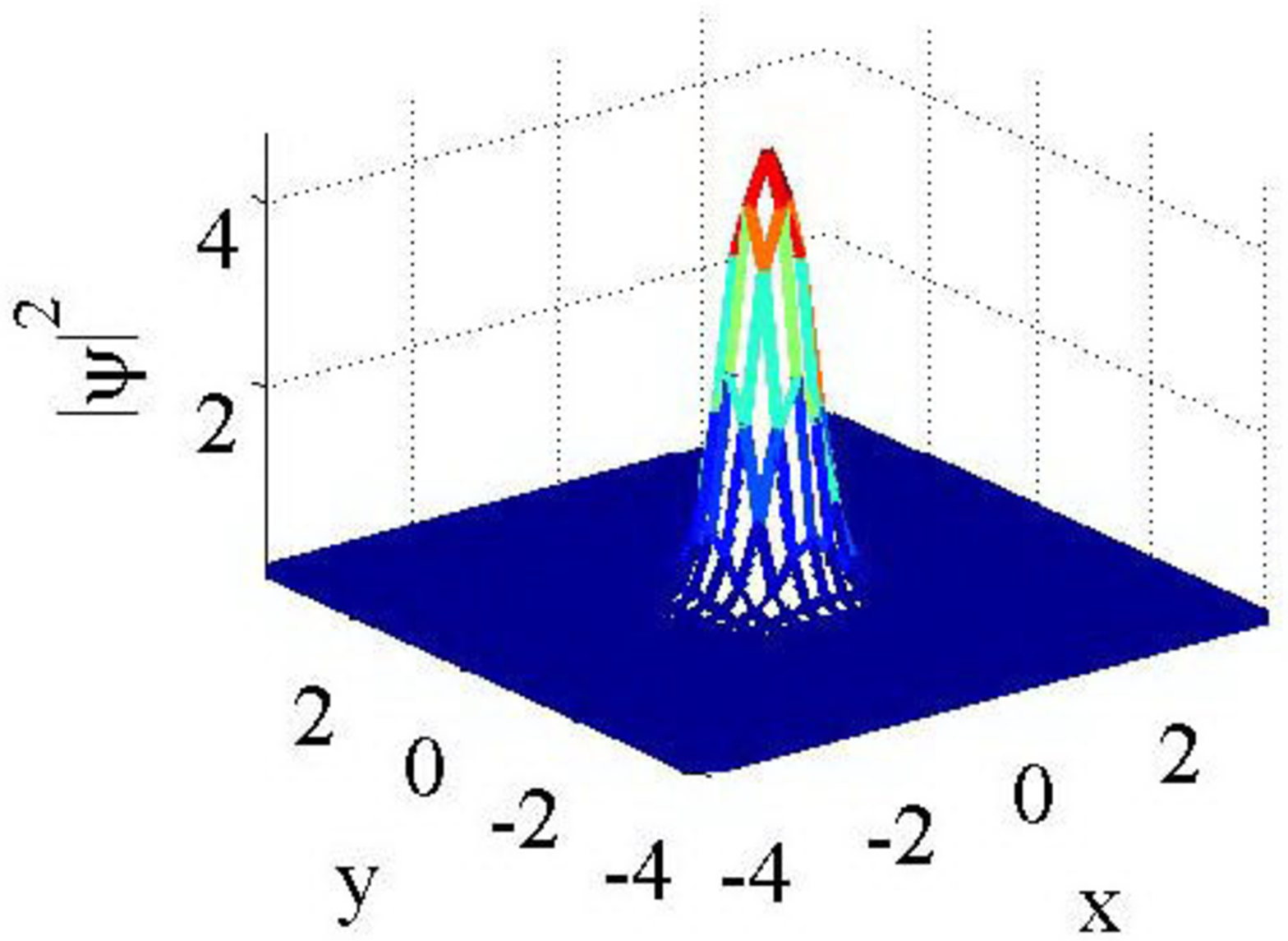}}
\subfigure[] {\label{fig5d-a}
\includegraphics[scale=0.3]{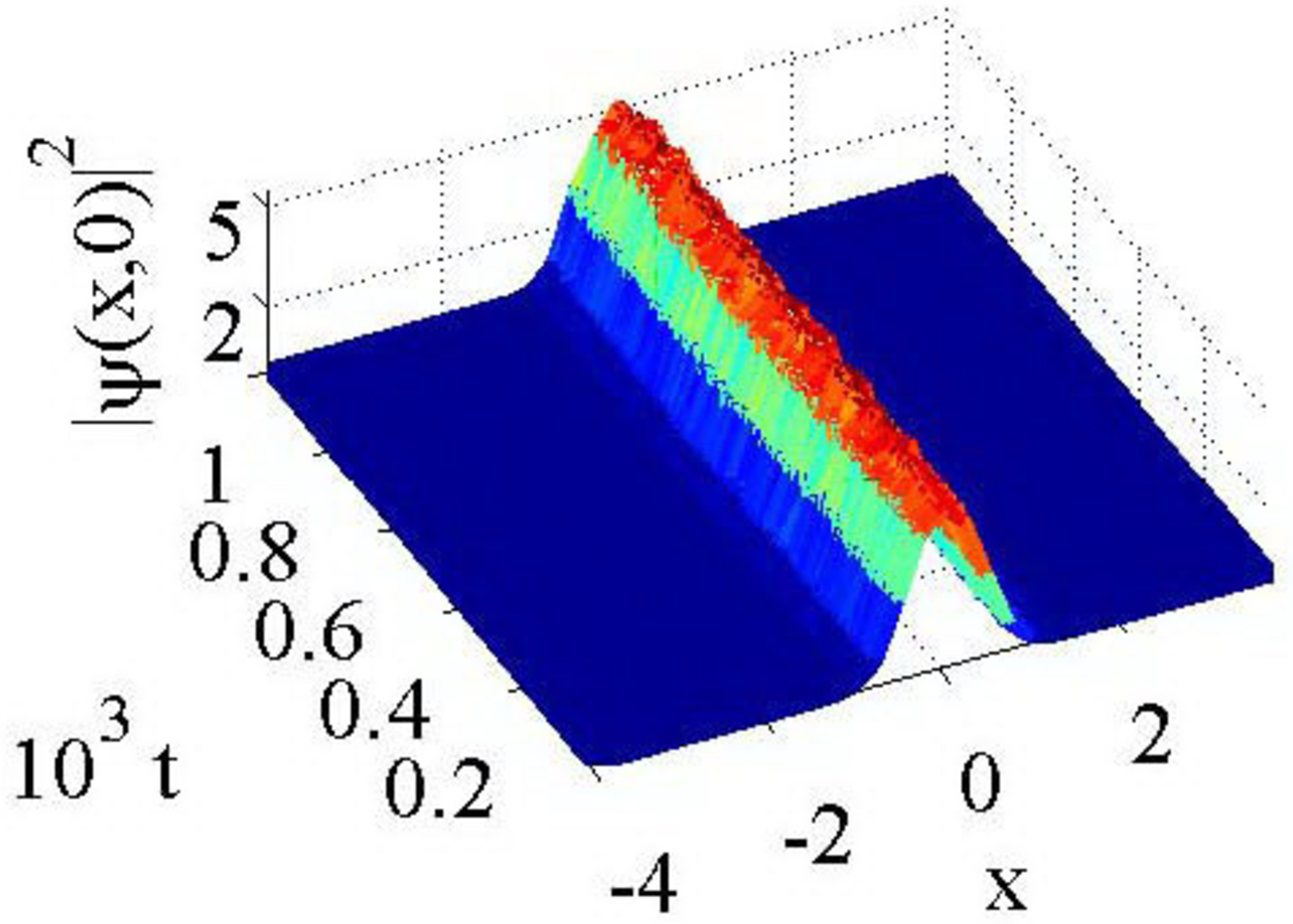}}
\caption{(Color online) (a) The 2D self-trapped mode with $\protect\chi =0.1$%
, $P=0.1$, and $g_{0}=0.2$. (b) Completely unstable perturbed evolution of
the mode from panel (a). (c) A strongly nonlinear 2D mode with $\protect\chi %
=0.1$, $P=5$, and $g_{0}=0.2$. (d) Stable perturbed evolution of the mode
from panel (c).}
\label{2DfundamentalVexp}
\end{figure}

Similar to the 1D case, the numerical solution of the 2D equation (\ref%
{scaledGPE}) was at first performed taking into regard the EP, $\chi >0$. As
Fig. \ref{2DfundamentalVexp} shows, it has been concluded that, like in the
1D model, the 2D modes with a sufficiently strong nonlinearity remain
robust, in agreement with the estimates presented at the end of Section I,
while weakly unstable modes are subject to strong instability.

\begin{figure}[tbp]
\centering\subfigure[] {\label{fig5a}
\includegraphics[scale=0.3]{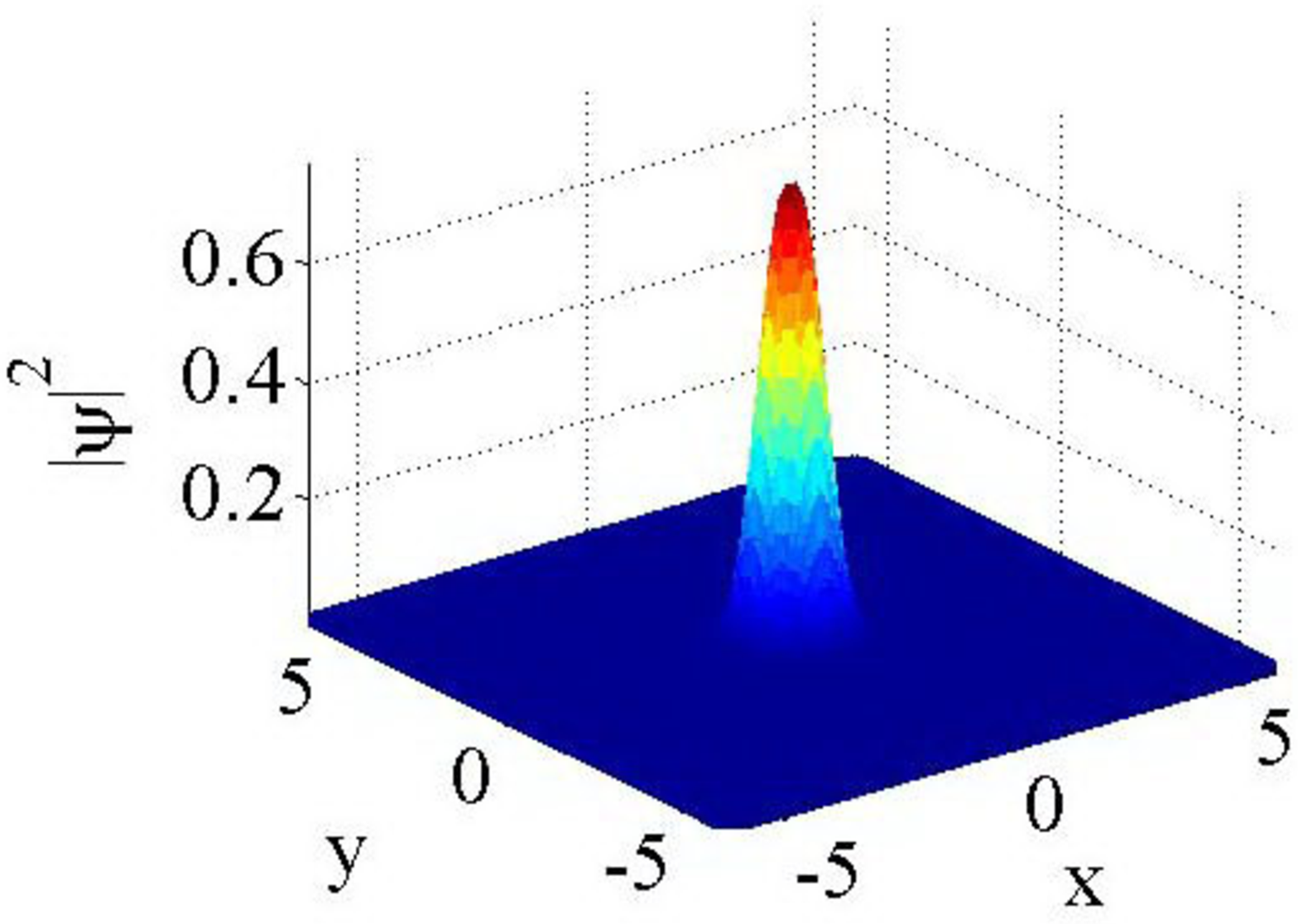}}%
\subfigure[] {\label{fig5b}
\includegraphics[scale=0.3]{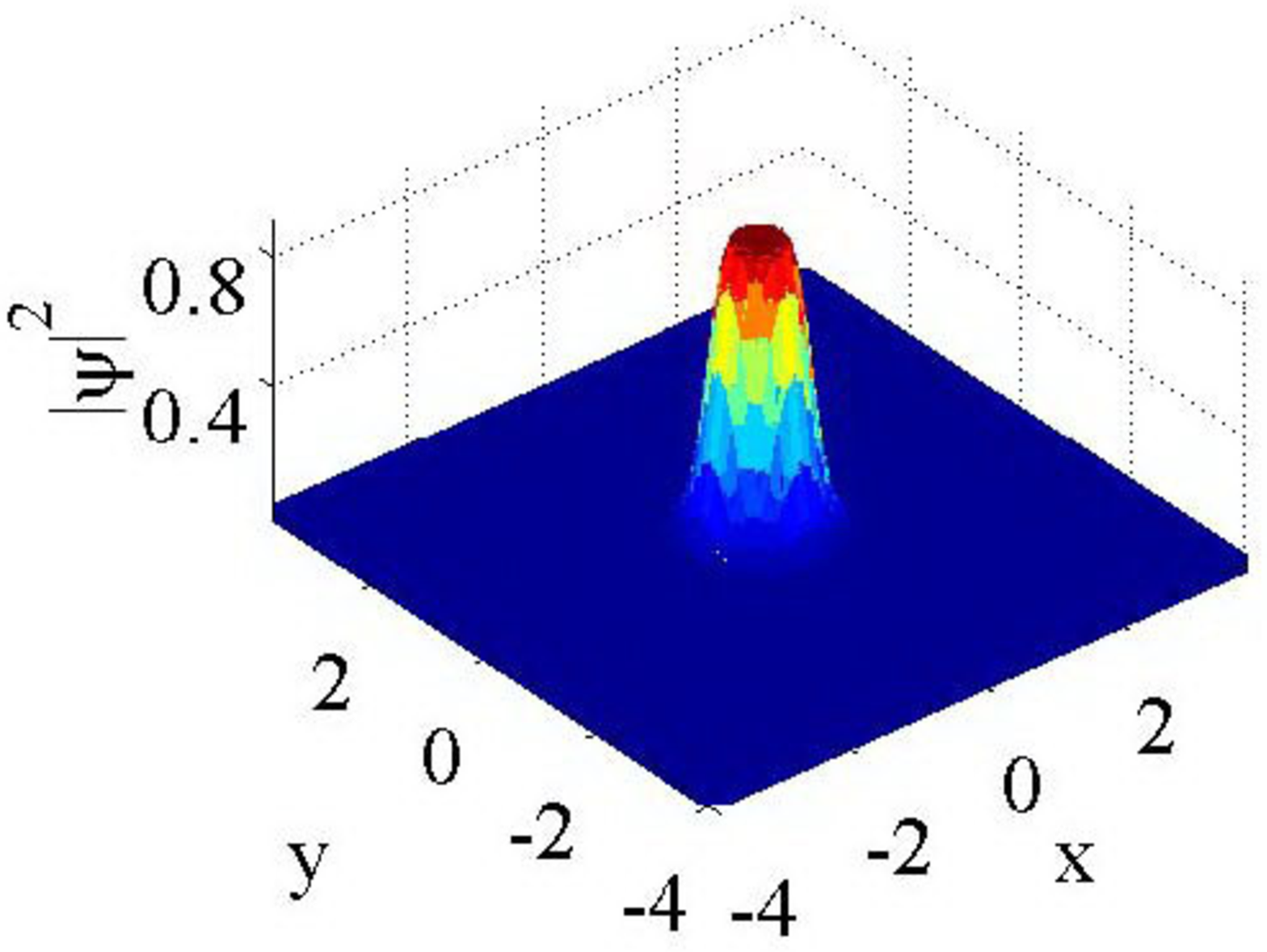}}
\subfigure[] {\label{fig5c}
\includegraphics[scale=0.22]{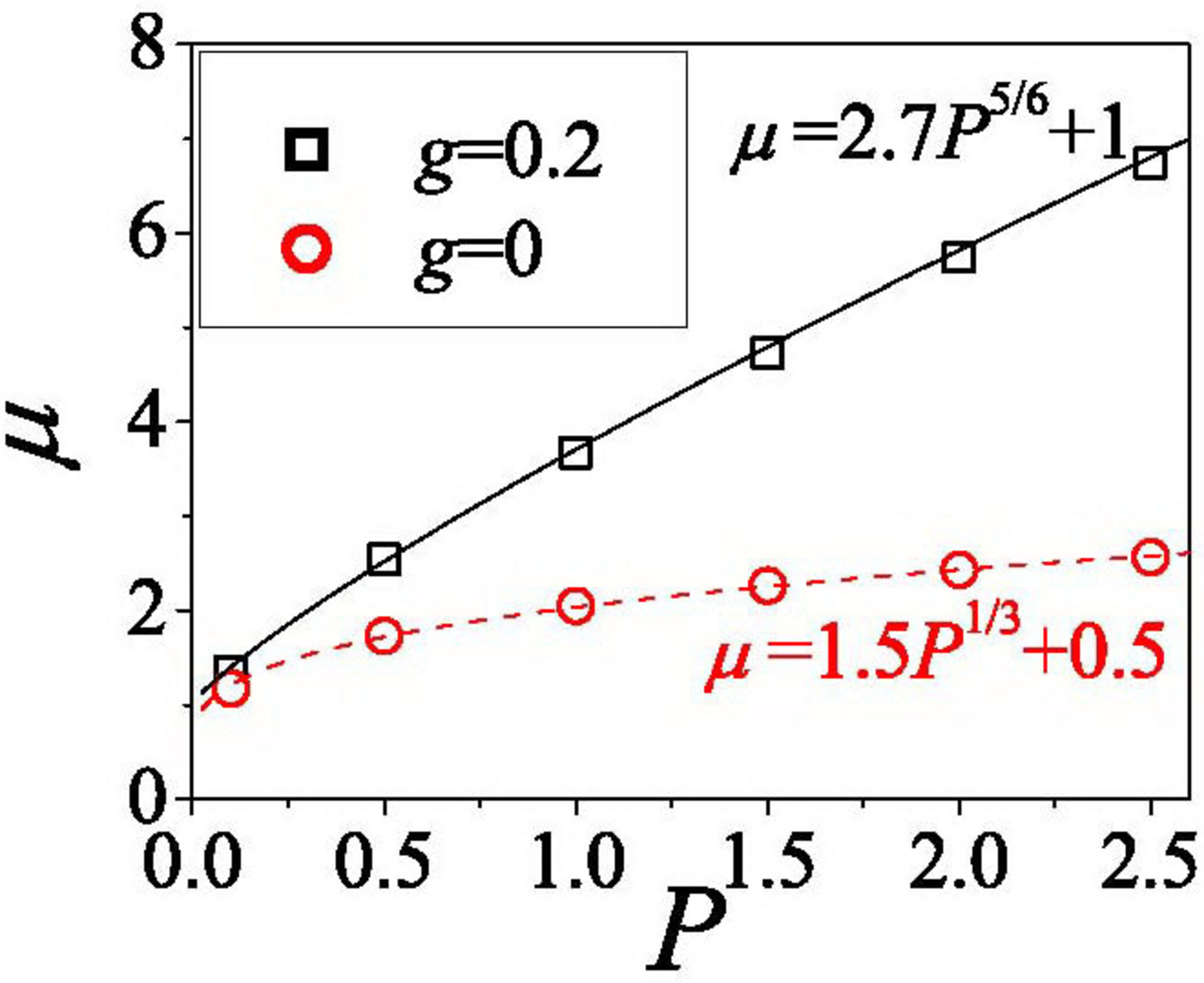}}
\subfigure[] {\label{fig5d}
\includegraphics[scale=0.22]{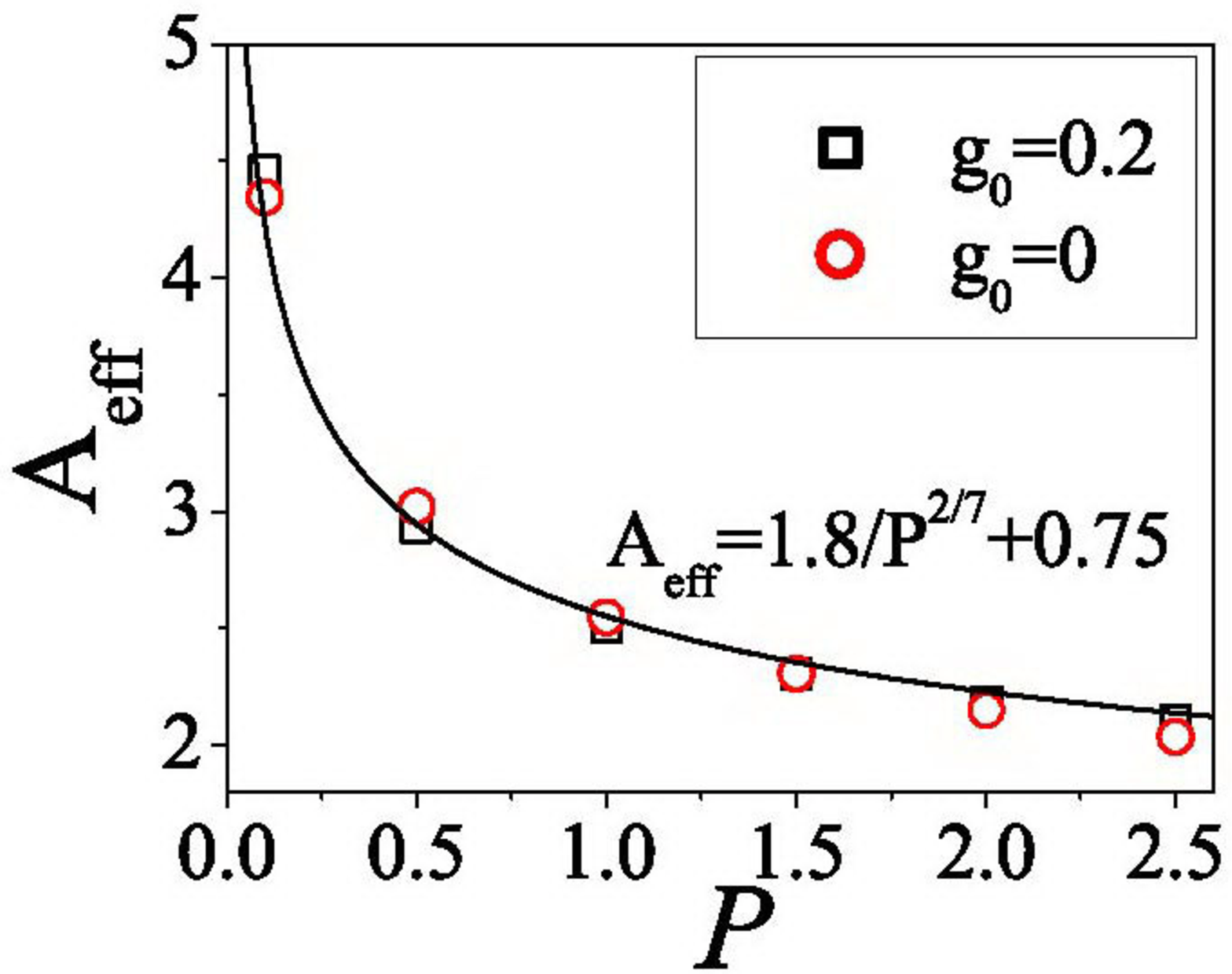}}
\subfigure[] {\label{fig6a}
\includegraphics[scale=0.3]{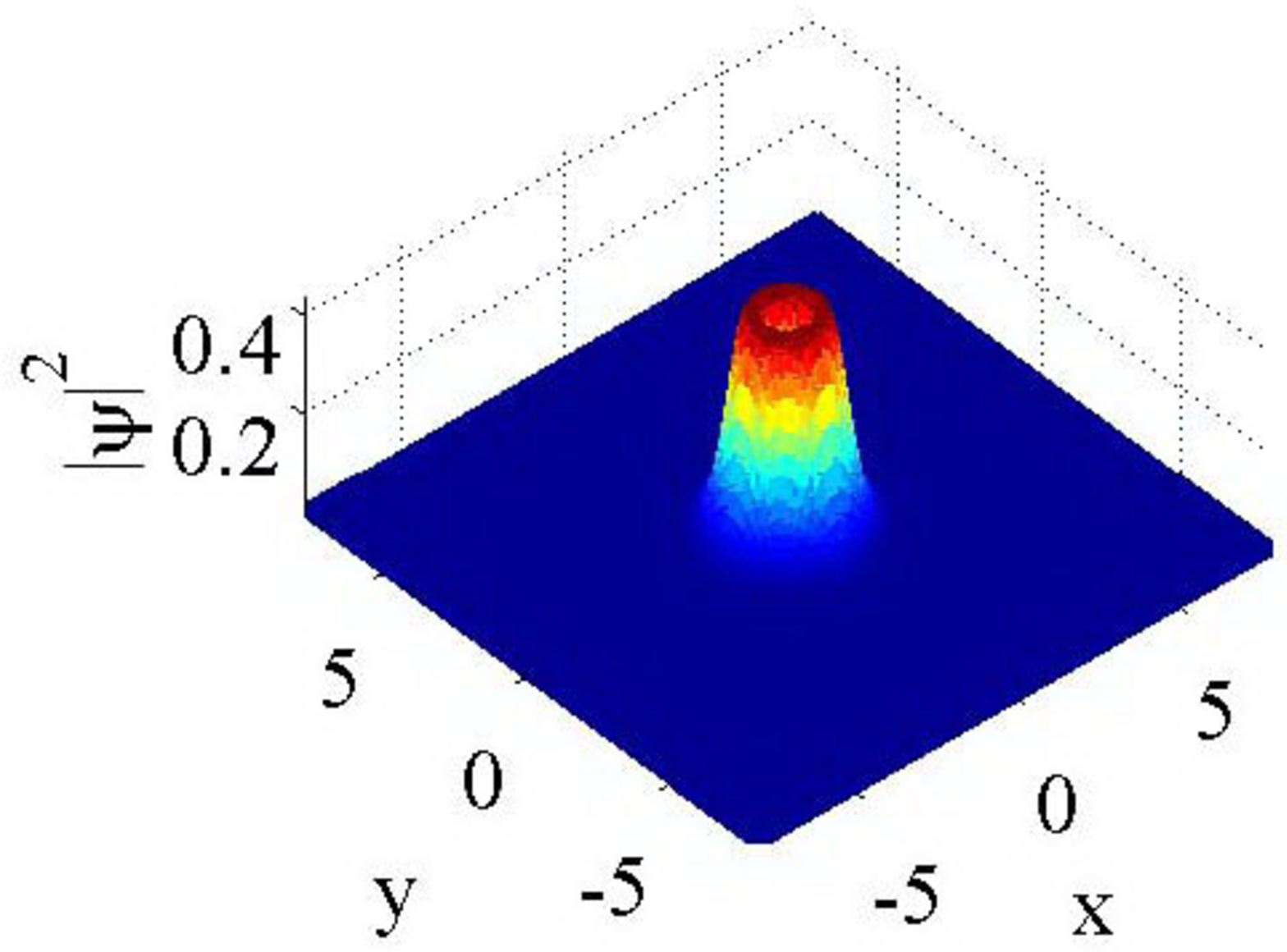}}%
\subfigure[] {\label{fig6b}
\includegraphics[scale=0.3]{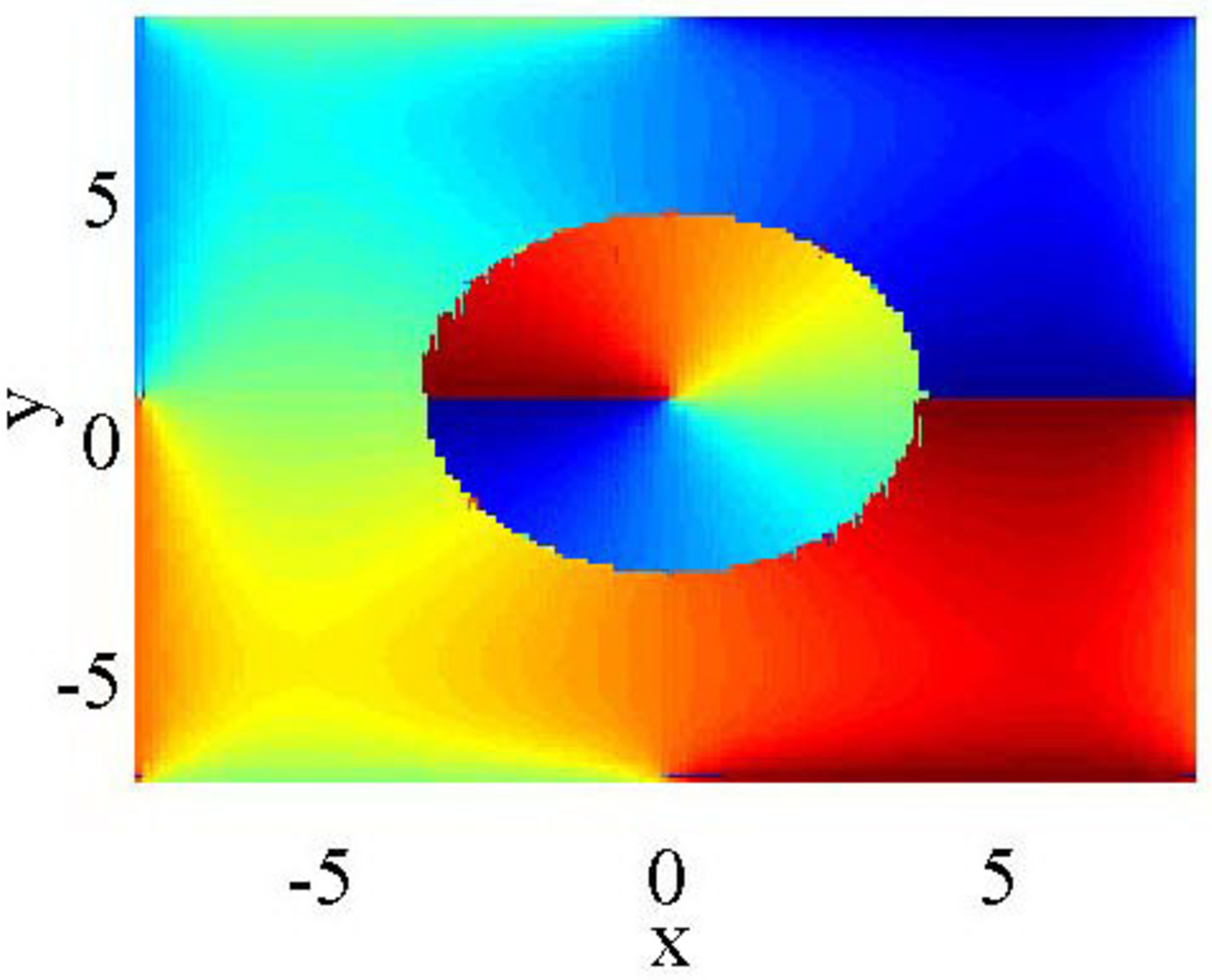}}
\subfigure[] {\label{fig6c}
\includegraphics[scale=0.22]{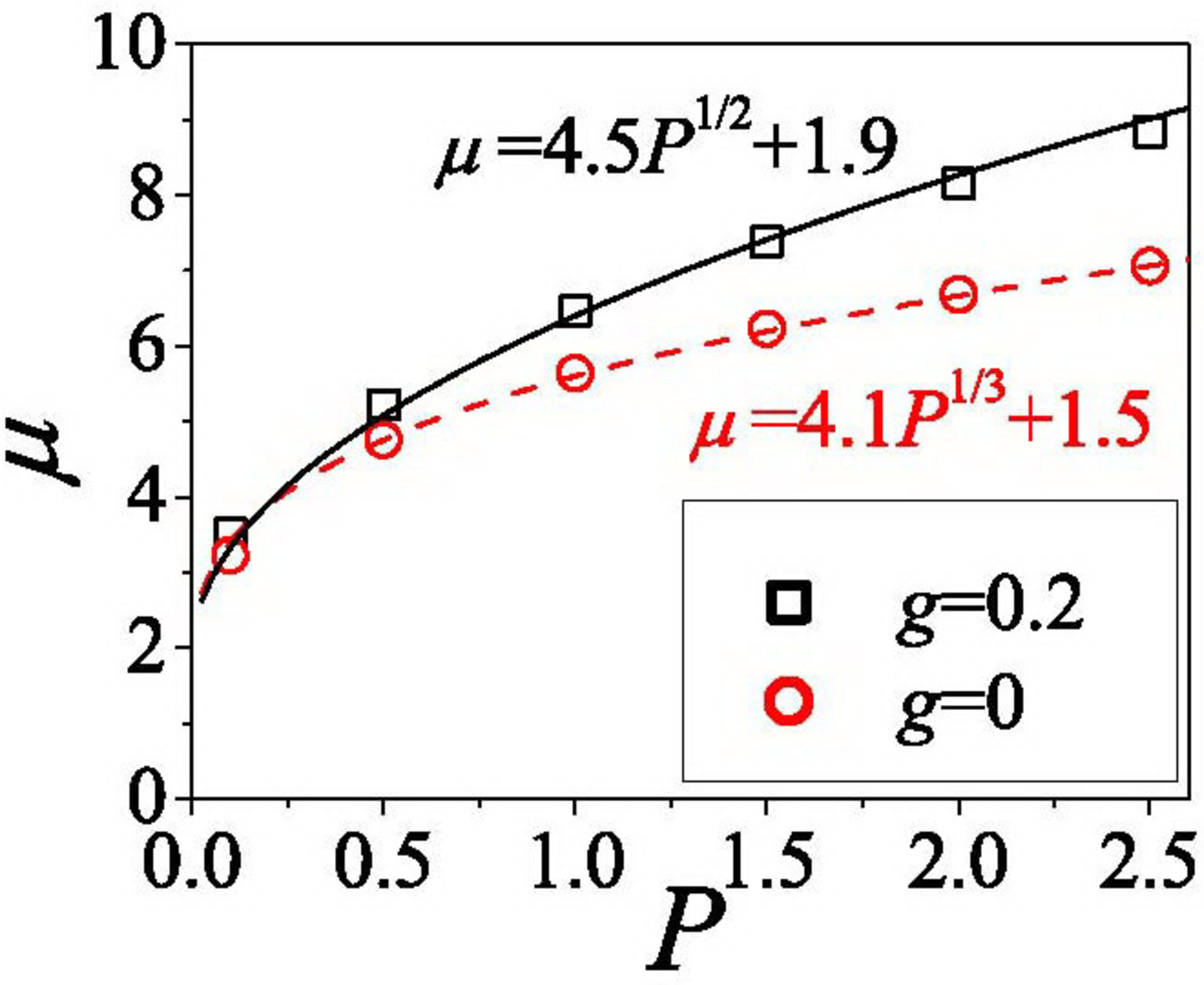}}
\subfigure[] {\label{fig6d}
\includegraphics[scale=0.22]{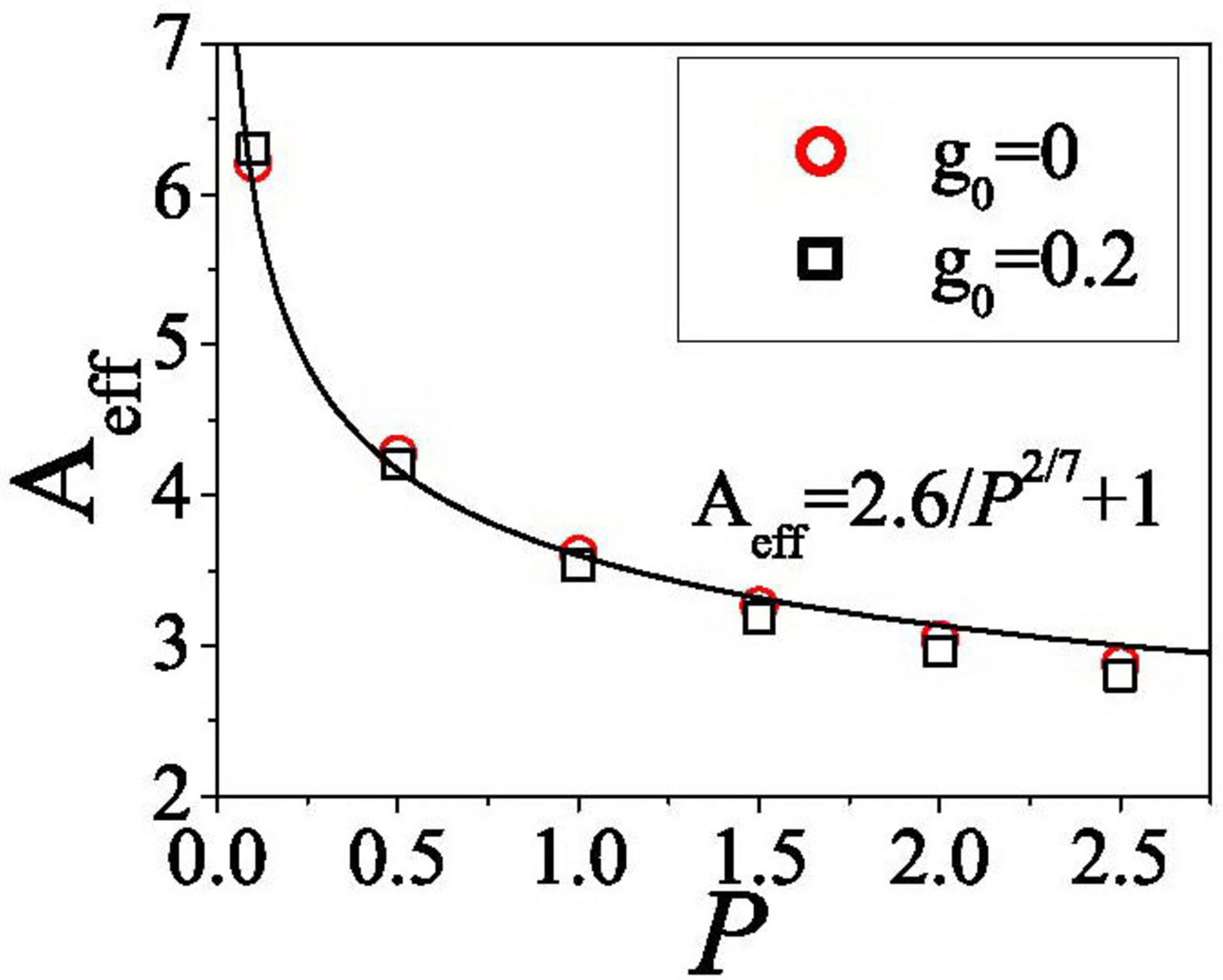}}
\caption{(Color online) 2D self-trapped modes obtained in the model without
the expulsive potential [$\protect\chi =0$ in Eqs. (\protect\ref{GPE}) and (%
\protect\ref{scaledGPE})]. (a) An example of the 2D fundamental soliton with
$P=1$ and $g_{0}=0.2$. (b) The Thomas-Fermi approximation for the same
soliton, obtained from Eq. (\protect\ref{scaledGPE}) with $m=10^{4}$. (c)
The chemical potential of the fundamental 2D solitons versus their total
norm at different values of $g_{0}$. (d) The effective soliton's area $%
\mathrm{A_{eff}}$ [see Eq. (\protect\ref{Aeff})] versus the total norm for
different values of $g_{0}$. (e) The amplitude profile of the 2D vortex
soliton for $P=1$, $g_{0}=0.2$, and vorticity $S=1$. (f) The phase
distribution in this vortex. (g) The chemical potential of the
vortex-soliton family versus the total norm for different values of $g_{0}$.
(h) The effective area, $\mathrm{A_{eff}}$, versus the total norm for the
same families of vortical solitons.}
\label{2DFundamental}
\end{figure}

Further, Eq. (\ref{scaledGPE}) with $\chi =0$ reveals families of stable
fundamental and vortical 2D solitons, which are displayed in Fig. \ref%
{2DFundamental}. Note that panel \ref{2DFundamental}(b) shows the TFA for
the 2D fundamental soliton, with a flat area at the center, which is a
typical feature of that approximation.

The vortices are produced, as usual, by the substitution of $\phi (r,\theta
)=\Phi (r)\exp \left( iS\theta \right) $ in Eq. (\ref{scaledGPE}), where $%
\left( r,\theta \right) $ are the polar coordinates in the 2D plane, $\Phi
(r)$ is a real amplitude function, and $S$\ is integer vorticity (we here
consider only $S=1$). Note that the asymptotic approximation (\ref{far})
applies, at $r\rightarrow \infty $, to the vortices as well as to the 2D
fundamental solitons.

The effective scaling of dependences $\mu (P)$ for $g_{0}=0$ in Figs. \ref%
{2DFundamental}(c) and \ref{2DFundamental}(g), as well as the scaling for
narrow solitons (large $P$) in the case of $g_{0}>0$ in Eq. (\ref{d}), which
is also presented in Figs. \ref{2DFundamental}(c) and \ref{2DFundamental}%
(g), is explained by the same relations (\ref{2/7}) and (\ref{2/3}) which
were derived above for the 1D case, as the derivation produces the results
which do not depend on the dimension (the dimension cancels out in the
process of the derivation), nor on the type of the soliton (fundamental or
vortical). In addition to that, a straightforward analysis of the scaling
for the effective area of the 2D solitons yields $A_{\mathrm{eff}}\sim
P^{-2/7}$, which also agrees well with the empiric scaling exponents
indicated in Figs. \ref{fig5d} and \ref{fig6d}.

Numerical tests of the perturbed evolution demonstrate that both the
fundamental and vortex soliton families are entirely stable. Comparison to
the model with the spatially modulated coefficient in front of the local
self-defocusing term \cite{Olga,B2} suggests that instability may arise for
higher-order vortices. This issue is beyond the scope of the present work.

\section{Mobility of the fundamental solitons}

As well as in the model with the spatially modulated strength of the
repulsive local nonlinearity \cite{B2}, it is relevant to consider motion of
stable solitons, which can be naturally initiated by a sudden shift of the
soliton from the central position, and/or by kicking it. In this section, we
study oscillatory and elliptic motion of 1D and 2D solitons, respectively,
in the absence of the EP in Eq. (\ref{GPE}), $\chi =0$.

\begin{figure}[tbp]
\centering\subfigure[] {\label{fig7a}
\includegraphics[scale=0.2]{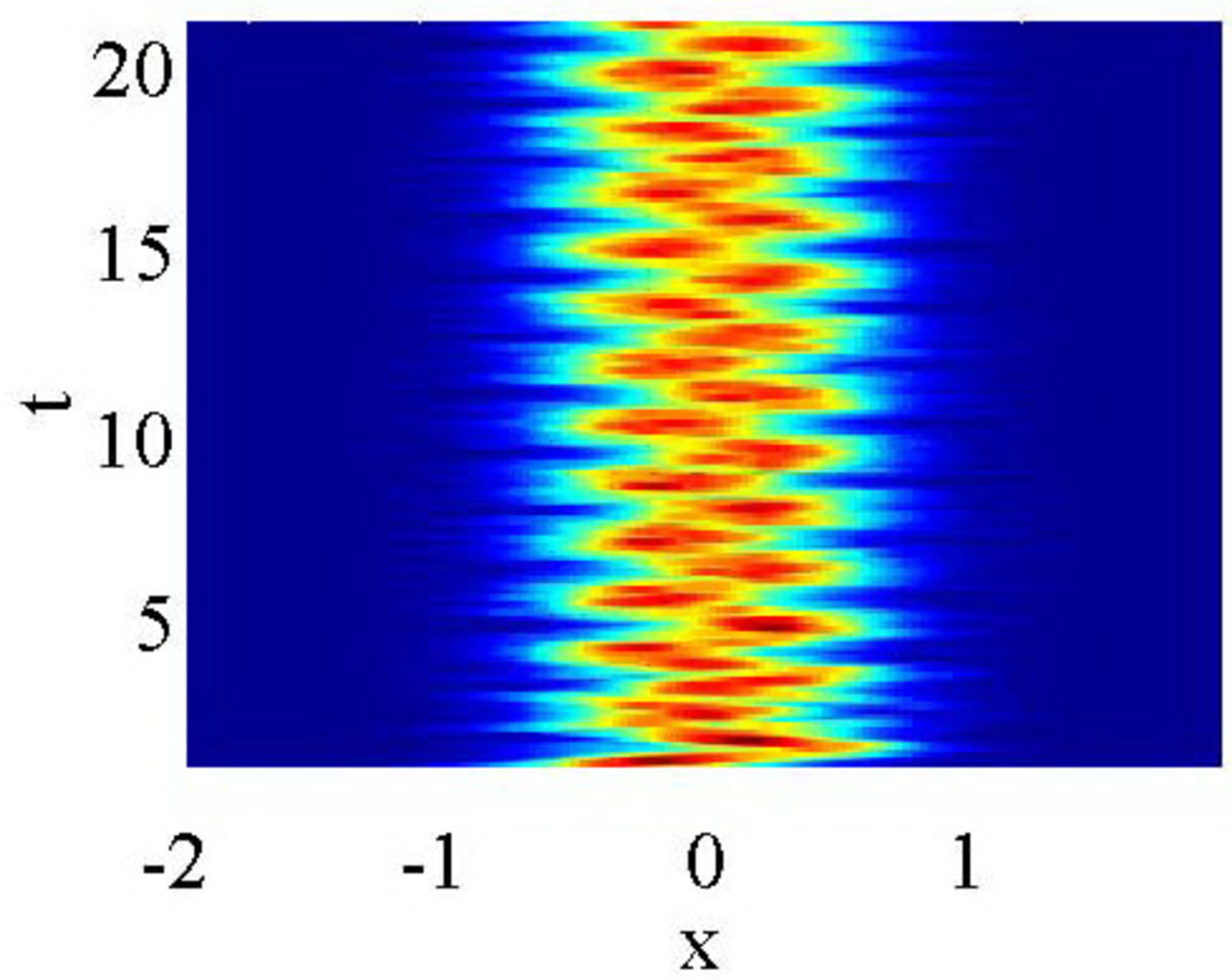}}%
\subfigure[] {\label{fig7b}
\includegraphics[scale=0.2]{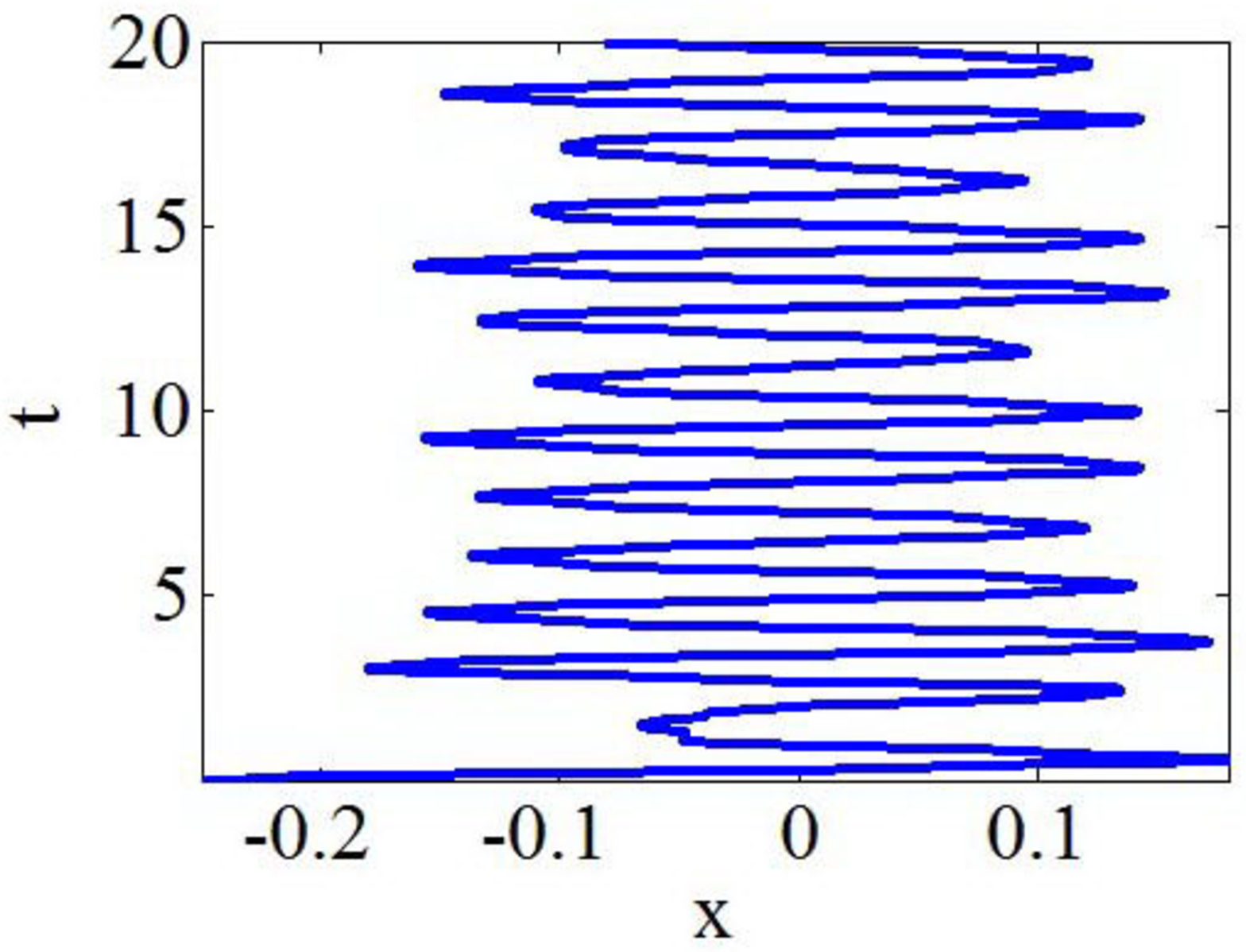}}
\subfigure[] {\label{fig7c}
\includegraphics[scale=0.2]{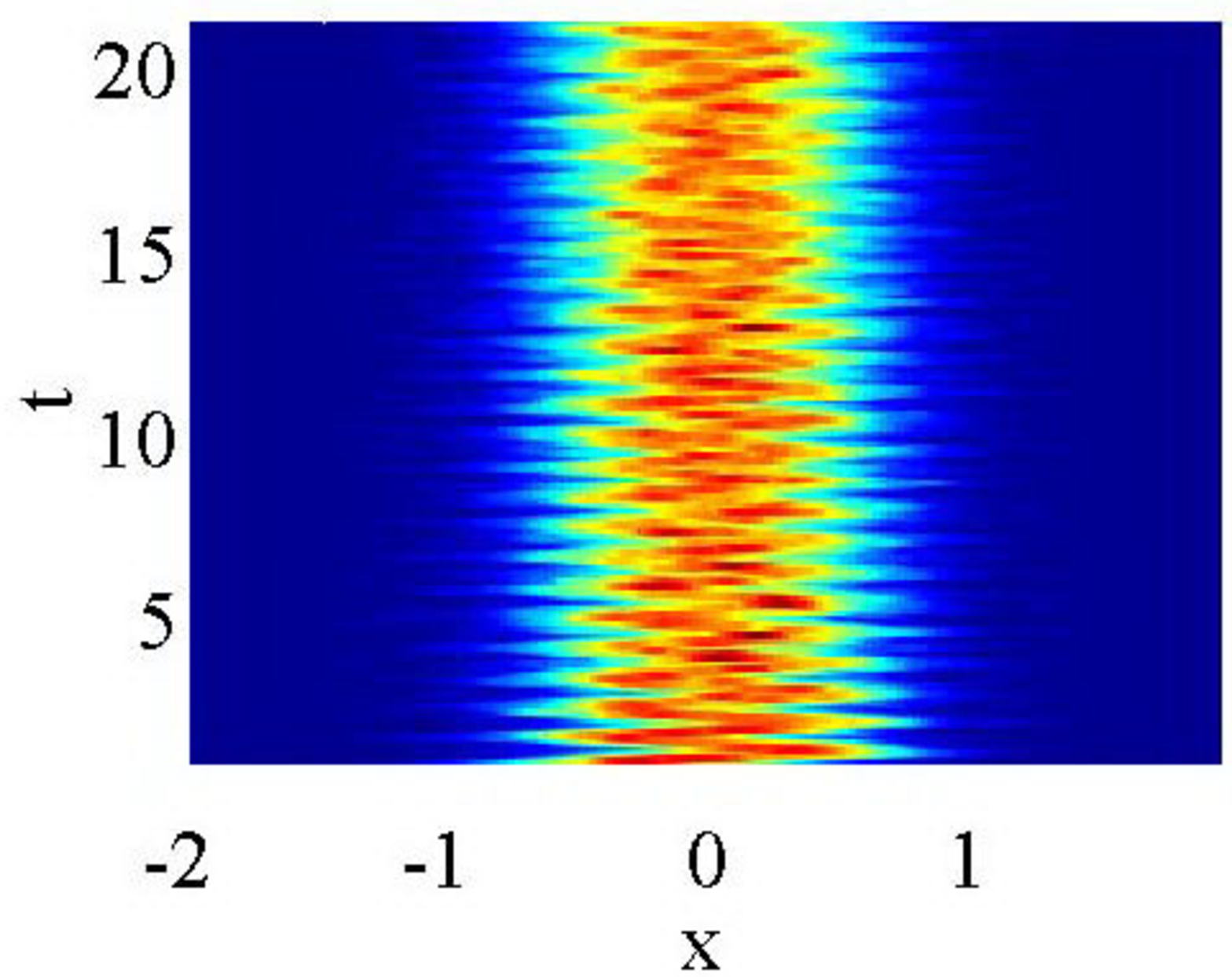}}
\subfigure[] {\label{fig7d}
\includegraphics[scale=0.2]{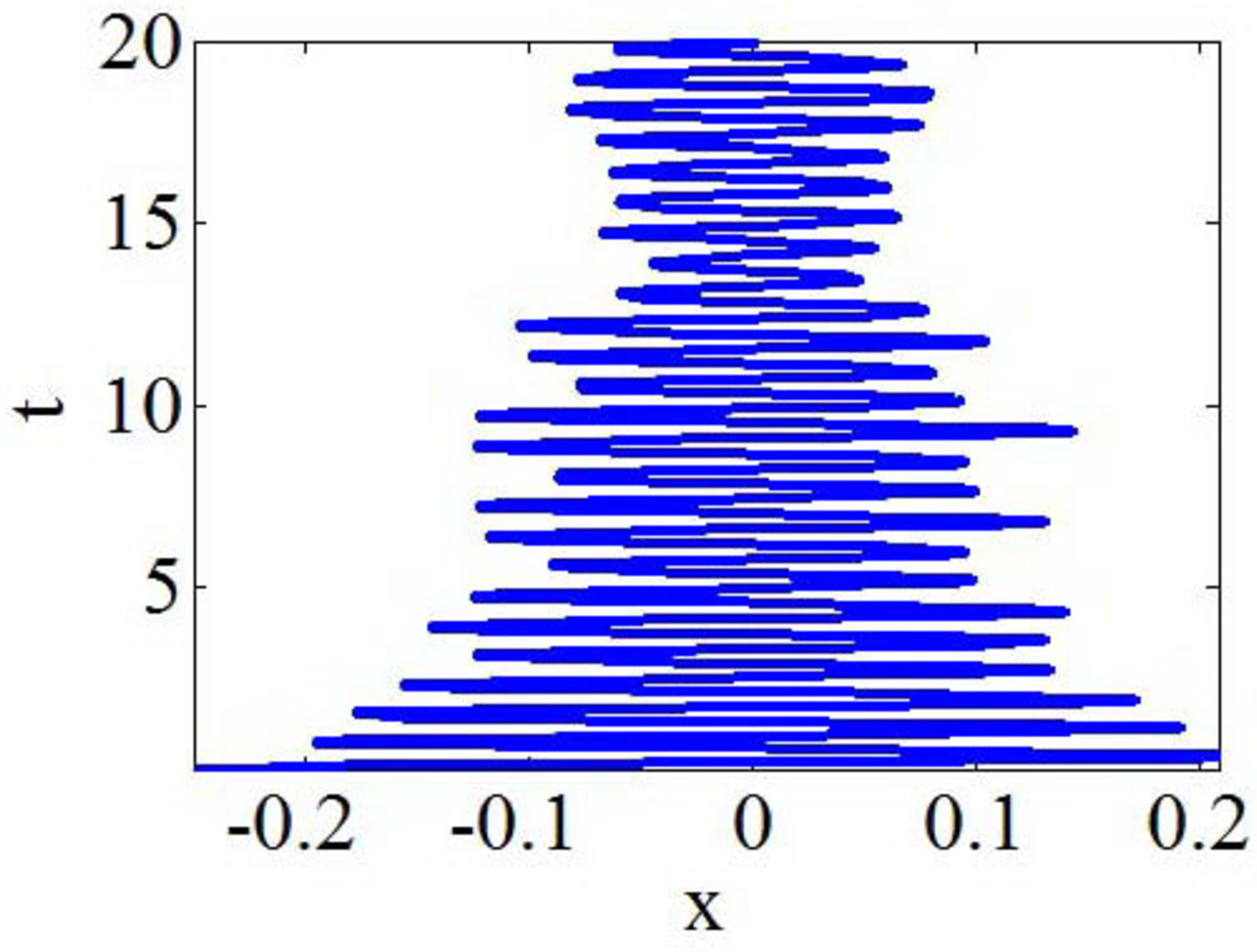}}
\subfigure[] {\label{fig7e}
\includegraphics[scale=0.2]{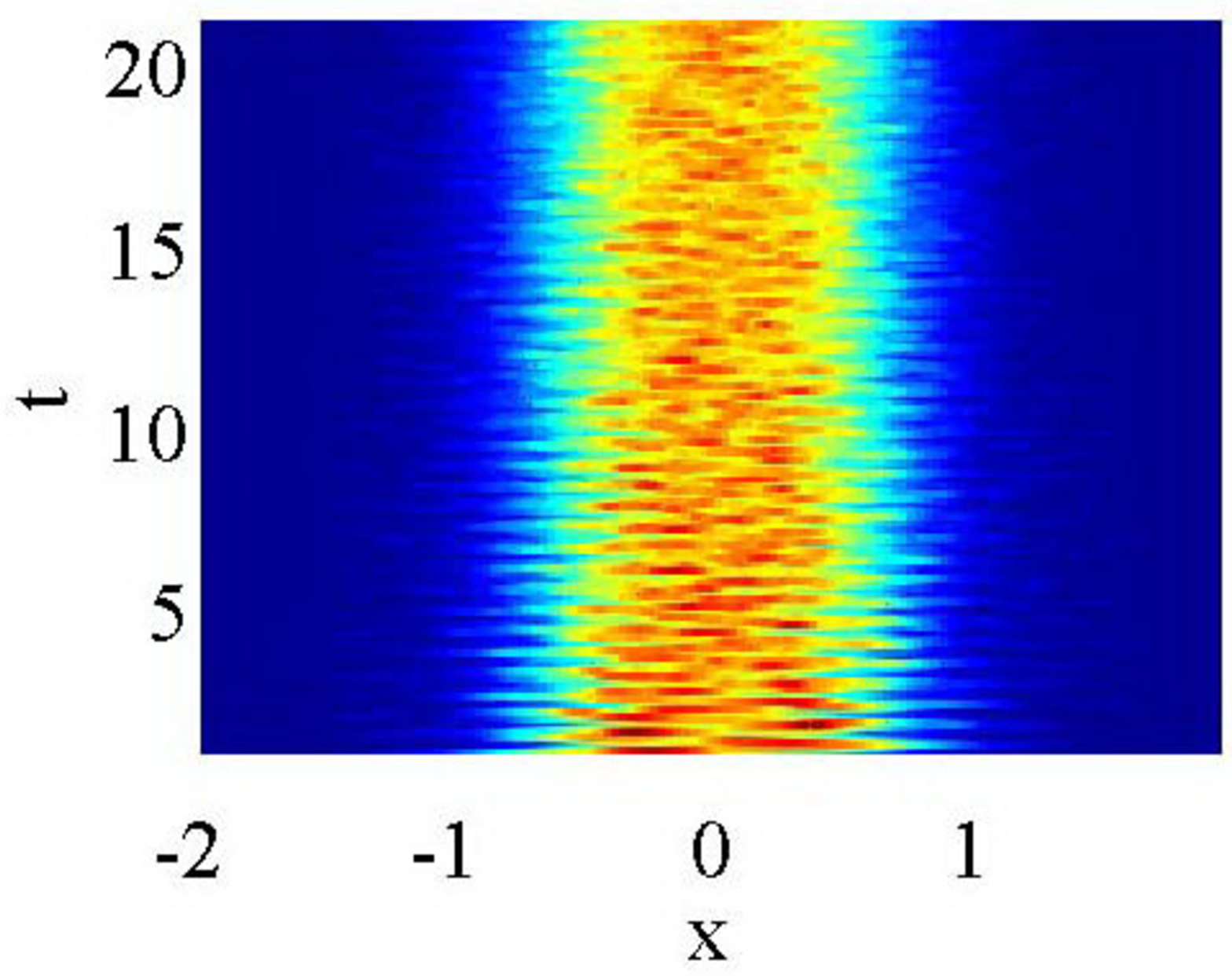}}
\subfigure[] {\label{fig7f}
\includegraphics[scale=0.2]{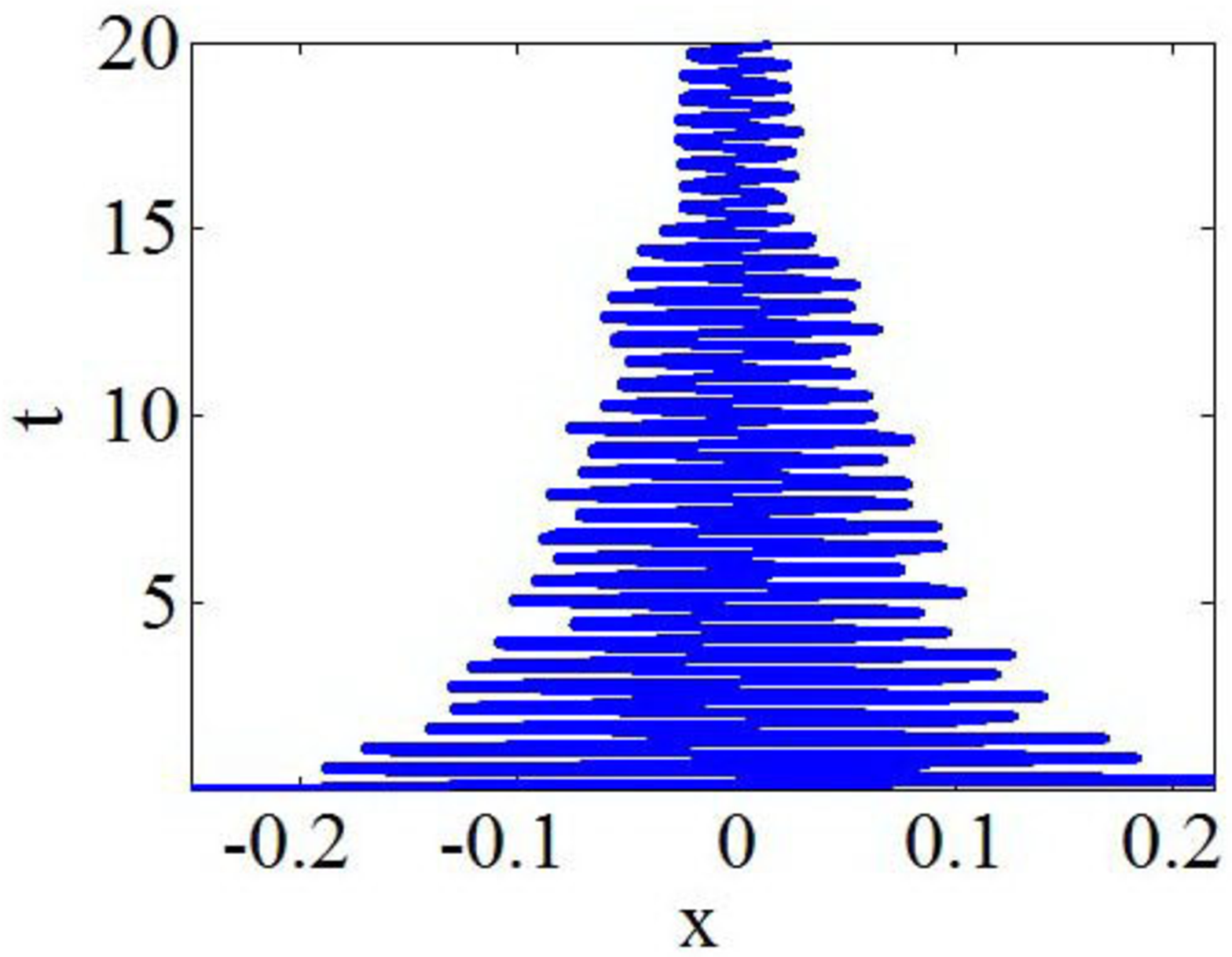}}
\subfigure[] {\label{fig7g}
\includegraphics[scale=0.2]{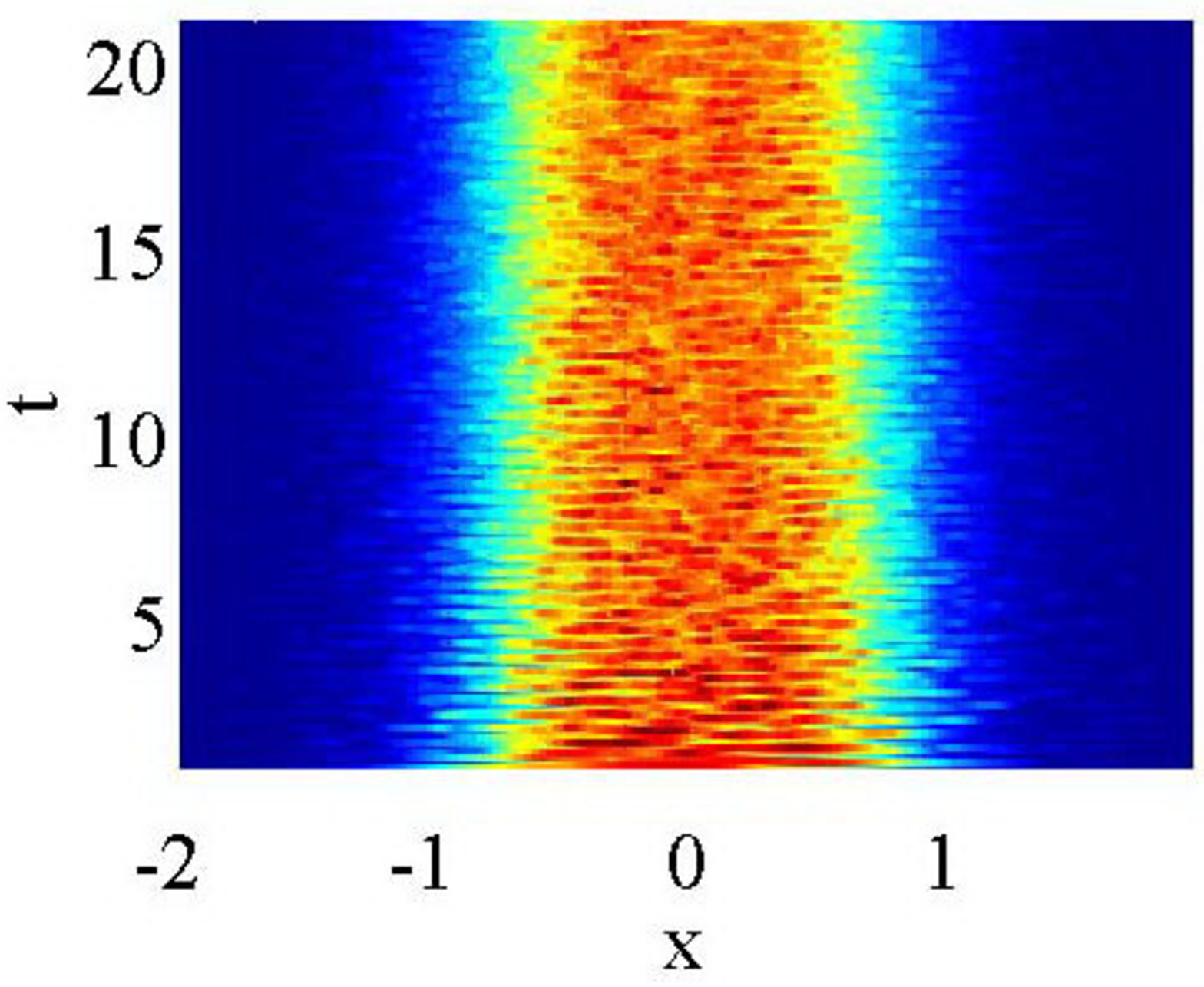}}
\subfigure[] {\label{fig7h}
\includegraphics[scale=0.2]{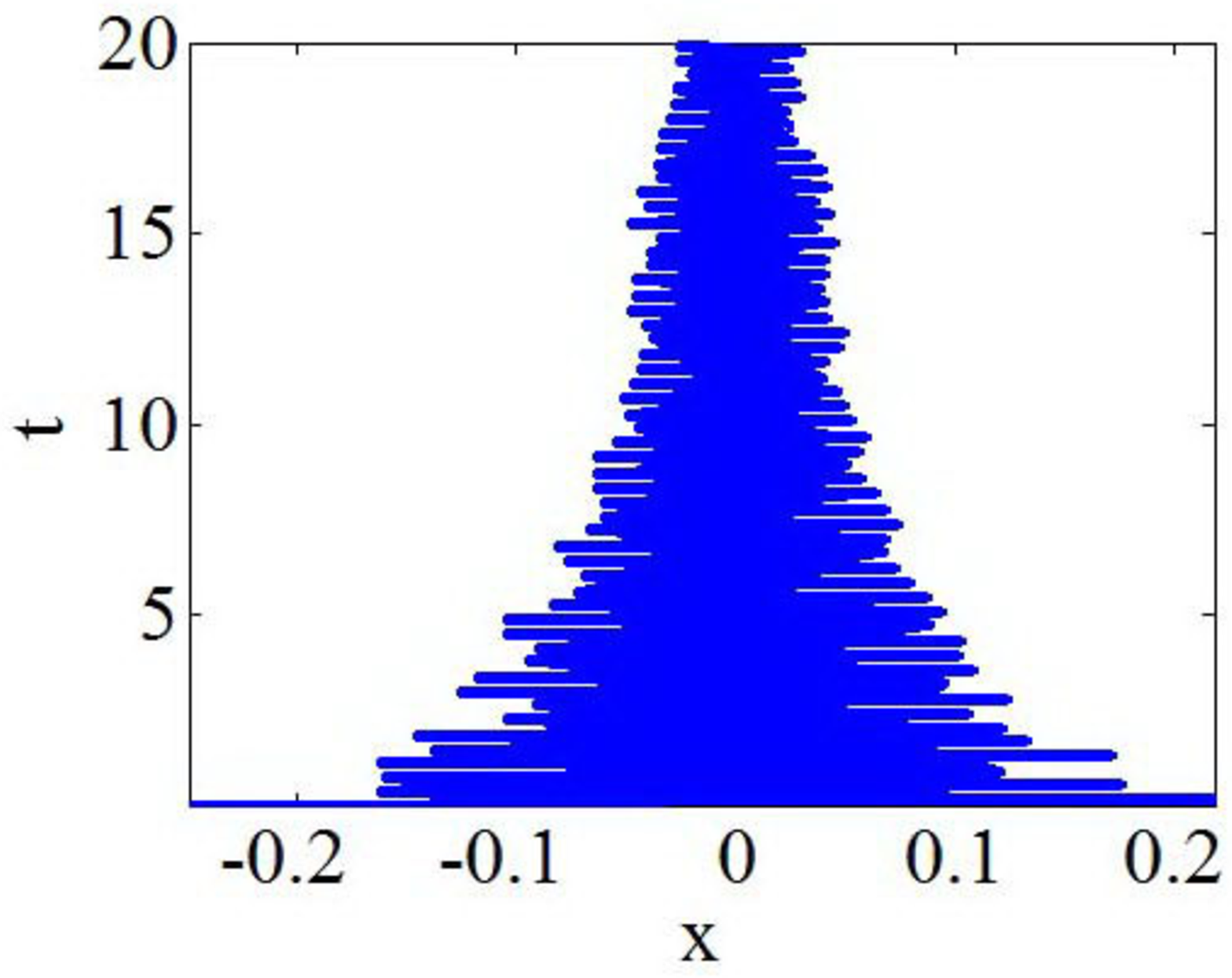}}
\caption{(Color online) The evolution of a 1D fundamental soliton with $P=1$
and $\protect\chi =0$, which was initially shifted off the center by $%
x_{0}=0.25$. Panels (a), (c), (e) and (g) display top views of the evolution
for $g_{0}=0,0.2,0.4$ and $0.8$, respectively. Panels (b), (d), (f) and (h)
are the center-of-mass trajectories of the moving solitons from panels (a),
(c), (e) and (g), respectively, with the center-of-mass coordinate defined
as $\mathrm{x}(t)\equiv P^{-1}\protect\int_{-\infty }^{+\infty }x|\protect%
\psi (x,t)|^{2}dx$.}
\label{1DOscillation}
\end{figure}

\begin{figure}[tbp]
\centering
\subfigure[]{\label{fig8xa}
\includegraphics[scale=0.2]{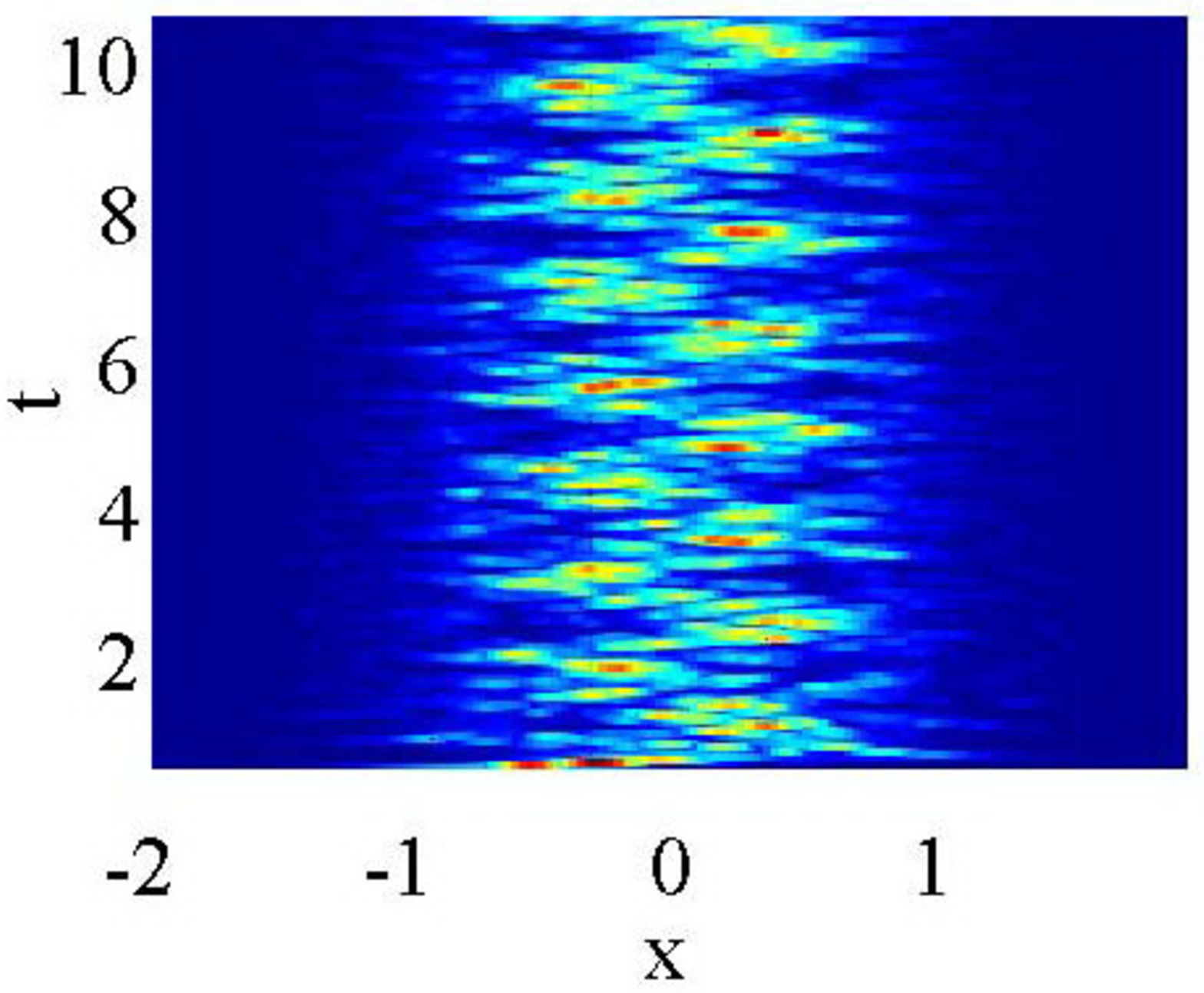}}%
\subfigure[] {\label{fig8xb}
\includegraphics[scale=0.2]{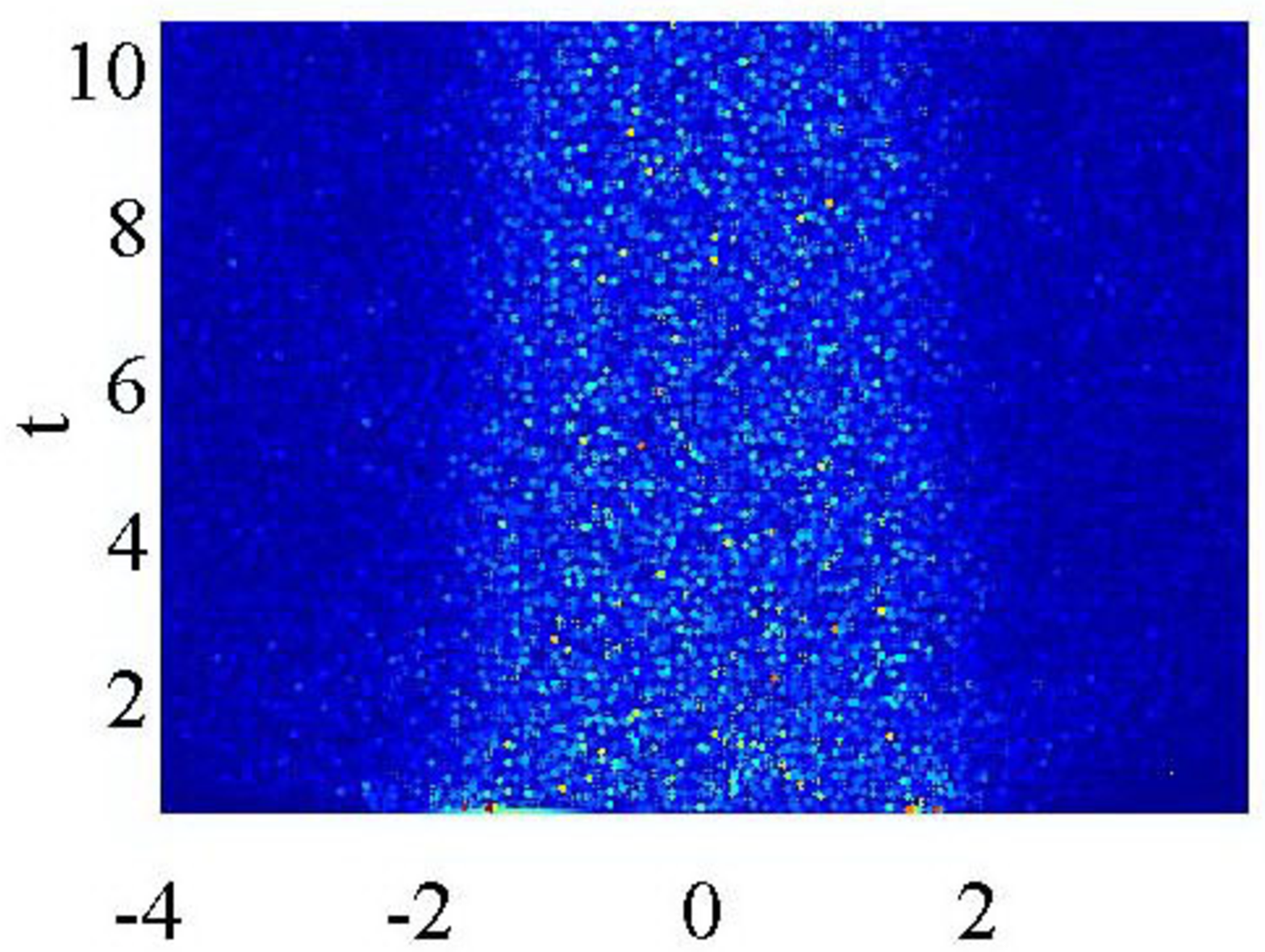}}
\caption{(Color online) (a) The evolution of the 1D fundamental soliton with
$P=1$ and $g_{0}=0$, which was initially shifted off the center by $%
x_{0}=0.625$. (b) Destruction of the soliton following the initial shift by $%
x_{0}=1.5$.}
\label{1DfuzzyOscillation}
\end{figure}

\subsection{The 1D motion: Oscillations and destruction of the soliton}

In the 1D case, the motion of the fundamental soliton was initiated by a
shift ($x_{0}$), which corresponds to initial condition $\psi (x,t=0)=\phi
(x+x_{0})$. Figure \ref{1DOscillation} displays generic examples of the
subsequent evolution of the shifted solitons, for different values of $g_{0}$%
. In Figs. \ref{fig7a} and \ref{fig7b}, pertaining to $g_{0}=0$, the soliton
initially {compresses itself, and }then approximately keeps its shape,
performing undamped, although apparently irregular, oscillations around the
center.{\ In Figs. \ref{fig7c} - \ref{fig7h} corresponding to $g_{0}\neq 0$ (%
$g_{0}=0.2,$ $0.4,$ and $0.8$), the frequency of the oscillatory motion
increases with $g_{0}$, while the amplitude of the center-of-mass
oscillations gradually decreases, and the soliton suffers a slow decay, in
the course of the evolution. The oscillating soliton features a gradual
decay too with the increase of the initial shift, see an example in Fig. \ref%
{fig8xa}. Eventually, the soliton is quickly destroyed if the initial shift
exceed a certain critical value, as shown in Fig. \ref{fig8xb}.}
\begin{figure}[tbp]
\centering{\label{fig8a} \includegraphics[scale=0.51]{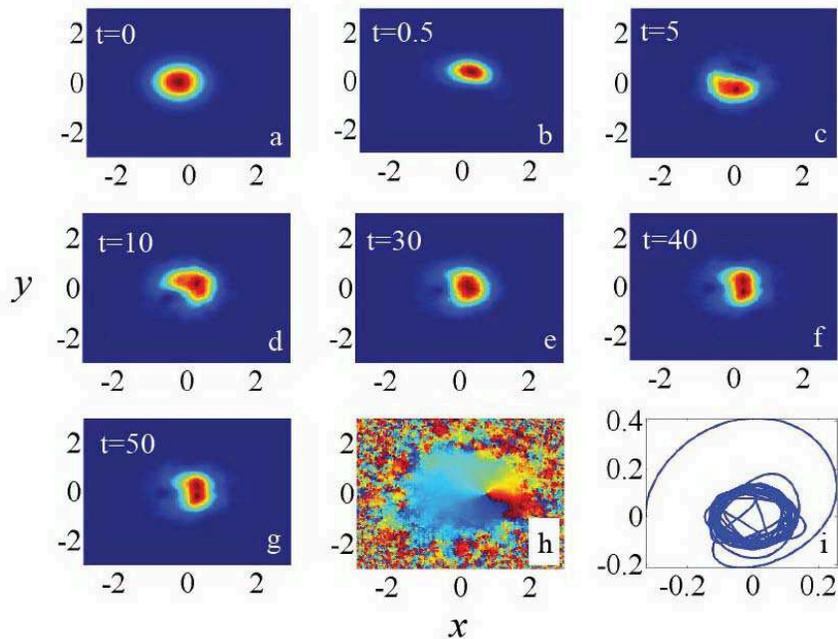}}
\caption{(Color online) The evolution of the 2D fundamental soliton (for $%
P=1 $, $g_{0}=0,~\protect\chi =0$) initially shifted by $x_{0}=0.3$ in the $x
$-direction, and kicked in the $y$-direction by factor $\exp \left( i\protect%
\eta y\right) $, with $\protect\eta =2$. Panels (a)-(g) display snapshots of
the amplitude distribution at indicated moments of time. Panel (h)
additionally displays the phase distribution at the last moment of time
shown, $t=50$. Panel (i) is the trajectory of the soliton's center of mass
over the time interval $0\leq t\leq 50$. The definition of the
center-of-mass' position is $\left\{ \mathrm{x}(t),\mathrm{y}(t)\right\}
=P^{-1}\protect\int \protect\int \left\{ x,y\right\} |\protect\psi %
(x,y,t)|^{2}dxdy$. }
\label{Elliptic1}
\end{figure}
\begin{figure}[tb]
\centering{\label{fig9a} \includegraphics[scale=0.51]{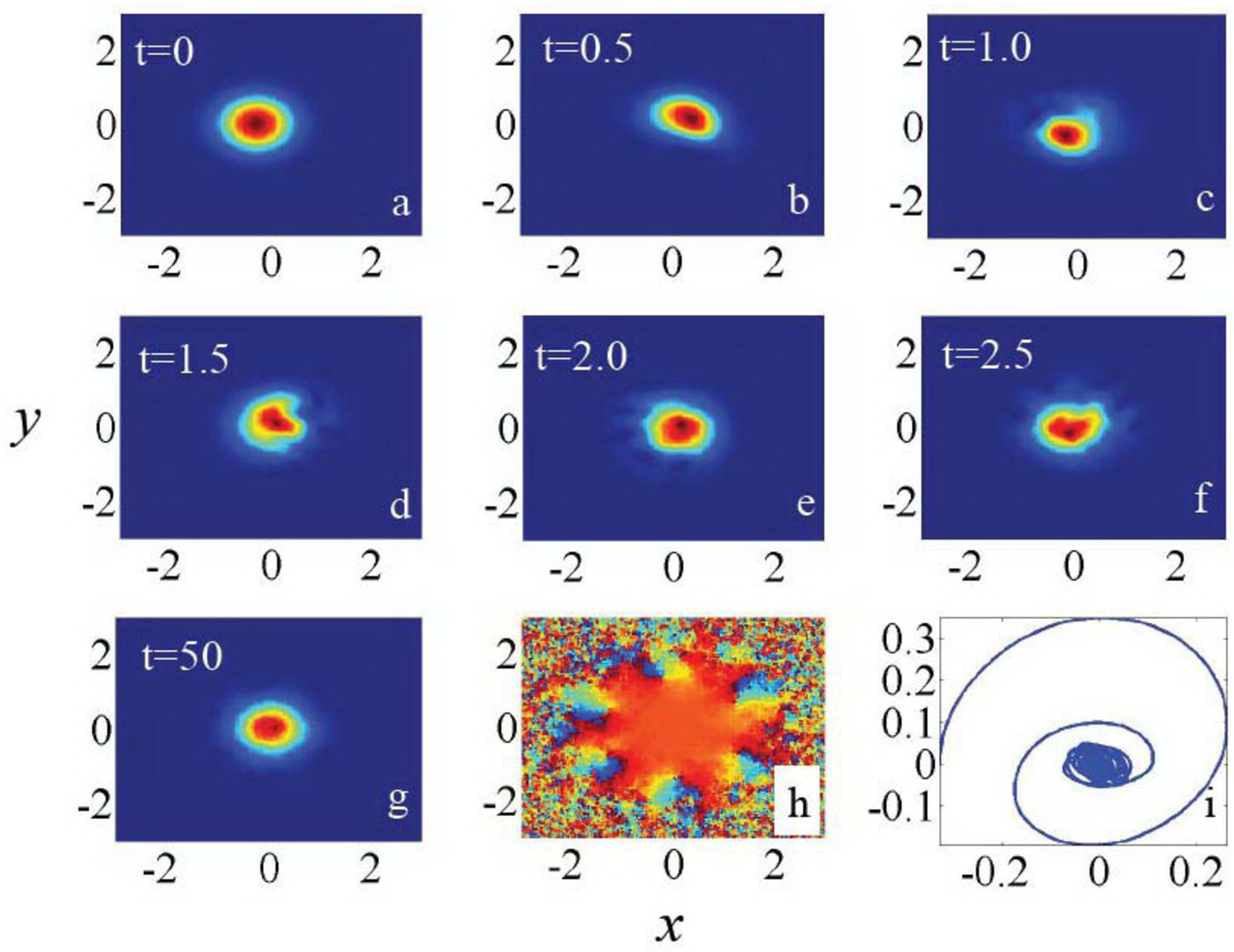}}
\caption{(Color online) The same as in Fig. \protect\ref{Elliptic1}, but for
$g_{0}=0.2$.}
\label{Elliptic2}
\end{figure}

\subsection{The motion of 2D solitons: spiral trajectories}

In the 2D case, one may expect the motion of a soliton, considered as a
quasi-particle, along an elliptic trajectory, which can be initiated by
shifting the soliton from the center along the $x$-direction (by distance $%
x_{0}$), and simultaneously kicking it (in other words, imparting some
velocity, $\eta $) in the $y$-direction, i.e., setting $\psi (\mathbf{r}%
,t=0)=\phi (\mathbf{r}-\mathbf{x}_{0})e^{i\eta y}$. Figures \ref{Elliptic1}
and \ref{Elliptic2} show results of the simulations at different values of $%
x_{0}$ and $\eta $.

Similar to the 1D case, the moving soliton maintains its shape for small $%
x_{0}$ and $\eta $, but splits into fragments if either $x_{0}$ or $\eta $
becomes too large. Therefore, we here discuss in detail only the case of
small $x_{0}$ and $\eta $. Figures \ref{Elliptic1} and \ref{Elliptic2} show
results of such simulations for $g_{0}=0$ and $0.2$.

As well as in the 1D setting, $g_{0}$ strongly affects the motion. In Fig %
\ref{Elliptic1}, panels (a)-(g) show that, at $g_{0}=0$, the soliton keeps
its shape and follows a stable elliptic trajectory for a relatively long
time, see panel \ref{Elliptic1}(i). On the other hand, for $g_{0}=2$ Fig. %
\ref{Elliptic2}(i) demonstrates that the soliton's trajectory is an
inward-winding spiral, rather than a closed ellipse, and in this case the 2D
soliton relatively quickly returns to the center.

The soliton which has returned to the central position maintains \textit{%
differential rotation} in its outer layer, which is necessary to conserve
the angular momentum lent to the system by the initial kick. This vortical
structure exists without appearance of a zero density at the center [see
Fig. \ref{Elliptic2}(g) and \ref{Elliptic2}(h)], which resembles known
regimes of the differential rotation in superfluids, see, e.g., Ref. \cite%
{differential}. 

\section{The double-well nonlinear potential}

A natural generalization of the single-well modulation profile (\ref{d1D})
is a double-well profile. In the 1D setting, it can be defined as
\begin{equation}
g(x)=\left( x^{2}-\sqrt{g_{0}}\right) ^{2},  \label{NDWP}
\end{equation}%
with two minima set at $x=\pm g_{0}^{1/4}$, where $g(x)$ vanishes. An
incentive for the study of this modulation shape is search for a possibility
of the spontaneous symmetry breaking between portions of the mean-field wave
function trapped in the two symmetric nonlinear potential wells, and also a
possibility of Josephson oscillations between them \cite{new-book}. Here we
briefly report results of this analysis performed in the framework of 1D
equation (\ref{GPE}) with $\chi =0$.

\begin{figure}[tbp]
\centering\subfigure[] {\label{fig10a}
\includegraphics[scale=0.3]{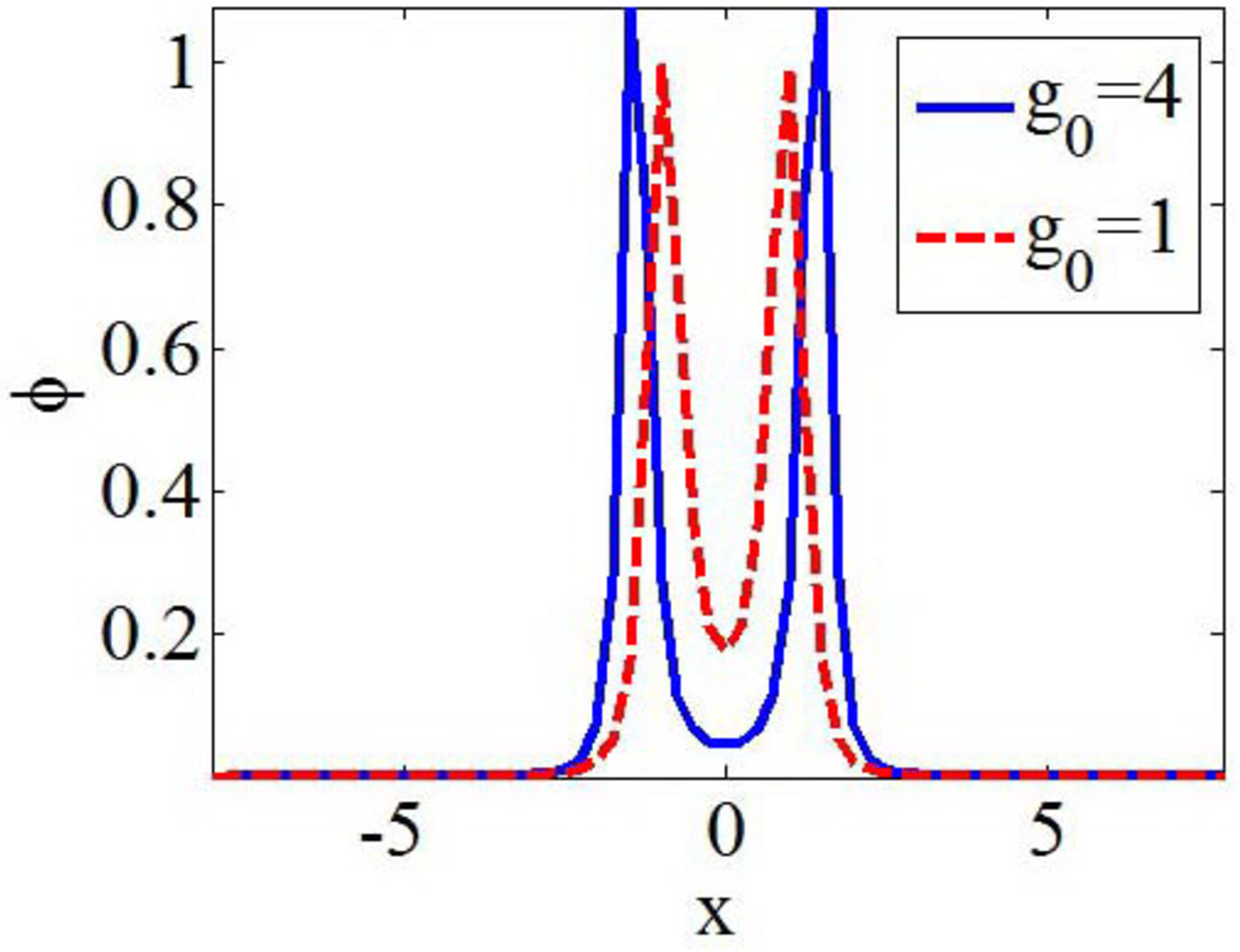}}%
\subfigure[] {\label{fig10b}
\includegraphics[scale=0.4]{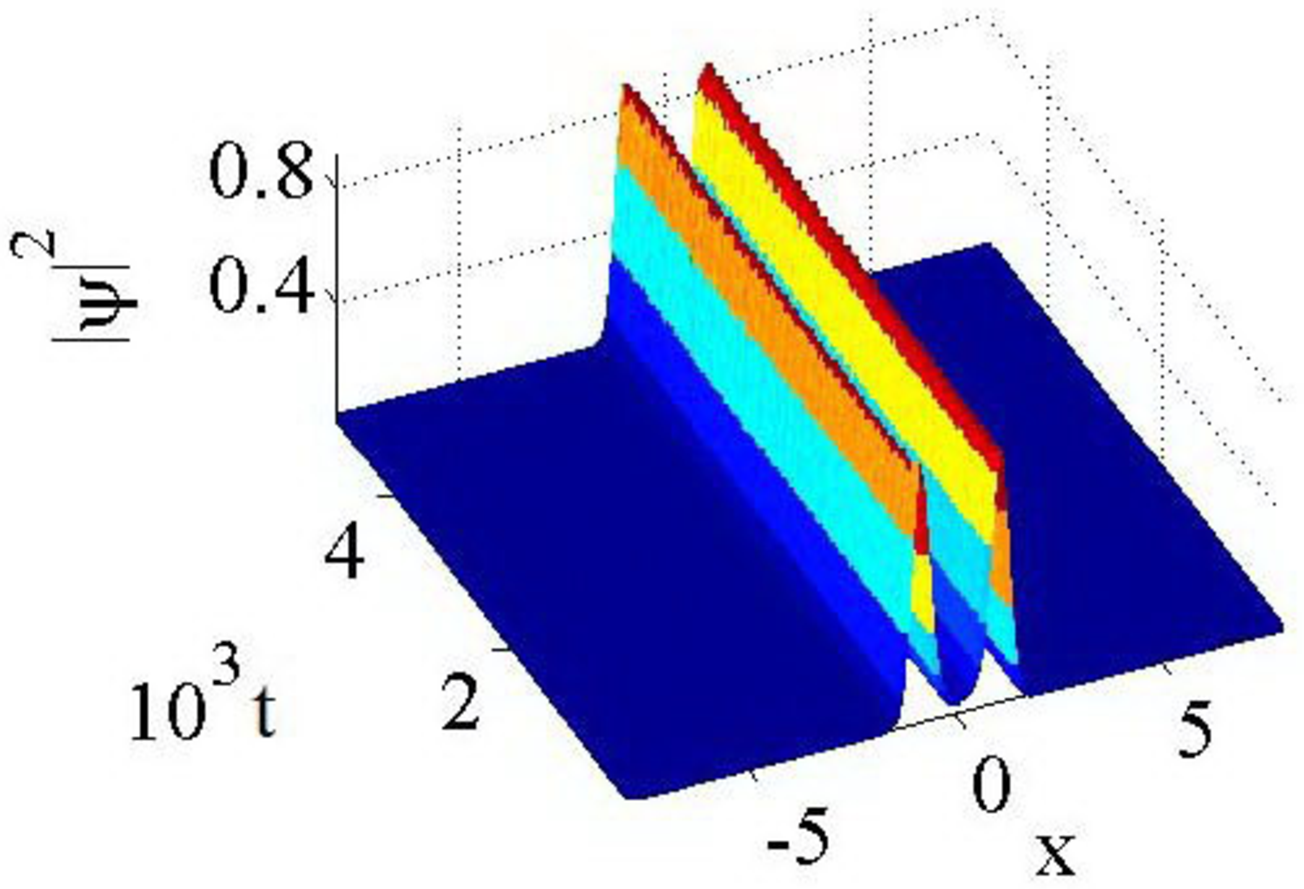}}
\subfigure[] {\label{fig10c}
\includegraphics[scale=0.3]{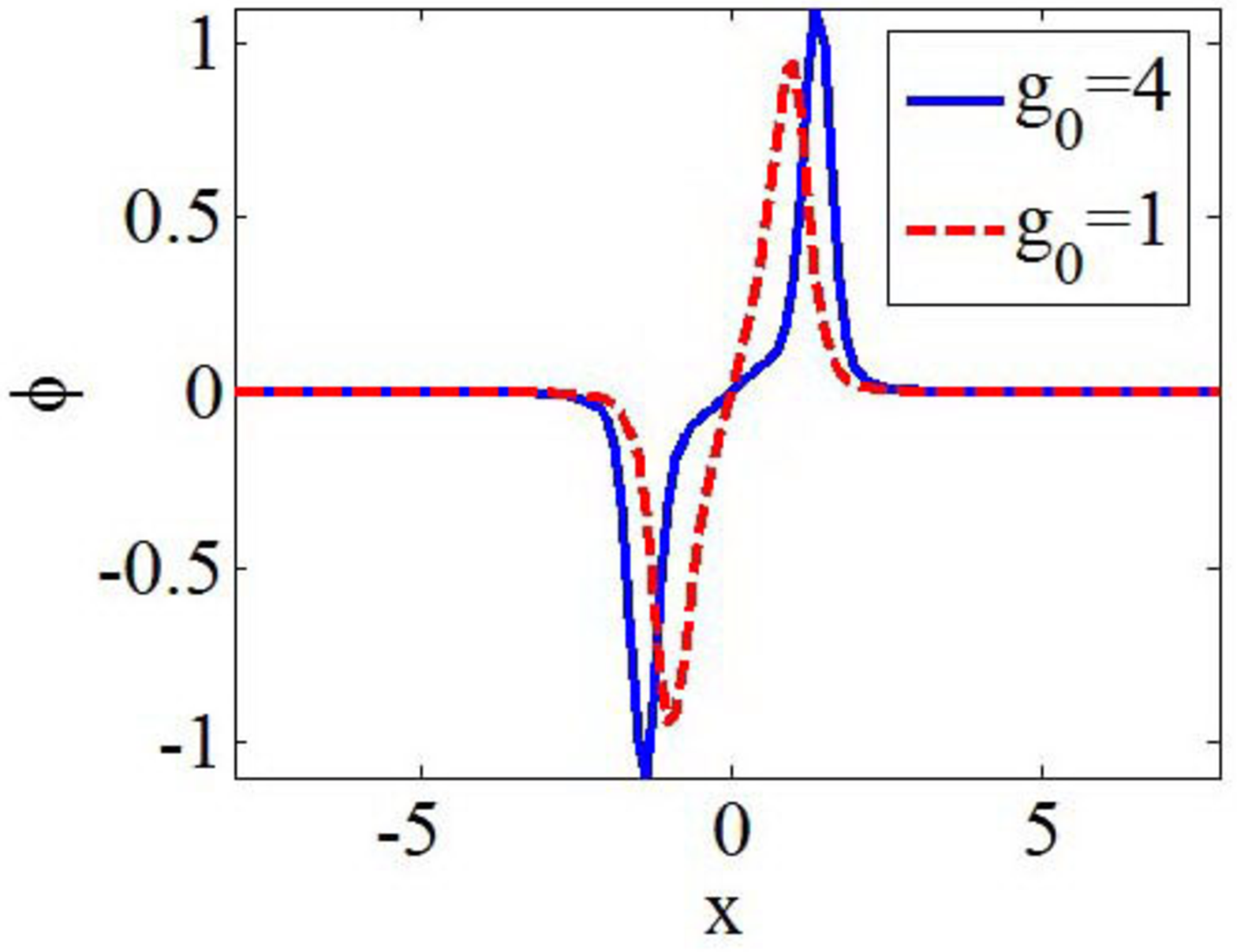}}
\subfigure[] {\label{fig10d}
\includegraphics[scale=0.4]{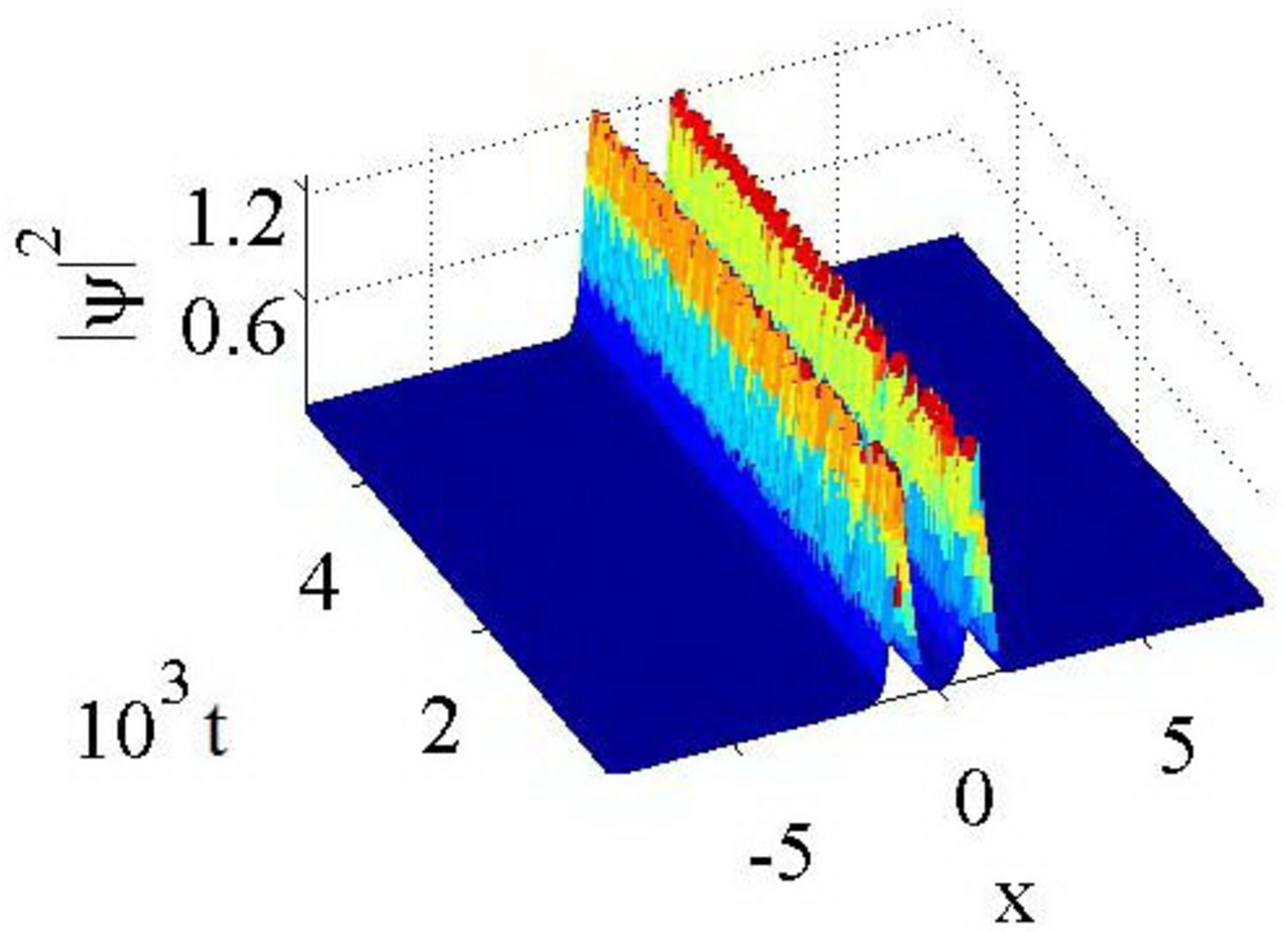}}
\caption{(Color online) (a) Examples of stable even (symmetric) 1D solitons
in the model with the double-well modulation function (\protect\ref{NDWP}),
for $P=1$, $\protect\chi =0$, and $g_{0}=1$ or $4$. (b) Simulations of the
perturbed evolution of the even soliton with $g_{0}=1$. (c) Stable odd
(antisymmetric) solitons with $P=1$ and $g_{0}=1$ or $4$. (d) Simulations of
the perturbed evolution of the odd soliton with $g_{0}=1$.}
\label{DWPEOS}
\end{figure}

\begin{figure}[tbp]
\centering%
\subfigure[] {\label{fig11a}
\includegraphics[scale=0.22]{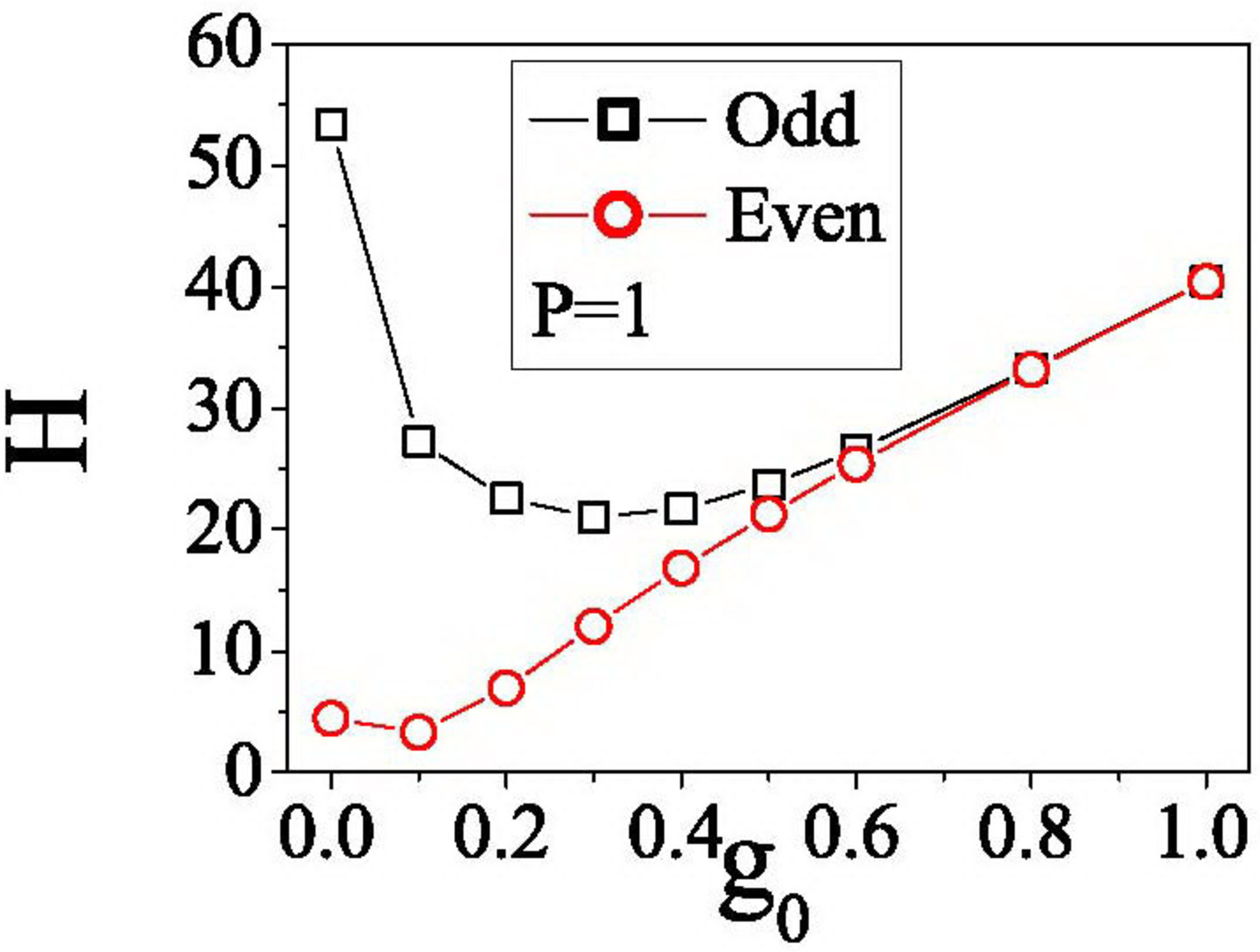}}%
\subfigure[] {\label{fig11b}
\includegraphics[scale=0.22]{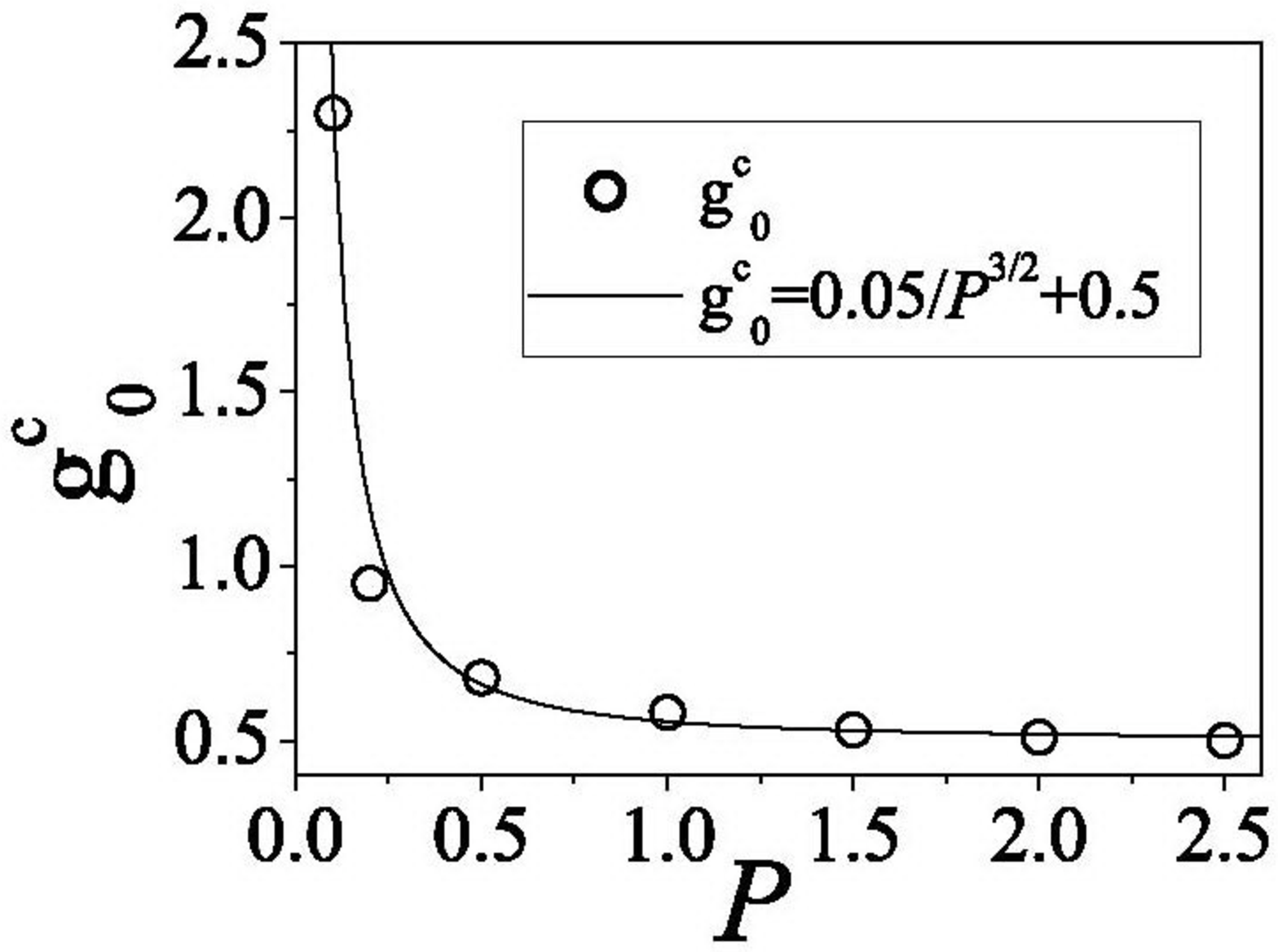}}
\caption{(Color online) (a) The Hamiltonian (energy) of the even and odd 1D
modes, trapped in the double-well modulation profile (\protect\ref{NDWP}),
versus $g_{0}$, for a fixed value of the total norm ($P=1$). (b) The
degeneration point, $g_{0}^{\mathrm{c}}$, at which energies of the even and
odd norms become virtually equal, as a function of $P$. The inset indicates
a fit of the dependence to a power-law approximation.}
\label{fig11}
\end{figure}

Numerical computations yield stable even and odd states trapped in the
double-well modulation profile (\ref{NDWP}). Typical examples of such
localized modes are shown in Fig. \ref{DWPEOS}. The comparison of the
corresponding values of Hamiltonian (\ref{Ham}), as a function of parameter $%
g_{0}$ in Eq. (\ref{NDWP}), is shown in Fig. \ref{fig11}(a). It is concluded
that the even mode represents the ground state, as it corresponds to a
minimum of the Hamiltonian, although the energies of the two states become
practically equal when $g_{0}$ exceeds a certain critical value, $g_{0}^{%
\mathrm{c}}$, see Fig. \ref{fig11}(a). The dependence of $g_{0}^{\mathrm{c}}$
on the total norm $P$ is displayed in Fig. \ref{fig11}(b).

\begin{figure}[tbp]
\centering\subfigure[] {\label{fig12a}
\includegraphics[scale=0.4]{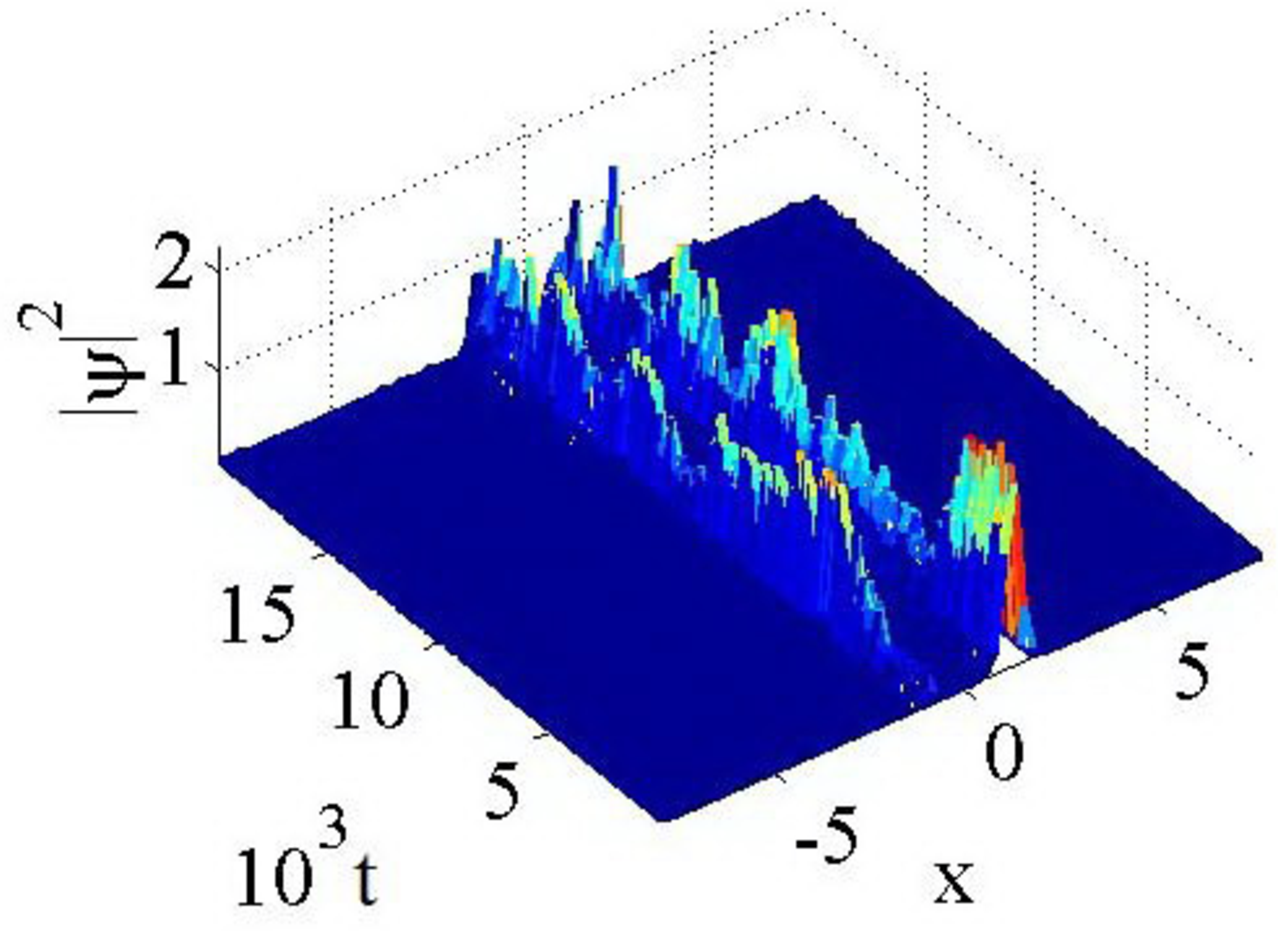}}%
\subfigure[] {\label{fig12b}
\includegraphics[scale=0.3]{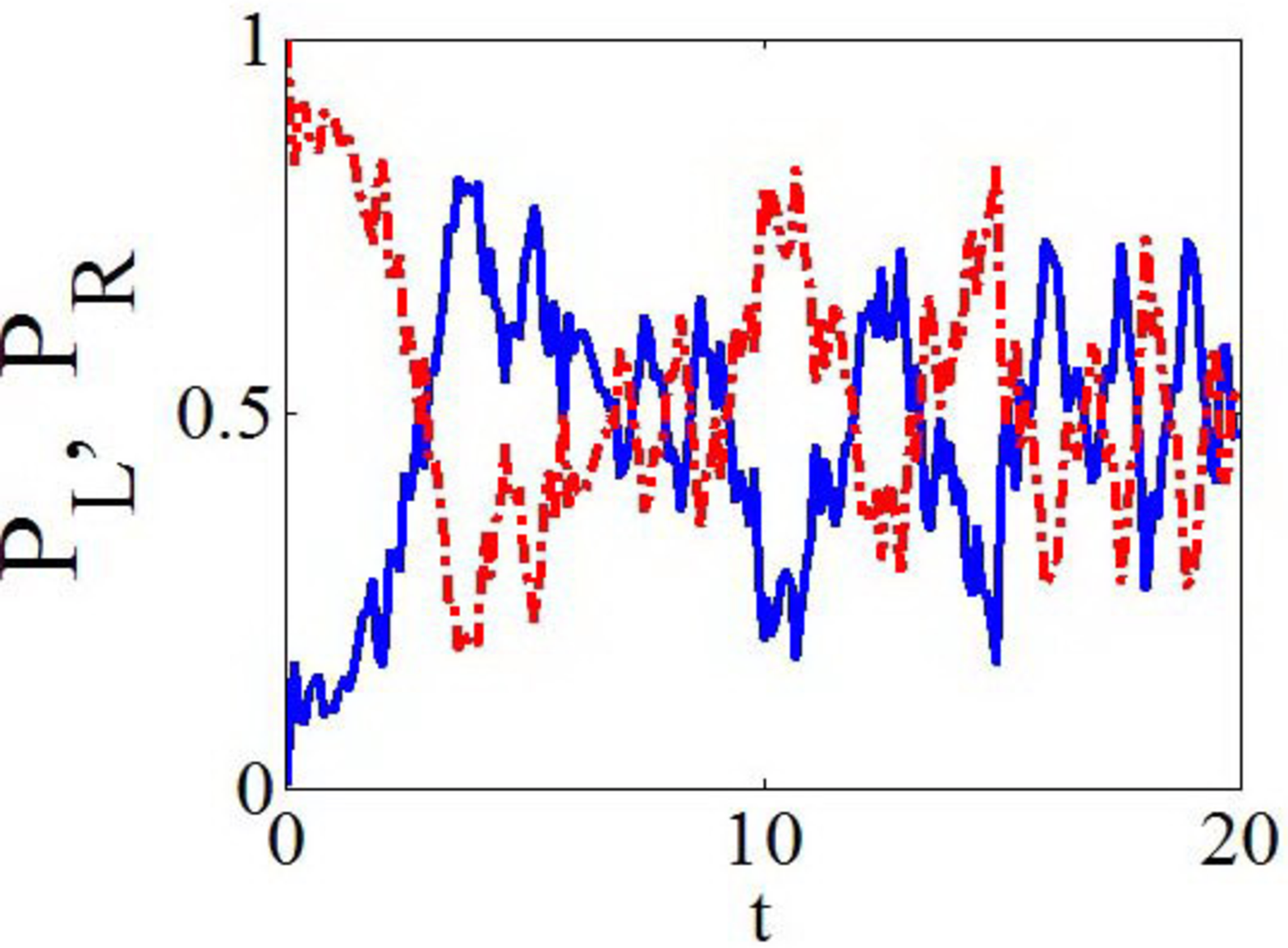}}
\caption{(Color online) Irregular Josephson oscillations, initiated by a 1D
wave packet originally placed around one minimum of the double-well
modulation profile (\protect\ref{NDWP}), with $\protect\chi =0$. (a) Results
of the simulations initialized by $\protect\psi \left( x,t=0\right) =\protect%
\sqrt{2}\mathrm{sech}\left( 4\left( x-g_{0}^{1/4}\right) \right) $, with $%
g_{0}=2$ and total norm $P=1$. (b) The corresponding evolution of the
half-norms, $P_{L}=\protect\int_{-\infty }^{0}|\protect\psi |^{2}dx$ and $%
P_{R}=\protect\int_{0}^{+\infty }|\protect\psi |^{2}dx$. }
\label{1DJO}
\end{figure}

Within the explored parameter region, no stationary states with broken
symmetry (or broken antisymmetry) have been found. On the other hand, robust
but seemingly irregular Josephson oscillations can be readily initiated by
placing the original matter-wave packet into one well. A typical example of
robust oscillations is displayed in Fig. \ref{1DJO}.

\section{Conclusion}

The objective of this work is to explore possibilities for the formation of
1D and 2D fundamental solitons and solitary topological modes in the
condensate of dipoles induced by spatially inhomogeneous polarizing fields.
Under physically relevant conditions, this can be realized as self-trapping
of bright solitons and vortices the under the action of repulsive DDIs
(dipole-dipole interactions) between the induced dipoles. Motivated by the
recent analysis reported for the model with contact repulsive interactions
\cite{Olga}-\cite{B4}, we have demonstrated that this counter-intuitive
result is possible if the local dipole moment of particles, induced by the
external fields perpendicular to the plane in which the condensate is
trapped grows from the center to periphery faster than $r^{3}$, for both
dimensions $D=1$ and $2$ (unlike the local model, where the growth rate must
be faster than $r^{D}$). The setting also includes the EP (expulsive
potential), due to the interaction of the induced dipoles with the
polarizing field. The EP can be eliminated by choosing an appropriate
relation between the dc and ac components of the field. Physical parameters
have been estimated for the realization of the setting by means of the
electric field. For modulation profile (\ref{d}) with $\alpha =4$, families
of fundamental 1D and 2D solitons, 1D dipole modes, and 2D vortices have
been found in a numerical form, and their scaling properties, which are
essentially different from what was found recently in the local models, were
explained analytically. It has also been demonstrated that the 1D and 2D
fundamental solitons remain robust in the presence of the EP, provided that
it is not too strong. The families of the trapped modes considered here are
entirely stable if the EP is eliminated. In addition, the TFA (Thomas-Fermi
approximation) was developed for the 1D and 2D fundamental solitons. The
character of the solitons' confinement is opposite to the character of the
repulsive nonlinearity: in the nonlocal model, the solitons are self-trapped
tightly (super-exponentially), while the self-trapping in the local model is
loose (algebraic).

A fundamental 1D soliton, shifted from the center, performs persistent
oscillations if the initial shift is small enough, while a large shift can
destroy it. Similarly, shifted and transversely kicked 2D solitons may
feature persistent motion along elliptic trajectories. The 1D double-well
modulation profile was considered too. In this case, both symmetric and
antisymmetric trapped modes are dynamically stable, the symmetric ones
realizing the ground state. In addition to that, the double-well structure
readily supports persistent, although irregular, Josephson oscillations
between the wells.

As an extension of this work, it may be interesting to study higher-order
(multipole) modes in the 1D setting, and higher-order vortices in 2D. The
effect of the EP on the twisted and vortical modes may be interesting too.
Another relevant extension may deal with the interplay of the modulated
repulsive DDIs and contact interactions. Furthermore, it should be quite
interesting to study patterns supported by spatially periodic modulations of
the polarizing field, which may be a specific ramification of the general
concept of nonlinear lattices which, thus far, were considered only in local
systems \cite{RMP}, expect for very recent work \cite{Brazil}, where bright
solitons were predicted in the 1D condensate of permanent dipoles under the
external field periodically changing its orientation along the coordinate.

\begin{acknowledgments}
This work was supported by Chinese agency CNNSF (grants No. 11104083,
11204089, 11205063), by the German-Israel Foundation through grant No.
I-1024-2.7/2009, and by the Tel Aviv University in the framework of the
``matching" scheme for a postdoctoral fellowship of Y.L.
\end{acknowledgments}

\end{document}